%% file: main.tex
\documentclass[review]{elsarticle}

\pdfoutput=1
\usepackage{graphicx}
\usepackage{amsfonts, amsmath, amssymb}
\usepackage{booktabs,siunitx}
\usepackage[svgnames,table]{xcolor}
\usepackage[tableposition=above]{caption}
\usepackage{pifont}
\usepackage{bm}
\usepackage{cancel}
\usepackage{caption}
\usepackage{color}
\usepackage{enumerate}
\usepackage{float}
\usepackage{hyperref}
\usepackage[english]{babel}
\usepackage[margin=1in]{geometry}
\usepackage{mathtools}
\usepackage{multirow}
\usepackage{setspace}
\usepackage{subfigure}
\usepackage{lineno}

\renewcommand \d [2]{\frac{{\rm d} #1}{{\rm d} #2}}

\newcommand{\V}[1]{\bm{\mathrm{#1}}}

\def \cH{\mathcal{H}}
\def \cT{\mathcal{T}}
\def \fe{f_\text{e}}
\def \fpto{f_\text{PTO}}
\def \fr{f_\text{r}}
\def \fw{f_\text{w}}
\def \fv{f_\text{v}}

\def \hr{h_\text{r}}
\def \Hr{H_\text{r}}
\def \kr{k_\text{r}}
\def \nr{n_\text{r}}

\def \omegan{\omega_\text{n}}
\def \omegap{\omega_\text{p}}
\def \omegaH{\omega_\text{H}}
\def \omegaL{\omega_\text{L}}
\def \rhow{\rho_\text{w}}
\def \deltap{\Delta_\text{p}}
\def \ap{a_\text{p}}

\definecolor{Gray}{gray}{0.9}

\newcounter{subsubsubsection}[subsubsection]
\def\subsubsubsectionmark#1{}

\def\subsubsubsection{\@startsection
      {subsubsubsection}{4}{\z@} {-3.25ex plus -1
      ex minus -.2ex}{1.5ex plus .2ex}{\normalsize\bf}}
\def\l@subsubsubsection{\@dottedtocline{4}{4.8em}
      {4.2em}}

\newcommand{\upperRomannumeral}[1]{\uppercase\expandafter{\romannumeral#1}}

\begin{document}

\begin{frontmatter}

\title{An adaptive and energy-maximizing control of wave energy converters using extremum-seeking approach}
\author[Polito]{Luca Parrinello}
\author[Polito]{Panagiotis Dafnakis}
\author[Polito]{Giovanni Bracco}
\author[SDSU]{Peiman Naseradinmousavi}
\author[Polito]{Giuliana Mattiazzo}
\author[SDSU]{Amneet Pal Singh Bhalla\corref{mycorrespondingauthor}}
\ead{asbhalla@sdsu.edu}

\address[Polito]{Department of Mechanical and Aerospace Engineering, Politecnico di Torino, Turin, Italy}
\address[SDSU]{Department of Mechanical Engineering, San Diego State University, San Diego, CA, USA}
\cortext[mycorrespondingauthor]{Corresponding author}

\begin{abstract}
\input{Abstract}
\end{abstract}


\end{frontmatter}


\section{Introduction}
\input{Introduction}


\section{Overview of extremum-seeking control and optimization} \label{sec_esc_overview}
\input{ESCOverview}


\section{Extremum-seeking control algorithms} \label{sec_esc_algos}
\input{ESCAlgorithms}

\subsection{Sliding mode extremum-seeking control} \label{sec_smesc_algo}
\input{SlidingMode}

\subsection{Relay extremum-seeking control} \label{sec_relay_algo}
\input{Relay}

\subsection{Least-squares extremum-seeking control} \label{sec_lsq_algo}
\input{LSQ}

\subsection{Self-driving extremum-seeking control} \label{sec_sd_algo}
\input{SelfDriving}

\subsection{Perturbation-based ES}
\input{Perturbation}


\section{Equations of motion} \label{sec_eqn_motion}

In this section we describe the equations of motion of energy-harvesting mechanical oscillators operating in air and in water environments. Although the objective of the current work is to maximize power extraction of point absorber devices using extremum-seeking control, an analytical solution to the former problem is available, which can be used to validate the optimal solution obtained using ES.  

\begin{figure}[]
    \centering
   \subfigure[] {
   		\includegraphics[scale=0.38]{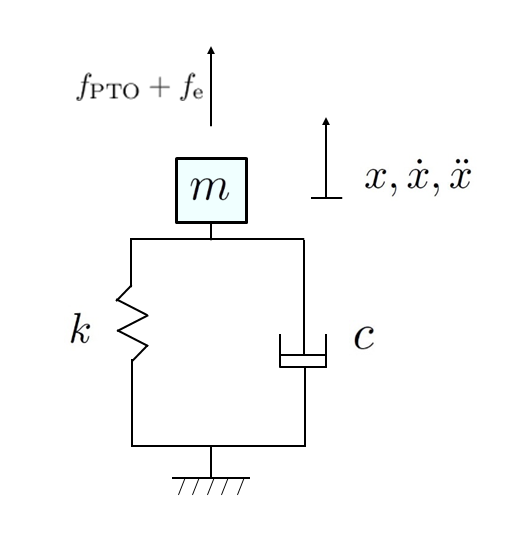}
		\label{fig:msd_scheme}
	}
    	\subfigure[]{
		\includegraphics[scale = 0.38]{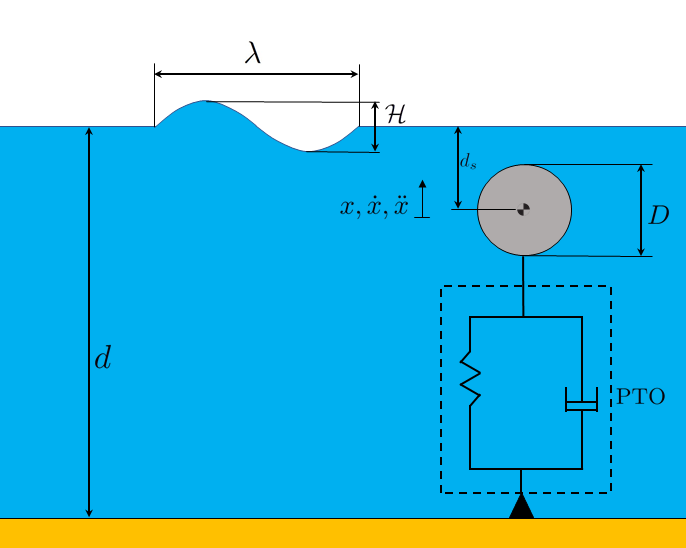}
		\label{fig:PA_scheme}
	}
    \caption{Schematic of an energy-harvesting~\subref{fig:msd_scheme} mass-spring-damper system, and~\subref{fig:PA_scheme} a fully-submerged point absorber wave energy converter.}
    \label{fig:mech_oscillator}
\end{figure}

\subsection{Bodies oscillating in air} \label{sec_msd}
\input{MSDModel}


\subsection{Bodies oscillating in water} \label{sec_pa}
\input{PointAbsorberModel}

\subsubsection{Device and wave characteristics} \label{sec_hull_waves}
\input{PAWaveCharacteristics}


\section{Results} \label{sec_results}
\input{IntroResults}

\subsection{Mass-spring-damper system} \label{sec_msd_results}
\input{MassSpringDamperResults}

\subsection{Cylindrical point absorber} \label{sec_cyl_results}
\subsubsection{ES performance in a regular sea} \label{sec_cyl_reg_results}
\input{CylinderRegularWavesResults}

\subsubsection{ES performance in an irregular sea} \label{sec_cyl_irreg_results}
\input{CylinderIrregularWavesResults}

\subsection{Spherical point absorber} \label{sec_sphere_results}
\input{SphericalPointAbsorberResults}

\section{Conclusions} \label{sec_conclusions}
\input{Conclusions}


\section*{Acknowledgements}
A.P.S.B and P.NAS acknowledge helpful discussions with Miroslav Krsti\'c over the course of this work. A.P.S.B acknowledges Nishant Nangia for carefully reading the manuscript and providing helpful comments. 

\appendix
\renewcommand\thesection{\Alph{section}}
\input{Appendix}

\newpage
\section*{Bibliography}
\begin{flushleft}
 \bibliography{ESCBibliography}
\end{flushleft}

\end{document}

%% file: Abstract.tex
In this paper, we systematically investigate the feasibility of different extremum-seeking (ES) control schemes to improve the conversion efficiency of wave energy converters (WECs). Continuous-time and model-free ES schemes based on the sliding mode, relay, least-squares gradient, self-driving, and perturbation-based methods are used to improve the mean extracted power of a heaving point absorber subject to regular and irregular waves. This objective is achieved by optimizing the resistive and reactive coefficients of the power take-off (PTO) mechanism using the ES approach. The optimization results are verified against analytical solutions and the extremum of reference-to-output maps. The numerical results demonstrate that except for the self-driving ES algorithm, the other four ES schemes reliably converge for the two-parameter optimization problem, whereas the former is more suitable for optimizing a single-parameter. The results also show that for an irregular sea state, the sliding mode and perturbation-based ES schemes have better convergence to the optimum, in comparison to other ES schemes considered here. The convergence of PTO coefficients towards the performance-optimal values are tested for widely different initial values, in order to avoid bias towards the extremum. We also demonstrate the adaptive capability of ES control by considering a case in which the ES controller adapts to the new extremum automatically amidst changes in the simulated wave conditions. Although not demonstrated here, a continuous-time and model-free ES could possibly be used within a nonlinear computational fluid dynamics framework, where typically evolution-based optimization strategies are used for performing black-box optimization. As demonstrated in our results, ES achieves an optimum within a single simulation, whereas evolutionary strategies typically require a large number of (possibly expensive) function evaluations. 

%% file: Introduction.tex
Renewable energy harvesting technologies have made tremendous progress over the last several decades, which are enabling us to reduce our current carbon footprint of energy production and consumption. In particular, technologies based on wind and solar power are now sufficiently mature and economically viable to be deployed at commercial and utility scales.  In contrast, wave energy conversion has yet to achieve a level of commercial success like solar and wind technologies, in spite of the concerted research efforts dating back since the early 1970s after the oil crisis~\cite{Evans1979}. 

The economy of scale model demonstrated for solar and wind farms, motivates future commercial wave farms. For example, the cost of a photovoltaic module dropped from \$66.1/W in 1976 to \$0.62/W in 2016. Similarly, the levelised cost of electricity generated from wind has significantly decreased over the years, with prices ranging from \$0.55/kWh in 1980 to \$0.05/kWh in 2012~\cite{DOE2013}. It is estimated that $2.11 \pm 0.05$ TW of wave energy is available globally~\cite{Gunn12}.  Moreover, wave power density is extremely high compared to wind and solar; compare 25 kW/m of crest width for wave energy against 1 kW/m$^2$ at peak insolation for solar energy or at wind speed of 12 m/s for wind energy~\cite{czech2012wave}. In spite of this, wave power is the most underutilized renewable energy resource. Nevertheless, significant progress has been made in the design and analysis of wave energy converter (WEC) devices, which convert the mechanical energy of the waves to electrical energy through a power take-off (PTO) system. 

One of the challenging aspects of making wave energy commercially profitable is designing an optimal control for a WEC device that maximizes its mean extracted power. Consequently, several optimal control formulations have been proposed in order to improve energy extraction from WECs. An extensive review on this topic can be found in Ringwood et al.~\cite{ringwood2014energy,ringwood2014control}. One such optimal control formulation is the model-free extremum-seeking (ES) method, which can be applied to both linear and nonlinear systems. ES is an adaptive control which tracks a maximum/minimum (extremum)  of a performance/cost function and then drives the output of this function to its extremum~\cite{krstic2000stability}. ES control has been used for a variety of applications, including but not limited to, reducing thermo-acoustic instabilities in gas turbines and rocket engines~\cite{banaszuk2000adaptive}, flight formation optimization~\cite{binetti2003formation}, control of thermo-acoustic coolers~\cite{li2005extremum}, autonomous vehicles~\cite{zhang2007extremum}, and robots~\cite{pnas2018}, and beam matching in particle accelerators~\cite{luo2009mixing}. ES control has also been widely used for wind~\cite{munteanu2009wind,shen2016sliding,hu2019sliding}, and  solar power applications~\cite{brunton2009solar,zazo2012mppt,kebir2017extremum}. However, only a limited number of ES studies are available for wave energy in the literature~\cite{garcia2012optimization,hals2011comparison}. The aim of the current study is to fill this gap by testing different ES control algorithms and demonstrating their feasibility for WECs.

ES control was conceived at the beginning of the twentieth century by Leblanc~\cite{leblanc1922}. The method first received considerable attention in the USSR in the 1940s~\cite{kazakevich1944extremum}, and then in the Western world in the 1950s and 1960s.
Although ES was one of the first forms of adaptive control, it was not until 2000 that a proof of stability for a generic plant was provided by Krsti\'c and Wang~\cite{krstic2000stability}. Soon after, many applications and variants of the algorithm followed in the literature. The first application of ES for WECs appeared in 2011 by Hals et al.~\cite{hals2011comparison}, in which various control strategies, including tuning of controller parameters using perturbation-based ES were compared through simulation results. Hals et al. defined a performance function based on low pass filters and knowledge of wave excitation forces for tuning linear damping or the threshold value of latching, depending on the controller.  The authors compared the controller parameters tuned through gain scheduling and ES strategies in their work~\cite{hals2011comparison}. In 2012, Garcia-Rosa et al.~\cite{garcia2012optimization} used a discrete-time ES scheme to obtain performance-optimal PTO coefficients of a hyperbaric point absorber converter. In their simulation results, both reactive and resistive coefficients were simultaneously optimized, without requiring the knowledge of wave excitation forces. Similar to Hals et al., the authors in~\cite{garcia2012optimization} also used a perturbation-based ES control.   

In this work, we use continuous-time ES control algorithms to optimize the resistive and reactive PTO coefficients for a heaving point absorber subject to regular and irregular waves. In particular, we test the sliding mode~\cite{pan2003stability}, relay~\cite{olalla2007analysis}, least-squares gradient~\cite{hunnekens2014dither}, self-driving~\cite{haring2016extremum}, and perturbation-based~\cite{krstic2000stability} ES schemes. The optimization results are verified against analytical solutions and the extremum of reference-to-output maps. The numerical results show that except for the self-driving ES algorithm, the other four ES schemes reliably converge for the two-parameter optimization problem, whereas the former is more suitable for optimizing a single-parameter. The results also show that for an irregular sea state, the sliding mode and perturbation-based ES schemes have better convergence to the optimum, in comparison to other ES schemes. The convergence of parameters towards the performance-optimal values are tested for widely different initial values, in order to avoid bias towards the extremum. We also demonstrate the adaptive capability of ES control by considering a case in which the ES controller adapts to the new extremum automatically amidst changes in the simulated wave conditions. 

The rest of the paper is organized as follows. We begin by stating the assumptions of  ES formulation in Sec.~\ref{sec_esc_overview}. Next, in Sec.~\ref{sec_esc_algos}, we define the performance function for a general energy-harvesting mechanical oscillator, and provide a brief overview and working principle of each ES scheme. The equations of motion for bodies oscillating in air and in water are described in Sec.~\ref{sec_eqn_motion}.  For some cases, analytical solution to the performance-optimal PTO coefficients are provided. The ES results for a simple mass-spring-damper system oscillating in air, and for cylindrical and spherical buoys heaving in regular and  irregular waves are provided in Sec.~\ref{sec_results}. Finally, conclusions are drawn in Sec.~\ref{sec_conclusions}.

%% file: ESCOverview.tex
Extremum-seeking (ES) control is an adaptive optimization technique that derives and maintains the input and output of the controlled plant to their respective extrema without requiring an explicit knowledge of the plant dynamics. ES control
can be applied to both linear and nonlinear systems, in which the extremum of a performance function is achieved and 
maintained by obtaining the gradient information with respect to the control inputs.  For the purpose of description, we 
consider a single-input single-output (SISO) nonlinear system with the following characteristics:

\begin{enumerate}
\item  an unknown dynamical plant $\dot{x} = f (x,u)$, with $x \in \mathbb{R}^n$ and $u \in \mathbb{R}$; 

\item  a performance function $J = h(x)$, with $J \in \mathbb{R}$; and  

\item  a state-feedback control law $u = \alpha (x,\vartheta )$, with $\vartheta \in \mathbb{R}$.

\begin{figure}[]
\centering
\includegraphics[scale=0.4]{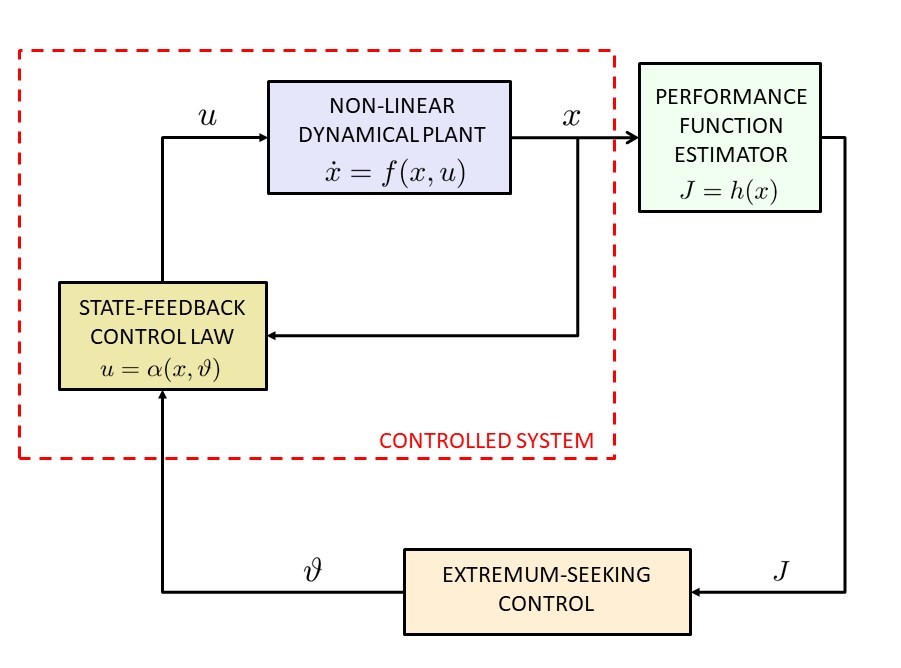}
\caption{Extremum-seeking control scheme for a general SISO nonlinear system~\cite{ariyur2003real,zhang2011extremum}.}
\label{fig:General_System_scheme}
\end{figure}

\end{enumerate}
\noindent Here, $x$ represents the state of the plant, $u$ is the control input, and $J$ is the performance (cost) metric/function which needs to be maximized (minimized) over a period of time. The function $f\left(x,u\right)$ governing the evolution of the plant dynamics is not explicitly required for the ES algorithm. Fig.~\ref{fig:General_System_scheme} shows a schematic representation of the plant with an embedded ES controller. The following assumptions are made for the extremum-seeking controlled plant~\cite{ariyur2003real,van2012extremum,haring2013extremum, pan2003stability,skafte2017introduction}:

\begin{itemize}
\item  \textbf{Assumption 1:} The control law $u=\alpha (x,\vartheta )$ is smooth, parametrized by $\vartheta$, and stabilizes the plant. 

\item  \textbf{Assumption 2:} There exists a smooth function $x_\text{eq}(\vartheta)$, such that

\begin{equation}
\dot{x} = f(x,\alpha (\vartheta ,x))=0\ \leftrightarrow x=x_\text{eq}(\vartheta). \label{eq_plant_equil}
\end{equation}

\item \textbf{ Assumption 3:} The functions $f:\mathbb{R}^n \times \mathbb{R}\longrightarrow \mathbb{R}^n$ and $h:\mathbb{R}^n\longrightarrow \mathbb{R}$ are smooth.

\item \textbf{ Assumption 4:} The static performance behavior of the system at the equilibrium point $x_\text{eq}(\vartheta)$ can be expressed by

\begin{equation}
J_\text{eq} = h (x_\text{eq}(\vartheta) ) = F(\vartheta),
\end{equation}
in which, $F(\vartheta)$ is a smooth function and admits a unique maximum or minimum at $\vartheta ={\vartheta }^*$.

\item \textbf{ Assumption 5:} The parameter $\vartheta$ evolves much slowly than the dynamics of the plant; the latter is assumed to be in equilibrium (Eq.~\ref{eq_plant_equil}). Thus, we can omit the dynamics of the plant while analyzing ES control. Given these operating conditions, we are aiming towards achieving a steady-state optimization of the plant. It then follows that the time derivative of the performance function of the \emph{stabilized plant} can be expressed as

\begin{equation}
    \dot{J}= \d{J}{t} = \d{J}{\vartheta} \d{\vartheta}{t} = \d{F(\vartheta)}{\vartheta} \dot{\vartheta}. 
\end{equation}

\end{itemize}

Under the same set of aforementioned assumptions, it is also possible to extend the ES analysis to a multiple-input single-output (MISO) system, in which $\vartheta \in \mathbb{R}^m$. The objective here is to optimize $m$ parameters simultaneously. For stability analysis of ES control of SISO system, we refer the readers to  Krsti{\'c} and Wang~\cite{krstic2000stability}, and of MISO system, the readers are referred to~\cite{rotea2000analysis,walsh2000application,teel2001solving,ariyur2002multivariable,azar2009real,zhang2007extremum,pnas2018}.

%% file: ESCAlgorithms.tex
In this work, we use \emph{model-free} ES (also referred to as \emph{black-box} ES) algorithms to enhance the power absorption
of periodically oscillating systems, such as wave energy converter devices. The ES algorithms used in this work belong to the class of \emph{derivative-based} optimization methods, which aim to determine the optimal value of the performance function $J$ by estimating the derivative d$J$/d$\vartheta$ to obtain the value $\vartheta^*$  that maximizes (minimizes) the performance (cost) function. Extremum-seeking control problems are typically formulated as unconstrained optimization problems, as done in this work and in prior works that used ES for WECs~\cite{garcia2012optimization,hals2011comparison}. In order to include inequality constraints on the plant parameters, the performance/cost function can be augmented with penalty functions~\cite{dehaan2005extremum,guay2015constrained}. However, constraints on the state variables and performance indicators cannot be imposed in general,  because the state of the plant and an explicit relationship between the plant parameters and performance indicators is not known a priori. Since a model-free ES algorithm is oblivious to the underlying system dynamics, its success depends upon the definition of the performance function, which the control designer is free to define. For energy harvesting systems, a natural choice of performance function is the amount of energy absorbed by the device over a period of time. An ES algorithm then finds the maxima of this concave (performance) function with respect to the PTO parameters. 

The performance/cost function defined for an ES algorithm should be time-invariant, and remain constant, if the parameter $\vartheta$ remains constant~\cite{haring2016extremum}. During the steady-state operation of a WEC device, the absorbed energy is a time-varying periodic function with a non-zero mean. Therefore, it is important to work with mean powers rather than instantaneous powers in the performance function, in order to satisfy Assumption 5 given in Sec.~\ref{sec_esc_overview}. The natural frequency of a typical point absorber device is higher than the wave frequency. This observation can be used to define the mean of the absorbed power over a couple of wave periods. There are two advantages of using this definition: (i) firstly, it ensures that the device dynamics are much faster than the performance function variation; and (ii) secondly, ES control remains adaptive in presence of changing wave conditions. The adaptive capability of ES is maintained because typically the wave climate changes over several hundred wave periods, while the performance function is (re-)defined over the last few (wave) periods.      

\begin{figure}[]
\centering
\includegraphics[scale=0.35]{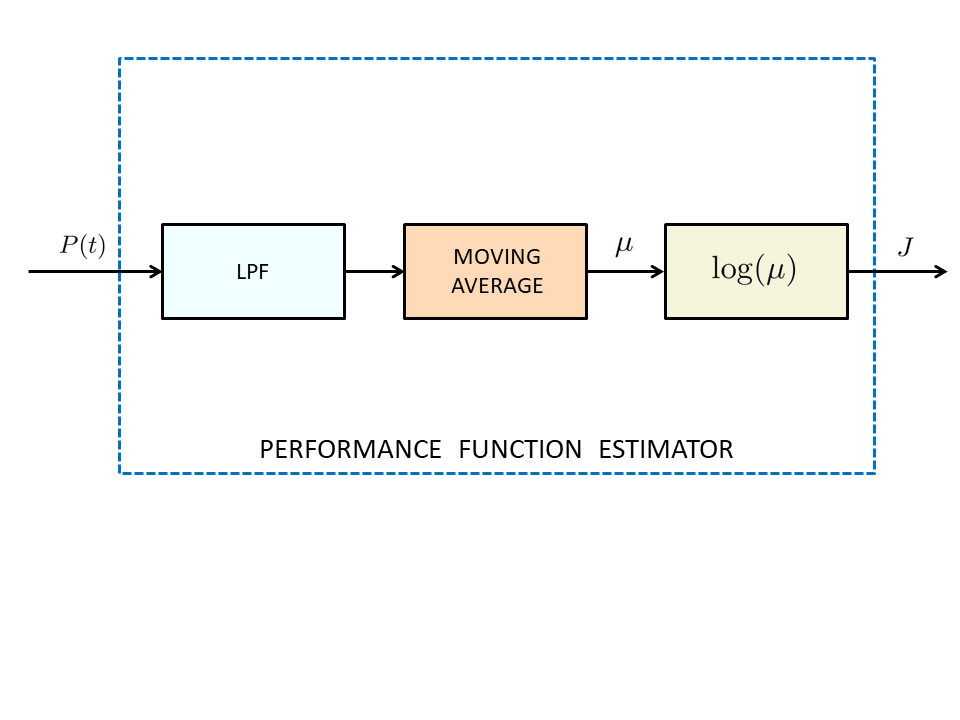}
\caption{Block-diagram scheme to obtain the performance function $J$ for ES algorithms using instantaneous power  $P(t)$ absorbed by the PTO unit. A first-order low-pass filter of the form $\frac{\omega_\text{L}}{s+\omega_\text{L}}$ is used, in which $\omega_\text{L}$ is the cut-off frequency.}
\label{fig:PF_scheme}
\end{figure}

Specifically in this work, the performance function $J$ for ES algorithms is defined to be a composite function of the instantaneous power $P(t)$ absorbed by the PTO unit as shown in Fig.~\ref{fig:PF_scheme}. The composition consists of (i) a low-pass filter (LPF); (ii) a moving average defined over last few wave periods; and (ii) a logarithmic function. Depending upon the type of the PTO machinery, the instantaneous power $P(t)$ could be only resistive in nature, or it could have both resistive and reactive power components. Irrespective of the PTO type, the power signal $P(t)$ is characterized by two main frequency contents: the first, which is at a higher value, is characterized by the fast dynamics of the plant itself, and the second, at a lower value, is characterized by the slow variation of the parameter $\vartheta$ provided by ES.  The purpose of the low-pass filter is to eliminate the higher frequency content from the performance metric, while that of moving average is to evaluate the steady-state performance of the plant (see Haring~\cite{haring2016extremum} for more discussion). The signal $\mu$ which is the output of the first two operators, can be used to assess the device performance as a function of $\vartheta$ parameter.  However, the plant or the device could be operating under wide variability of conditions (such as changing wave heights or periods), and the resulting power can vary in different orders of magnitude. As suggested by Ciri et al.~\cite{ciri2019evaluation}, the purpose of the logarithmic function is to limit the variation of the performance metric drastically, which avoids re-tuning ES hyper-parameters for changing operating conditions of the plant.
 
Next, we discuss and describe different ES algorithms used in this paper, and make some recommendations on the hyper-parameters of the ES algorithms.

%% file: SlidingMode.tex
\begin{figure}[]
\centering
\includegraphics[scale=0.4]{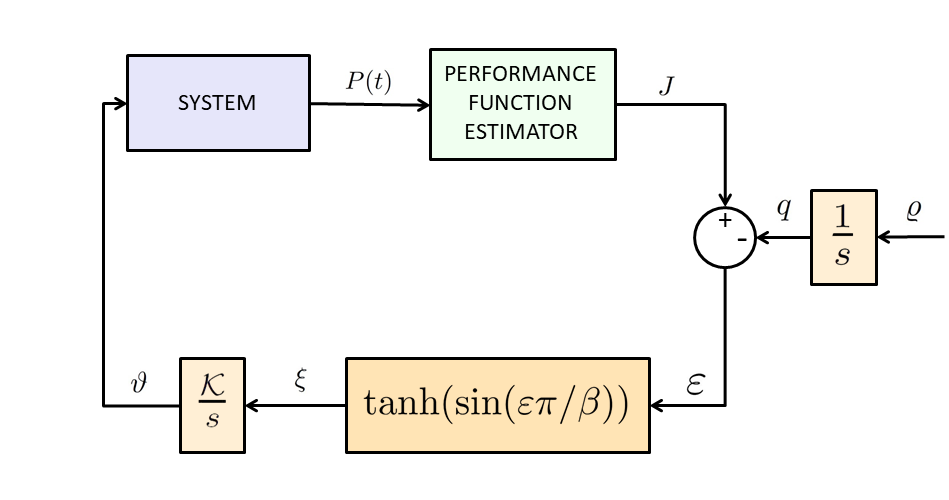}
\caption{Block diagram of sliding mode extremum-seeking control system~\cite{pan2003stability}.}
\label{fig:SM_scheme}
\end{figure}

The basic idea of the sliding mode extremum seeking control (SM-ES) is to make the performance function $J$ follow an increasing function  
of time $q(t)$, irrespective of the unknown gradient d$J$/d$\vartheta$. The error signal $\varepsilon = J(\vartheta) - q(t)$ is then kept at a constant value by a proper choice of $\dot{\vartheta}$ given by

\begin{align*}
	\dot{\varepsilon} & =  \d{J}{\vartheta} \dot{\vartheta} - \dot{q}(t), \\
	\dot{\vartheta}     & = \mathcal{K} \tanh \left( \sin \left(\varepsilon \pi /\beta \right) \right), \quad \mathcal{K} > 0.
\end{align*}

Fig.~\ref{fig:SM_scheme} shows a block diagram of sliding mode ES control system. Equivalently, the SM-ES can be described by the following set of equations:
\begin{align}     \label{eqn_smesc}
\text{SM-ES} & = 
\begin{cases}
    \dot{q}=\varrho, & \varrho > 0  \\ 
    \varepsilon =J-q, \\ 
    \xi = \tanh \left( \sin \left(\varepsilon \pi /\beta \right) \right), & \beta > 0 \\
    \dot{\vartheta }=\mathcal{K}\xi. & \mathcal{K} > 0
\end{cases}    
\end{align}
For more details on SM-ES, including its stability analysis and recommendations for tuning its hyper-parameters, we refer the readers to~\cite{pan2003stability,olalla2007analysis,chen2017switching}. Below we list some key recommendations followed in this work:
\begin{itemize}
    \item The ratio $\beta \mathcal{K}/2\varrho$ affects the convergence rate to attain the performance-optimum value $\vartheta^*$. A value too small slows down the convergence rate, whereas an extremely large value can be detrimental for the system stability. 
    
    \item $\varrho/\mathcal{K}$ and $\beta$ have to be chosen small enough to ensure the sliding modes.
    
    \item To satisfy the Assumption 5 of Sec.~\ref{sec_esc_overview}, i.e., the variation of the parameter $\vartheta$ needs to be much slower than the dynamics of the plant, the hyper-parameters $\mathcal{K}$ and $\varrho$ need to be smaller than $\beta$. Our empirical tests suggest that the ratio $\beta/\varrho$ should be approximately equal to $10^2$ to satisfy this condition.
\end{itemize}

%% file: Relay.tex
\begin{figure}[h]
\centering
\includegraphics[scale=0.4]{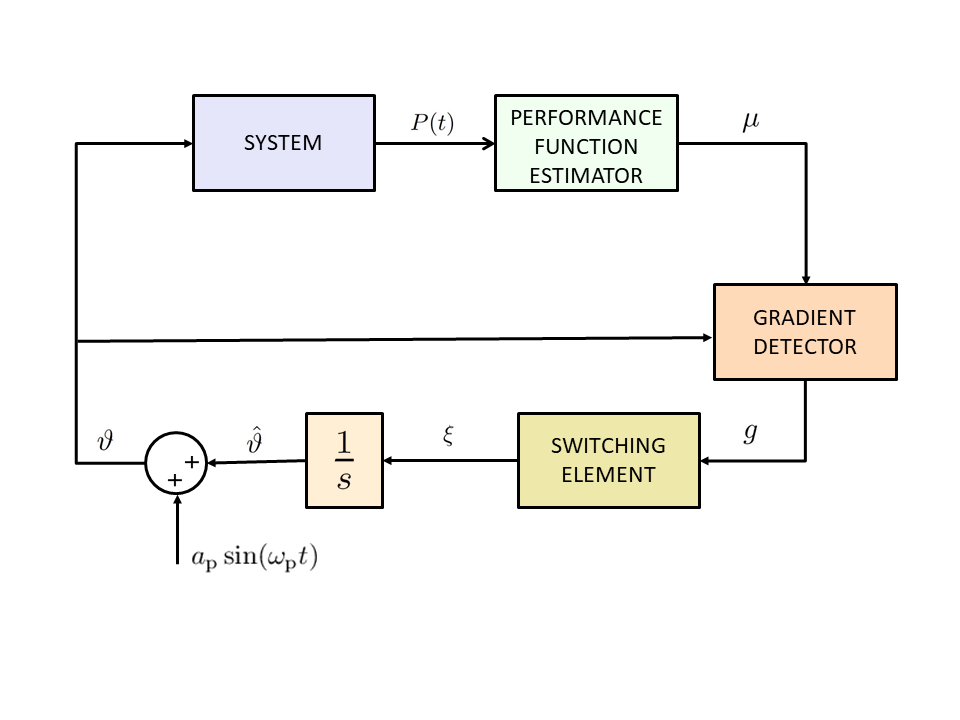}
\caption{Block diagram of relay extremum-seeking control system~\cite{olalla2007analysis}.}
\label{fig:Relay_scheme}
\end{figure}

\begin{figure}[]
\centering
\includegraphics[scale=0.4]{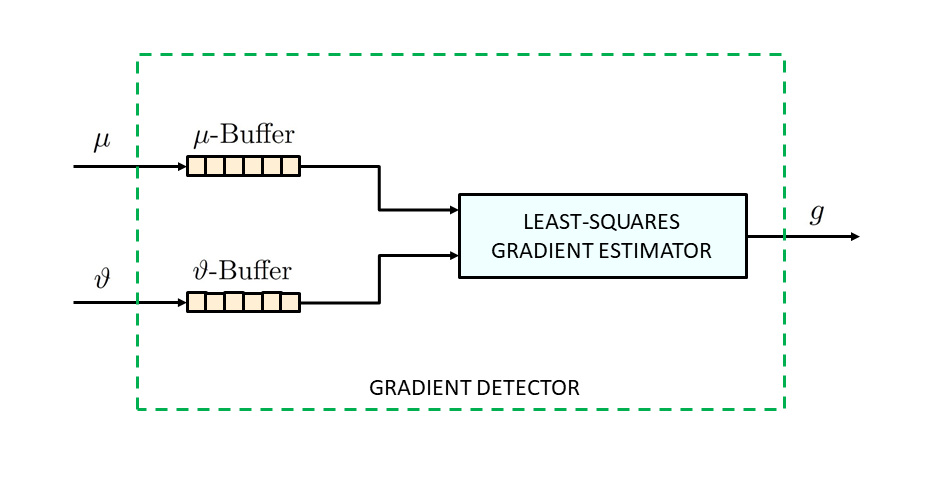}
\caption{Least-squares gradient estimation for relay ES.}
\label{fig:Relay_gradient}
\end{figure}

A relay extremum-seeking control~\cite{olalla2007analysis} provides an estimation of plant optimum based on the sign of the 
gradient. This feature is particularly useful when the controller is applied to a plant operating under widely varying conditions, which can cause the gradient value to vary in different orders of magnitude, as discussed in Sec.~\ref{sec_esc_algos}. In this case, using a logarithmic function in the functional composition of the performance metric is redundant, and the performance metric $\mu$ (instead of $J$) can be used directly for the relay ES. However, a drawback of discarding the gradient magnitude is that the parameter $\vartheta$ keeps oscillating in the neighborhood of the optimal value $\vartheta^*$, where the gradient norm is $|\text{d}\mu/\text{d}\vartheta|_{\vartheta^*} \approx 0$. Nevertheless, acceptable oscillations can be achieved with a proper tuning of the relay ES hyper-parameters~\cite{olalla2007analysis}. Fig.~\ref{fig:Relay_scheme} shows a block diagram of relay ES control system, which is described by the following set of equations:

\begin{align} \label{eqn_relay_esc}
\text{Relay ES} & = 
\begin{cases}
    g = \frac{\delta\mu}{\delta\vartheta},  \\ 
    \xi =\xi_0 \; \textrm{signum}(g), & \xi_0 > 0 \\ 
    \dot{\hat{\vartheta }} = \xi,   \\
    \vartheta = \hat{\vartheta} + \ap \sin(\omegap t), & \ap > 0
 \end{cases}
\end{align}
in which, $\delta \mu/\delta\vartheta$ is the estimation of the gradient $\text{d}\mu/\text{d}\vartheta$ obtained using a least-squares gradient estimation method, and the quantity $\deltap(t) = \ap \sin(\omegap t)$ is a small perturbation/dither added to the parameter $\vartheta$ in order to avoid  numerical issues related to least-squares gradient estimation. The gradient $g$ estimation is performed by generating two buffers of length $n_{\text{buff}}$, which contain the last $n_{\text{buff}}$ values of performance metric $\mu$ and parameter $\vartheta$, as shown in Fig.~\ref{fig:Relay_gradient}. Given these two buffers, it is possible to estimate $\text{d}\mu/\text{d}\vartheta$ through least-squares method, as explained in Hunnekens et al.~\cite{hunnekens2014dither}. A proof of stability for relay ES is provided in Leyva et al.~\cite{leyva2006mppt}.

Since the gradient estimation of the performance function is performed without a dither signal (as done in
classical \emph{perturb and observe} ES methods), the convergence speed of relay ES does not depend on the time scale of the external perturbation signal, which potentially allows for a faster convergence of relay ES systems. 

In our empirical testing, we observed that adopting an excessively small buffer size makes the controller more reactive, but also much more sensitive to the selection of hyper-parameters. On the contrary, using large buffer size makes the controller less reactive, and leads to a poor estimation of the gradient. The size of the buffer also depends on the time step size adopted for acquiring data;  if the step size is extremely small, the buffers need to be larger and vice-a-versa. In our tests, we maintain a buffer size sufficient enough to store the history of system performance over the last few (wave) periods. We remark that for the algorithm to converge during the initial phase of the simulation, we initialize $\vartheta$-buffer with some non-constant values, and keep the controller inactive. The $\mu$-buffer is filled as the simulation proceeds forward within this initial phase. 

The hyper-parameters $\ap$ and $\omegap$ of the sinusoidal perturbation, should be such that the time scale of the perturbation is larger than the time scale of the plant; $\omegap$ must be chosen small enough for this reason. The perturbation amplitude $\ap$ is also kept small to preserve the stability of the ES algorithm, as well as to reduce the oscillations around the performance-optimal value $\vartheta^*$. In contrast, if $\ap$ value is too small, then inaccuracies stemming from least-squares gradient estimation can cause a numerical instability.

The hyper-parameter $\xi_0$ is selected large enough to grant an appreciable variation of the parameter $\vartheta$, which accelerates the convergence of the algorithm. In contrast, an extremely large value of $\xi_0$ causes excessive oscillations, and  possibly numerical instabilities. Hence, a suitable value should be found empirically.

%% file: LSQ.tex
\begin{figure} []
    \centering
    \includegraphics[scale=0.4]{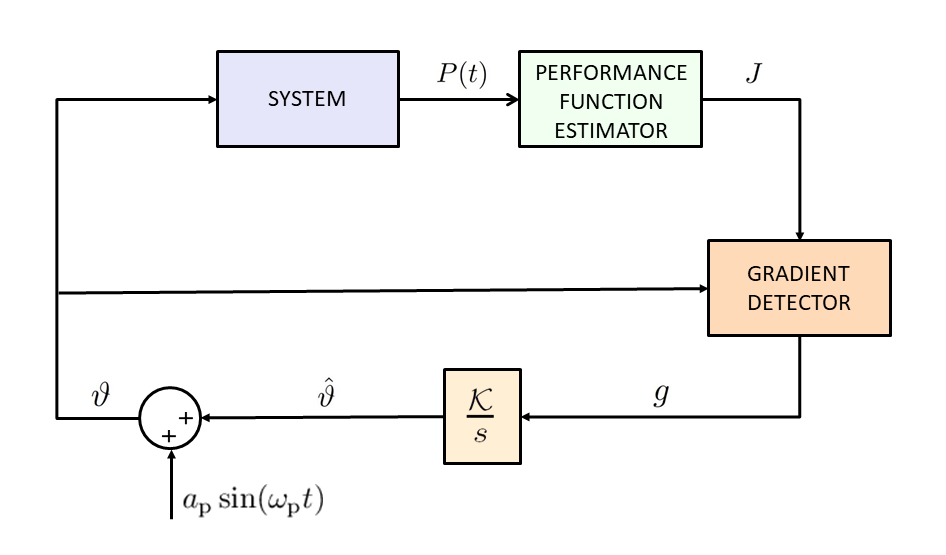}
    \caption{Block diagram of least-squares gradient estimation-based extremum-control system~\cite{hunnekens2014dither}.}
    \label{fig:lsq_scheme}
\end{figure}

The  least-squares gradient estimation-based extremum-seeking control algorithm (LSQ-ES) is an extension of the relay ES, in which both the sign and magnitude of the gradient d$J$/d$\vartheta$ is used to derive $\vartheta$ to its performance-optimal value. A block diagram of LSQ-ES control system is shown in Fig.~\ref{fig:lsq_scheme}. The algorithm is equivalently described by the following set of equations:

\begin{align} \label{eqn_lsq_esc}
\text{LSQ-ES} & =
\begin{cases}
    g = \frac{\delta J}{\delta\vartheta},  \\ 
    \dot{\hat{\vartheta }} = \mathcal{K} g,  & \mathcal{K} > 0 \\
    \vartheta = \hat{\vartheta} + \ap \sin(\omegap t), & \ap > 0
\end{cases}    
\end{align}
in which, $\delta J/\delta \vartheta$ is the least-squares gradient estimation of the performance function d$J$/d$\vartheta$, as described in Hunnekens et al.~\cite{hunnekens2014dither}. Here, we use $J$ instead of $\mu$ as a performance metric, for reasons discussed previously. 

The stability analysis of LSQ-ES performed in~\cite{hunnekens2014dither}, shows that the parameter $\mathcal{K}T$ affects the algorithmic 
performance; $\mathcal{K}T$ should be chosen small enough to guarantee the convergence and stability of the algorithm. Here, the hyper-parameter $\mathcal{K}$ is the gain of the integrator as written in Eq.~\eqref{eqn_lsq_esc}, and $T$ is the time period over which the input and output signals are stored in the data buffers.  Therefore, $\mathcal{K}$ should be adjusted according to the duration of the time period $T$. As LSQ-ES is based on the relay ES algorithm, the remaining hyper-parameters should be selected as per Sec.~\ref{sec_relay_algo}.

%% file: SelfDriving.tex
\begin{figure}[]
    \centering
    \includegraphics[scale=0.4]{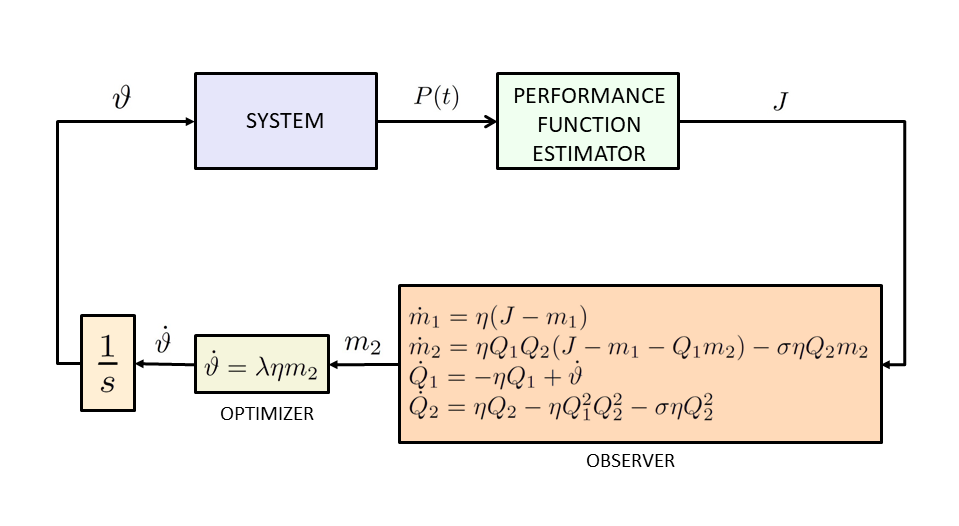}
    \caption{Block diagram of self-driving extremum-seeking control system~\cite{haring2016extremum}.}
    \label{fig:Self_driving_scheme}
\end{figure}

A self-driving extremum-seeking control scheme, like the sliding mode extremum-seeking control, does not require perturbations to estimate the gradient of the performance function. As no perturbations are used, the algorithm avoids the time scale associated with the perturbations, 
which may potentially allow for a faster convergence towards the optimum. Although self-driving systems were part of ES algorithms surveyed by Sternby~\cite{Sternby80} in 1980, they have since not gained much popularity compared to the perturbation-based ES methods. However, lately they are receiving a renewed attention in the literature; see Haring~\cite{haring2016extremum} and references therein. Here, we follow Haring to present a block diagram of a self-driving ES control system, as shown in Fig.~\ref{fig:Self_driving_scheme}. The controller can be equivalently described by the following set of equations:

\begin{align} \label{eqn_sd_esc}
     \text{Self-driving ES} & = 
 \begin{cases}
    \dot{m}_1=\eta (J-m_1), & \eta > 0  \\ 
    \dot{m}_2 =\eta Q_1 Q_2 (J-m_1-Q_1m_2)-\sigma\eta Q_2 m_2, & \sigma \ge 0 \\ 
    \dot{Q}_1=-\eta Q_1+\dot{\vartheta}, \\
    \dot{Q}_2 =\eta Q_2 -\eta Q_1^2 Q_2^2 -\sigma \eta Q_2^2, \\
    \dot{\vartheta}=\lambda \eta m_2, & \lambda > 0
\end{cases}
\end{align}

Succinctly, the \emph{observer} block estimates the gradient of the performance function, $m_2 \approx$ d$J$/d$\vartheta$, and the \emph{optimizer} block steers the parameter $\vartheta$ towards the performance-optimal value $\vartheta^*$.  As discussed in Haring~\cite{haring2016extremum}, if $m_2(t = 0)$ and $Q_1( t = 0)$ are equal to zero, the parameter $\vartheta$ will remain constant, and not converge to the optimal value. In our simulations, we initialize  $m_2(0)$ and $Q_1( 0)$ with $\mathcal{O}(1)$ values, and evolve them for a sufficiently long period of time, during which phase the controller is kept inactive. Also, to accelerate the convergence rate, higher values of the hyper-parameters $\lambda$ and $\eta$ should be chosen. However, if these are excessively high, stability issues may occur. Numerical instabilities may also arise due to high values of $Q_2$ during the simulation; it is recommended to regularize $Q_2$ through a non-zero $\sigma$ value as shown in the set of Eqs.~\eqref{eqn_sd_esc}. Since larger values of $\sigma$ can compromise the accuracy of the performance-optimal $\vartheta^*$, we set $\sigma \approx 10^{-11}$ in our simulations. 

Through our empirical tests done for the cases studied in this work, we find that the self-driving algorithm as is described in this section does not converge well for MISO systems. However, for SISO systems, the convergence towards the optimum solution is reasonably fast, and free of  oscillations.

%% file: Perturbation.tex
\begin{figure}[]
    \centering
    \includegraphics[scale=0.4]{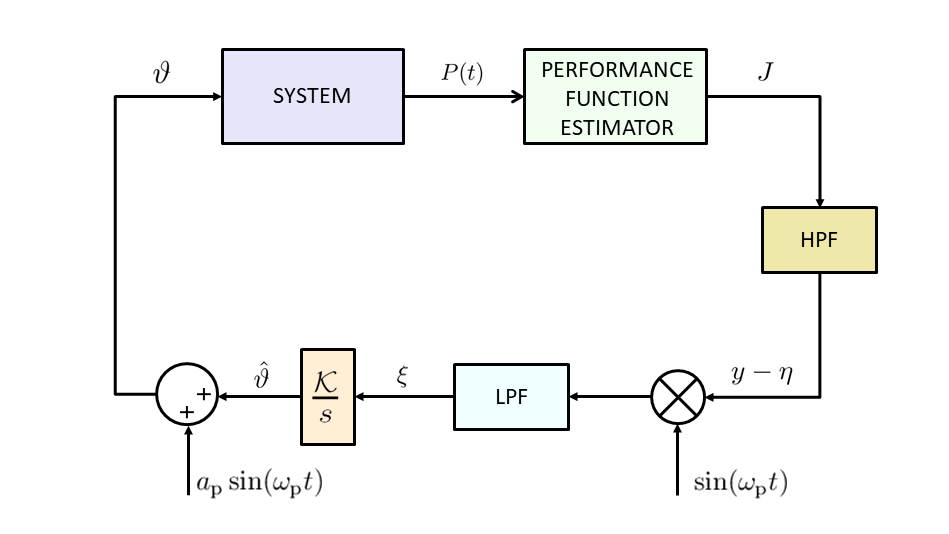}
    \caption{Block diagram of perturbation-based extremum-seeking control system~\cite{ariyur2003real}. First-order low-pass and high-pass filters of the form $\frac{\omega_\text{L}}{s+\omega_\text{L}}$ and $\frac{s}{s+\omega_\text{H}}$, respectively, are used.}
    \label{fig:perturbation_scheme}
\end{figure}

Perturbation-based extremum-seeking control is historically the first ES control algorithm. In 1922, Leblanc first applied a perturbation-based ES to maximize the power transfer from an overhead electrical transmission line to a tram car~\cite{leblanc1922}. The method underwent several extensions and modifications till 1960s, after which it had lost popularity in lieu of other control schemes at that time. This was mainly because of the difficulty to generalize the algorithm for a large class of plants. In 2000, Krsti\'{c} and Wang~\cite{krstic2000stability} proved the stability of perturbation-based extremum-seeking control scheme for a generic plant satisfying certain properties, as listed in Sec.~\ref{sec_esc_overview}. This publication has reignited the interest of the research community in ES methods, and since then, many variants and applications based on extremum-seeking control algorithms have been considered --- some of which were discussed in the previous sections. 
Fig.~\ref{fig:perturbation_scheme} shows a block diagram of perturbation-based ES control system. The scheme can be equivalently described by the following set of equations:

\begin{align} \label{eqn_perturbation_esc}
\text{Perturbation-based ES} & = 
\begin{cases}    
	\dot{\eta} = \omegaH (J-\eta), & \omegaH > 0  \\ 
    	\dot{\xi} = \omegaL \left ( (J-\eta) \sin (\omegap t) -\xi \right), & \omegaL > 0 \\
    	\dot{\hat{\vartheta }} = \mathcal{K} \xi,  & \mathcal{K}  > 0 \\
	\vartheta = \hat{\vartheta} + \ap \sin(\omegap t). & \ap > 0
\end{cases}      
\end{align}

An intuitive working principle of perturbation-based ES can be explained as follows. A perturbation signal $\deltap(t) = \ap \sin(\omegap t)$ is added to the current estimate of the parameter $\vartheta$ and passed into the plant. The plant's performance $J$ is  measured/calculated for the updated parameter. If the output signal $J$ can be linearized around the current estimate of $\vartheta$, then the change in $J$ due to the perturbation signal can be obtained using a Taylor series expansion
\begin{equation}
	J - \eta \approx \d{J}{\vartheta} \deltap,
\end{equation}
in which, $\eta$ is the DC component of the signal that can be subtracted from $J$ by passing it through a high-pass filter (HPF) of cut-off frequency $\omegaH$. The change in the performance function due to the added perturbation is then multiplied by another perturbation signal to yield the following quantity:
\begin{equation}
	(J - \eta) \deltap \approx \d{J}{\vartheta} \deltap^2.
\end{equation}
The resultant of this operation is passed through a low-pass filter (LPF) of cut-off frequency $\omegaL$, which extracts the DC component of the performance gradient $\xi \approx \text{d}{J}/\text{d}{\vartheta}$ with the correct sign~\footnote{As $\deltap^2$ is a positive quantity, the sign of the signal is decided by the gradient of the performance function $\text{d}{J}/\text{d}{\vartheta}$.}. Following the gradient estimation, an integrator with gain $\mathcal{K}$ updates the $\vartheta$ parameter, and the optimization process is repeated until convergence is obtained. 

The amplitude $\ap$ of the perturbation, and the gain $\mathcal{K}$ of the integrator should be chosen small enough to guarantee convergence, as discussed in~\cite{powell2017investigation,krstic2000stability}. Moreover, smaller value of $\ap$ reduces oscillation of $\vartheta$ in the neighborhood of performance-optimal value $\vartheta^*$.  The angular frequency $\omegap$ of the dither signal should be lower than the plant frequency, in order to satisfy Assumption 5 given in Sec.~\ref{sec_esc_overview}.

%% file: MSDModel.tex
The equation of motion of a single degree of freedom spring-mass-damper system oscillating in air, and subject to an external periodic 
force of frequency $\omega$ reads as

\begin{equation} \label{eqn_msd}
    m\ddot{x}(t) + c\dot{x}(t) + kx(t) = \fe(t) + \fpto(t),
\end{equation}
in which, $m$ is the mass of the oscillator, $c$ is the damping coefficient, $k$ is the spring stiffness coefficient, and $x$ is the upward heaving direction; see Fig.~\ref{fig:msd_scheme}. The external periodic force $\fe$, and the control force $f_\text{PTO}$ applied through the power-take off unit, are taken to be of the form
\begin{align}
	 \fe(t) &= f_0 \sin(\omega t),   \label{eqn_fext}  \\
	 \fpto(t) &= -K x(t) - C\dot{x}(t), \label{eqn_fpto}
\end{align} 
in which, $f_0$ is the amplitude of the external force, and a proportional-derivative (PD) control law is used for the control force.  Using impedance-matching or complex-conjugate control analysis,  it can be shown that the average-power 
\begin{equation} \label{eqn_p_msd}
    \bar{P} =\frac{1}{\cT} \int_{t}^{t+ \cT}  C \dot{x}^2 \, \text{d}t, 
\end{equation}
extracted from the system over a time period $\cT = 2\pi/ \omega$ is maximal, when the system is in resonance with the external forcing. In other words, when the natural frequency of the system $\omegan$ equals the external frequency  $\omega$, extracted power is maximized. Since it is inconvenient  or often times impossible to change the inherent characteristics of the system ($m,c,k$), the PD control law  allows the control designer to adjust the reactive and resistive PTO coefficients, $K$ and $C$ respectively, to optimize the system performance for varying external forces and disturbances. The energy-maximizing PTO parameters (in absences of disturbances) can be found analytically as~\cite{korde2016book}
\begin{align}
	K_\text{opt} &= \omega^2 m - k, \label{eqn_kopt_msd} \\
	C_\text{opt} &= c. \label{eqn_copt_msd}
\end{align}
In Sec.~\ref{sec_msd_results} we make use of Eqs.~\eqref{eqn_kopt_msd}-\eqref{eqn_copt_msd} to verify the optimal solutions $K(\vartheta_1^*)$ and/or $C(\vartheta_2^*)$ obtained using ES. 

We remark that in some other optimal control formulations, for example in model predictive control of converters~\cite{faedo2017optimal}, the objective is to find the energy-maximizing control force $\fpto$ directly. The average-power extracted by the system over a time period $\cT$ is expressed as
 \begin{equation} \label{eqn_p_mpc}
    \bar{P} =\frac{1}{\cT} \int_{t}^{t+ \cT}  -\fpto \, \dot{x} \, \text{d}t.
\end{equation}
For a PD control law, Eq.~\eqref{eqn_p_mpc} is equivalent to Eq.~\eqref{eqn_p_msd}
 \begin{align} \label{eqn_p_mpc_pd}
    \bar{P} &= \frac{1}{\cT} \int_{t}^{t+ \cT}  -\fpto \, \dot{x} \, \text{d}t           \nonumber                          \\         
                &= \frac{1}{\cT} \int_{t}^{t+ \cT}  \left(K x + C \dot{x}\right) \dot{x} \, \text{d}t      \nonumber \\
                & =  \frac{1}{\cT} \int_{t}^{t+ \cT}  K \d{}{t}\left(\frac{x^2}{2}\right) \, \text{d}t + \frac{1}{\cT} \int_{t}^{t+ \cT} C \dot{x}^2 \, \text{d}t,
\end{align}
as the first term in Eq.~\eqref{eqn_p_mpc_pd} vanishes under the time-periodic motion of the device. Appendix~\ref{sec_appendix} numerically verifies that the inclusion of the reactive component of power in the performance function, does not affect the final optimized values of PTO coefficients using ES. Therefore, we use only the resistive component of power or Eq.~\eqref{eqn_p_msd} for defining the performance function.

%% file: PointAbsorberModel.tex
The equation of motion of a fully-submerged point absorber with a  single degree of freedom (see Fig.~\ref{fig:PA_scheme}) reads as
\begin{equation} \label{eqn_PA}
    m \ddot{x} (t)  = \fw(t) + \fr(t) + \fv(t) + \fpto(t),  
\end{equation}
in which, $m$ is the mass of the point absorber, $\fw$ is the Froude-Krylov wave excitation force, $\fr$ is the radiation force, $\fv$ is the viscous drag force, and $\fpto$ is the PD control force applied by the PTO mechanism, as given in Eq.~\eqref{eqn_fpto}.  The physical origin of radiation force stems from the energy dissipation mechanism of a moving body that emanates waves during its motion in water. Using the Cummins equation~\cite{Cummins1962}, the radiation force is expressed as
\begin{equation}
  \fr(t) = - A_\infty \ddot{x} - \int_{0}^{t} \hr(t-\tau) \dot{x}(\tau) \,  \text{d}\tau, \label{eqn_fr}
\end{equation} 
in which,  $A_\infty$  is the infinite-frequency added mass, and $\hr(t)$ is the radiation impulse response function that contains the fluid-memory effect. The radiation force can also be obtained from the inverse Fourier transform of $\Hr(j \omega)$
\begin{align}
\Hr(j \omega) &= B(\omega) + j \omega \left(  A(\omega) - A_\infty \right),  \label{eqn_Hr} \\
\hr(t) &= \frac{1}{\pi} \int_{0}^{\infty} \Hr(j\omega) e^{j \omega t} \, \text{d} \omega,  \label{eqn_hr}
\end{align}
in which, the frequency-dependent added mass $A(\omega)$, and the frequency-dependent radiation damping $B(\omega)$, can be 
obtained using boundary element method (BEM)-based codes like WAMIT~\cite{Lee1995} or ANSYS AQWA~\cite{Ansys2014}. For computational efficiency, as well as for representational convenience, the radiation convolution integral in Eq.~\ref{eqn_fr} can be evaluated by an equivalent state-space 
formulation~\cite{FDIToolboxA,FDIToolboxB}
\begin{align} \label{eqn_ss_convol}
\kr(t) = \int_{0}^{t} \hr(t-\tau) \dot{x}(\tau) \,  \text{d}\tau  \simeq 
\begin{cases}
	\dot{\V \zeta}_\text{r}(t) =\V{\mathcal{A}}_\text{r} \V{\zeta}_\text{r}(t) + \V{\mathcal{B}}_{\rm r} \dot{x}(t),  & \V{\mathcal{A}}_\text{r} \in \mathbb{R}^{\nr \times \nr},  \V{\zeta}_\text{r} \in \mathbb{R}^{\nr \times 1} \\
    	\kr(t) = \V{\mathcal{C}}_{\rm r} \V{\zeta}_{\rm r}(t), & \V{\mathcal{C}}_{\rm r} \in \mathbb{R}^{1 \times \nr}
\end{cases}
\end{align}  
in which, $\V{\mathcal{A}}_\text{r}, \V{\mathcal{B}}_\text{r}$, and $\V{\mathcal{C}}_\text{r}$ are the state-space matrices, and $\nr$ is the approximation-order of $\Hr(j \omega)$ in the frequency-domain or $\hr(t)$ in the time-domain. In this work, we also follow the state-space approach to approximate the radiation convolution integral. Combining Eqs.~\eqref{eqn_PA} and~\eqref{eqn_ss_convol}, the system of equations for the fully-submerged point absorber heaving in $x-$direction reads as
\begin{align}
(m + A_\infty) \ddot{x} (t) +   \V{\mathcal{C}}_{\rm r} \V{\zeta}_{\rm r} (t) & = \fw(t) + \fv(t) - K x(t) -C \dot{x}(t),  \label{eqn_pa_td}\\
\dot{\V \zeta}_\text{r}(t) & = \V{\mathcal{A}}_\text{r} \V{\zeta}_\text{r}(t) + \V{\mathcal{B}}_{\rm r} \dot{x}(t).
\end{align}
The viscous drag force in Eq.~\eqref{eqn_pa_td} is modeled as 
\begin{equation}
\fv(t) = -\frac{1}{2} \rhow C_{\rm d} S_x |\dot{x}| \dot{x}, 
\end{equation}
in which, $\rhow = 1025$ kg/m$^3$ is the density of water, $C_{\rm d}$ is the drag coefficient, and $S_x$ is the planar cross-section area normal to the force.

We use linear potential flow theory to compute the wave excitation force $\fw$ on the submerged point absorber. Both regular and irregular sea states are considered in this work. The regular waves are characterized by a single wave frequency $\omega$ or time period $\cT = 2 \pi/\omega$, wave height $\cH$ or amplitude $a = \cH/2$, and wavelength $\lambda$ or wavenumber $\kappa = 2 \pi/\lambda$. The wave frequency and the wavenumber satisfy the dispersion relation~\cite{book_Waves_in_oceanic_and_coastal_waters}
\begin{equation}
	\omega^2 = g\kappa \tanh(\kappa d), \label{eqn_dispersion}
\end{equation}
in which, $g = 9.81$ m/s$^2$ is the acceleration due to gravity, and $d$ is the mean depth of water. In contrast, an irregular sea state consists of a large number of regular wave components, each having its own wave amplitude $a_i$, angular frequency $\omega_i$ (or equivalently wavenumber $\kappa_i$ obtained from the dispersion relation given by Eq.~\eqref{eqn_dispersion}), and a random phase $\theta_i$ that is uniformly distributed in the range $[0, 2\pi]$.  The linear superposition of regular wave components implies that the energy carried by an irregular wave is the sum of the energy transported by individual wave components. When the number of wave components tends to infinity, a continuous wave spectral density function $S(\omega)$ is used to describe the energy content of the wave components in an infinitesimal frequency bandwidth d$\omega$. In this work we use JONSWAP~\cite{book_offshore_hydromechanics} spectrum to generate the irregular sea state. The JONSWAP spectrum is characterized by two statistical parameters: significant wave height $\cH_{\rm s}$, and peak period $\cT_{\rm p}$. The amplitude of each wave component is related to the spectral density function by
\begin{equation}
a_i = \sqrt{2 \cdot S(\omega_i) \cdot \Delta \omega}.
\end{equation}

In a regular sea, the point absorber system is mathematically equivalent to the mechanical oscillator of the previous section, if one replaces the damping coefficient $c$ with the radiation damping $B(\omega)$, mass of the oscillator $m$ with the total mass $\left( m + A(\omega) \right)$, and the spring stiffness $k$ with the hydrostatic stiffness $k_{\rm hydro}$, which is zero for a fully-submerged body. Therefore, using the impedance-matching control theory, the energy-maximizing PTO parameters can be found as
\begin{align}
	K_\text{opt} &= \omega^2 \left( m + A(\omega) \right), \label{eqn_kopt_pa} \\
	C_\text{opt} &= B(\omega). \label{eqn_copt_pa}
\end{align}
We remark that if an additional viscous drag force is included in the equations of motion, as done in this work, then the optimal PTO resistive coefficient $C_\text{opt}$ would be higher than $B(\omega)$. We make use of Eqs.~\eqref{eqn_kopt_pa} and \eqref{eqn_copt_pa} to verify the results of ES algorithms.

For an irregular sea state, the optimal PTO parameters cannot be found analytically as multiple frequencies are present in the point absorber velocity $\dot{x}$ (and other state variables), which is used to evaluate the $\bar{P}$ expression  in Eq.~\eqref{eqn_p_msd}. Therefore, to verify the optimal solution obtained from ES algorithms, we create a performance map of the system using a brute-force search of parametric space.

%% file: PAWaveCharacteristics.tex
Motivated by our prior work on numerical modeling of a fully-submerged axisymmetric point absorber device~\cite{dafnakis2019fully}, we simulate a two-dimensional cylindrical, and a three-dimensional spherical buoy to perform extremum-seeking control simulations. Both devices have the same diameter $D = 0.16$ m, and a homogeneous mass density of $\rho_{\rm s} = 922.5$ kg/m$^3$. Their depth of submergence is taken to be $d_{\rm s} = 0.25$ m, and the mean depth of water is taken as $d = 0.65$ m. Table~\ref{tab:sea_states} tabulates the regular and irregular sea states simulated in this work. These wave characteristics are chosen based on the scale of the device; see Dafnakis et al.~\cite{dafnakis2019fully} and Khedkar et al.~\cite{khedkar2020inertial} for discussion.

\begin{table}[]
 \centering
 \caption{Sea states.}
  \rowcolors{2}{}{gray!10}
            \begin{tabular}{c c c | c c c}
            \toprule
            Regular sea ID  & $\cT$ (s) & $\cH$ (m)  & Irregular sea ID  & $\cT_{\rm p}$ (s) & $\cH_{\rm s}$ (m)\\
            \midrule
            Reg.1       &   0.625     &  0.01      	   & Irreg.1       &   0.625     &  0.01      	\\
            Reg.2       &   0.8         &  0.02          & Irreg.2       &   0.8         &  0.02          \\ 
            Reg.3       &   1.0         &  0.0075      & Irreg.3       &   1.0         &  0.0075      \\
            \bottomrule
 \end{tabular}
 \label{tab:sea_states}
\end{table}

%% file: IntroResults.tex
In this section, results are presented for energy-maximizing $K$ and $C$ PTO coefficients using different ES schemes. 
For each considered system, we present results for a two-parameter optimization problem, in which both $K$ and $C$ parameters are simultaneously optimized. In some cases, results for a single-parameter optimization problem, in which either $K$ or $C$ are optimized are also presented.

%% file: MassSpringDamperResults.tex
\begin{table}[]
 \centering
 \caption{Mass-spring-damper parameters.}
  \rowcolors{2}{}{gray!10}
            \begin{tabular}{c c}
            \toprule
            Parameter  & Value \\
            \midrule
            $m$              & 18.55 kg               \\ 
            $k$                & 200 N/m              \\ 
            $c$               & 15 N$\cdot$s/m    \\ 
            $\cT$             & 0.5 s   		      \\ 
            $f_0$             & 10 N    		     \\ 
            $K_\text{opt}$  & 2729 N/m     \\ 
            $C_\text{opt}$  & 15 N$\cdot$s/m    \\ 
            \bottomrule
 \end{tabular}
 \label{tab:msd_parameters}
\end{table}

\begin{figure}[]
 \centering
        \subfigure[Sliding mode ES]{
        	        \includegraphics[scale=.33]{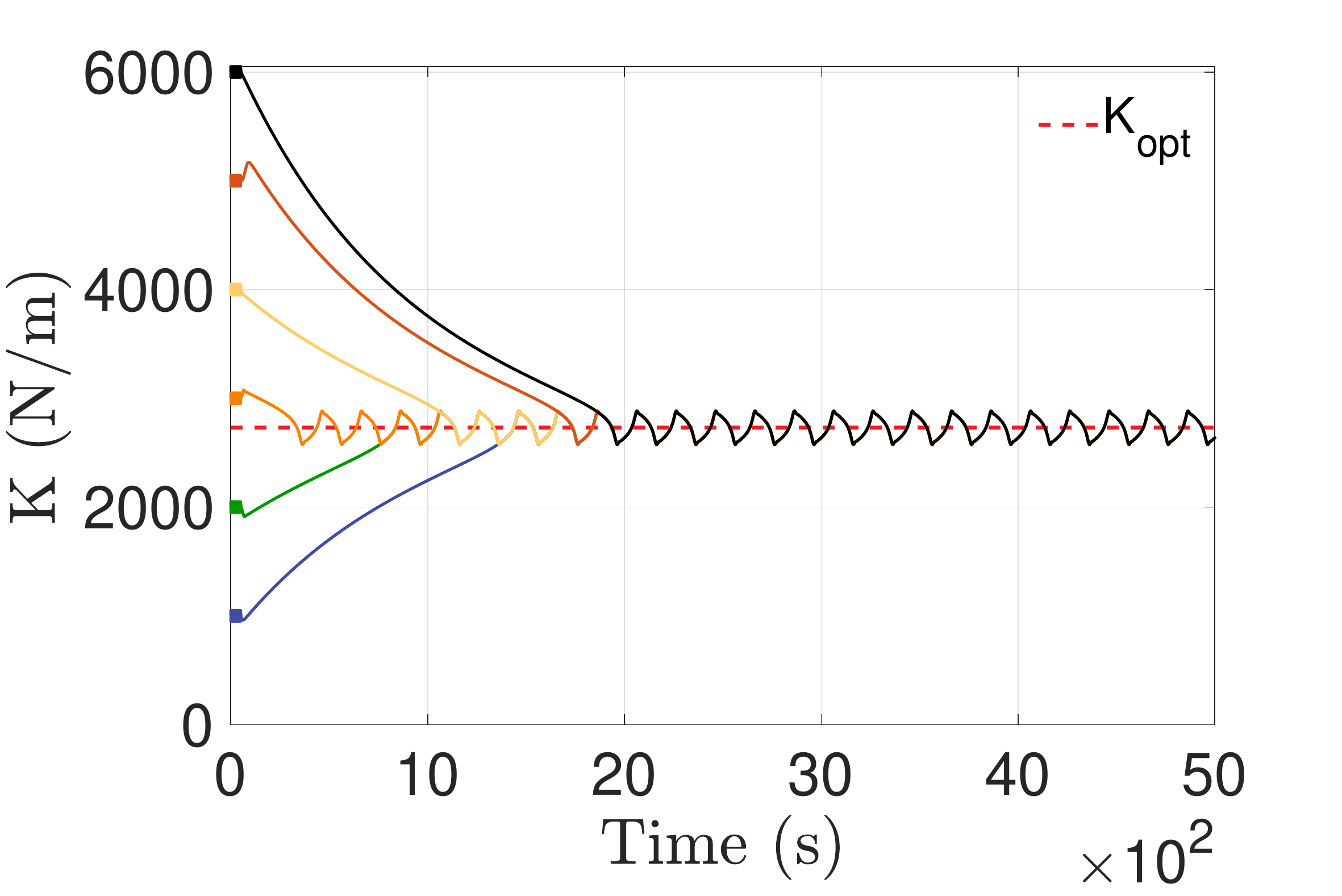}
	        \label{subfig:smesc_msd_k}
	  } 
      \subfigure[Self-driving ES]{
      	       \includegraphics[scale=.33]{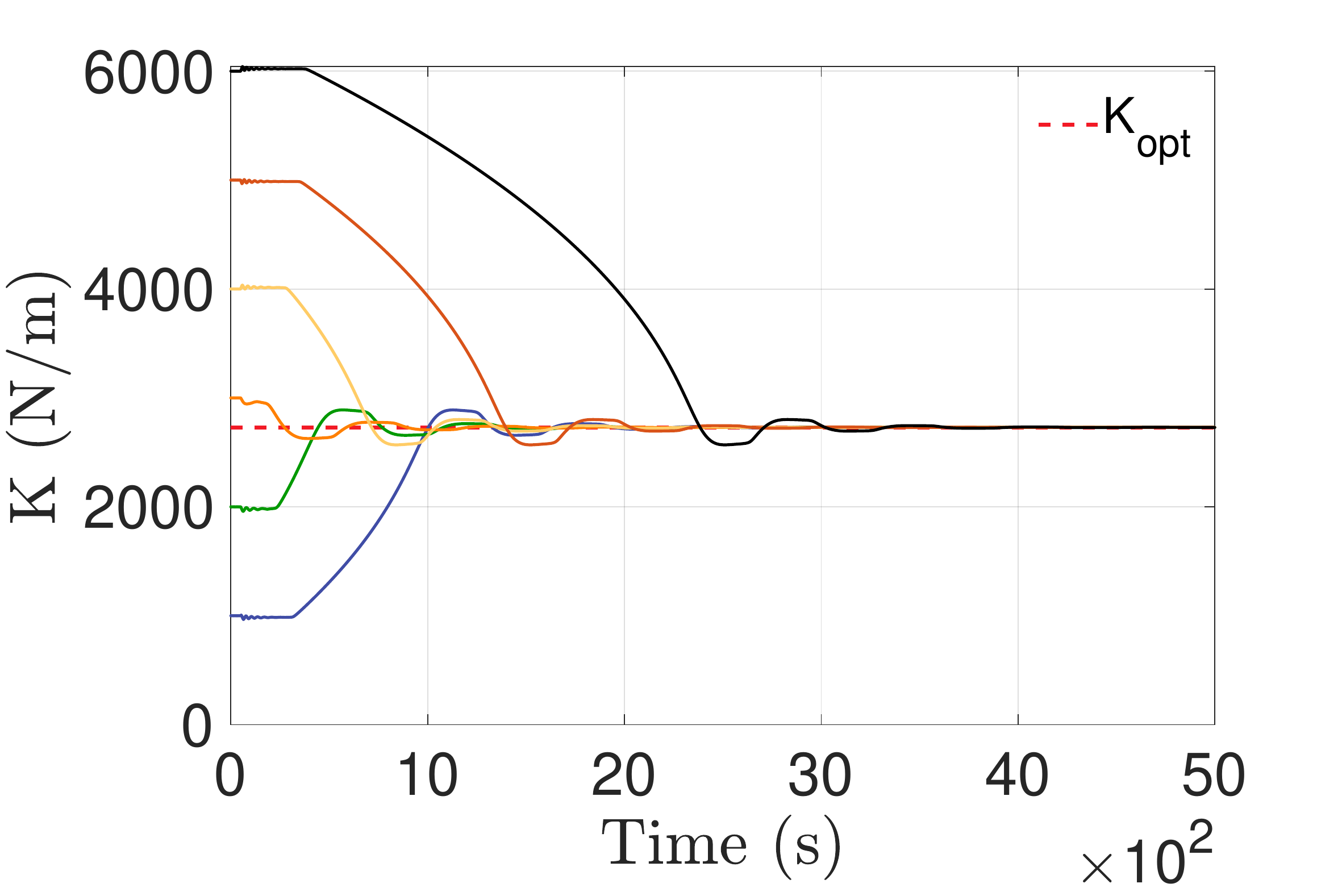}
	 }
        \subfigure[Relay ES]{
        	      \includegraphics[scale=.33]{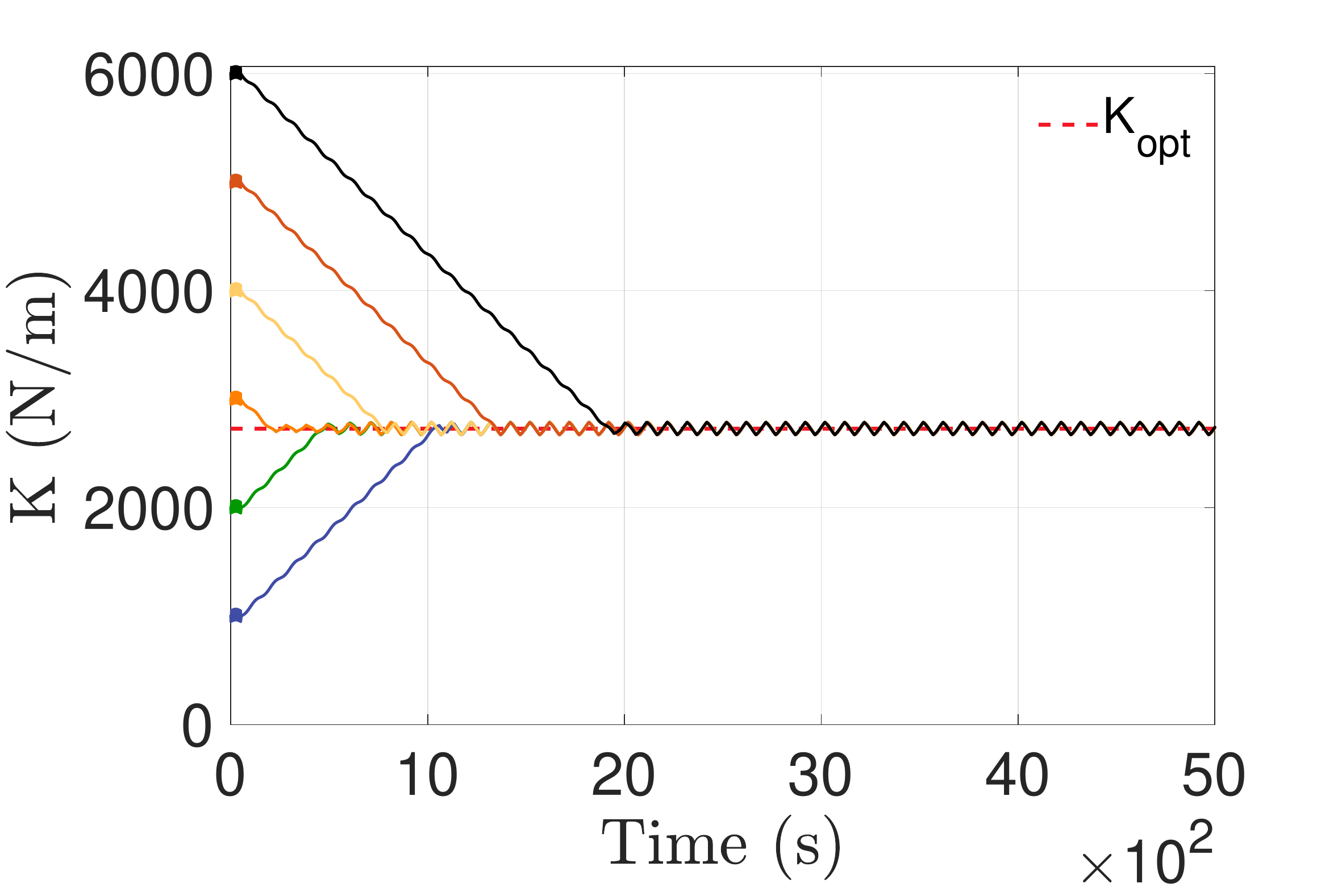}
	 } 
       \subfigure[LSQ-ES]{
       	      \includegraphics[scale=0.33]{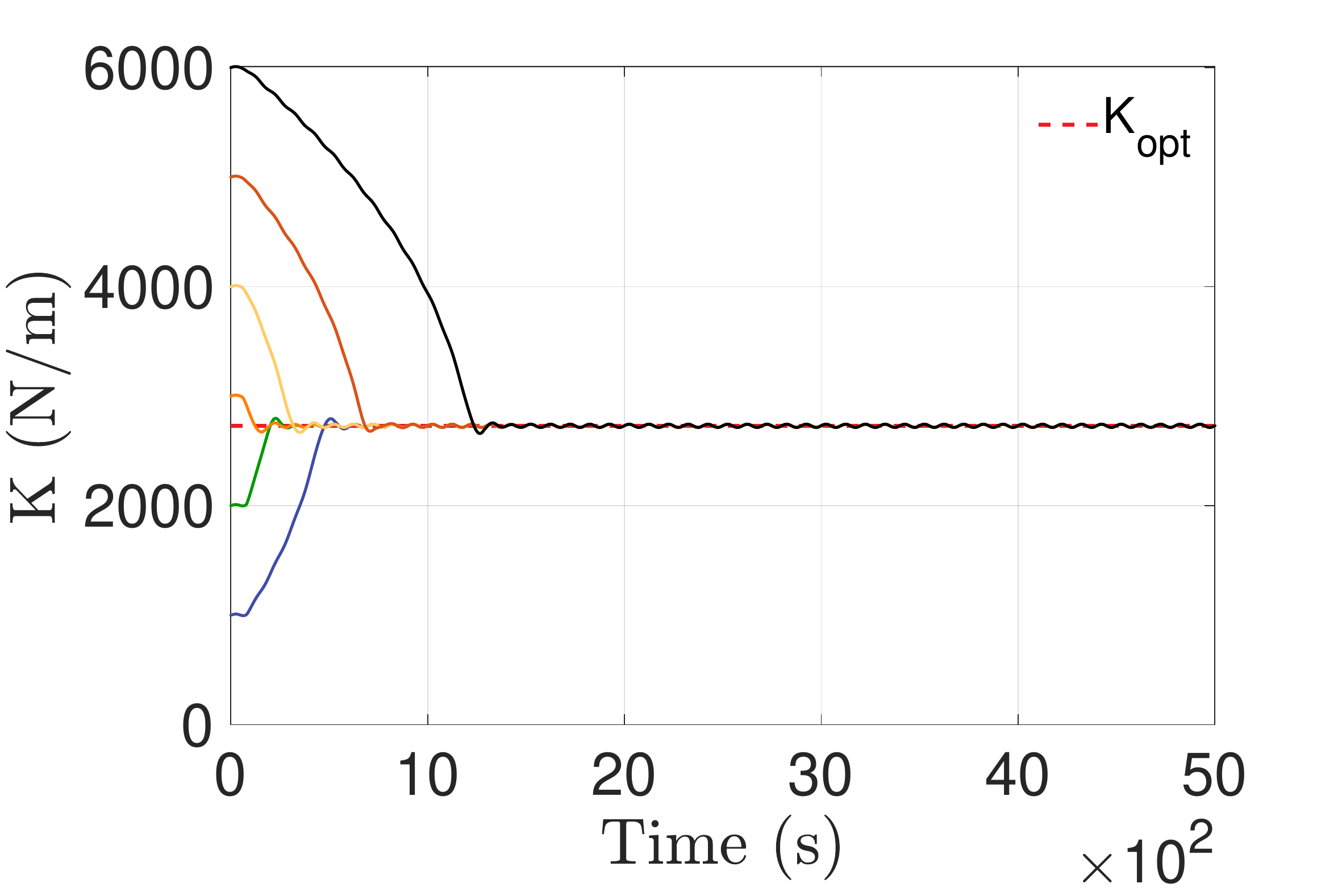}
       }
        \subfigure[Perturbation-based ES]{
        	     \includegraphics[scale=0.33]{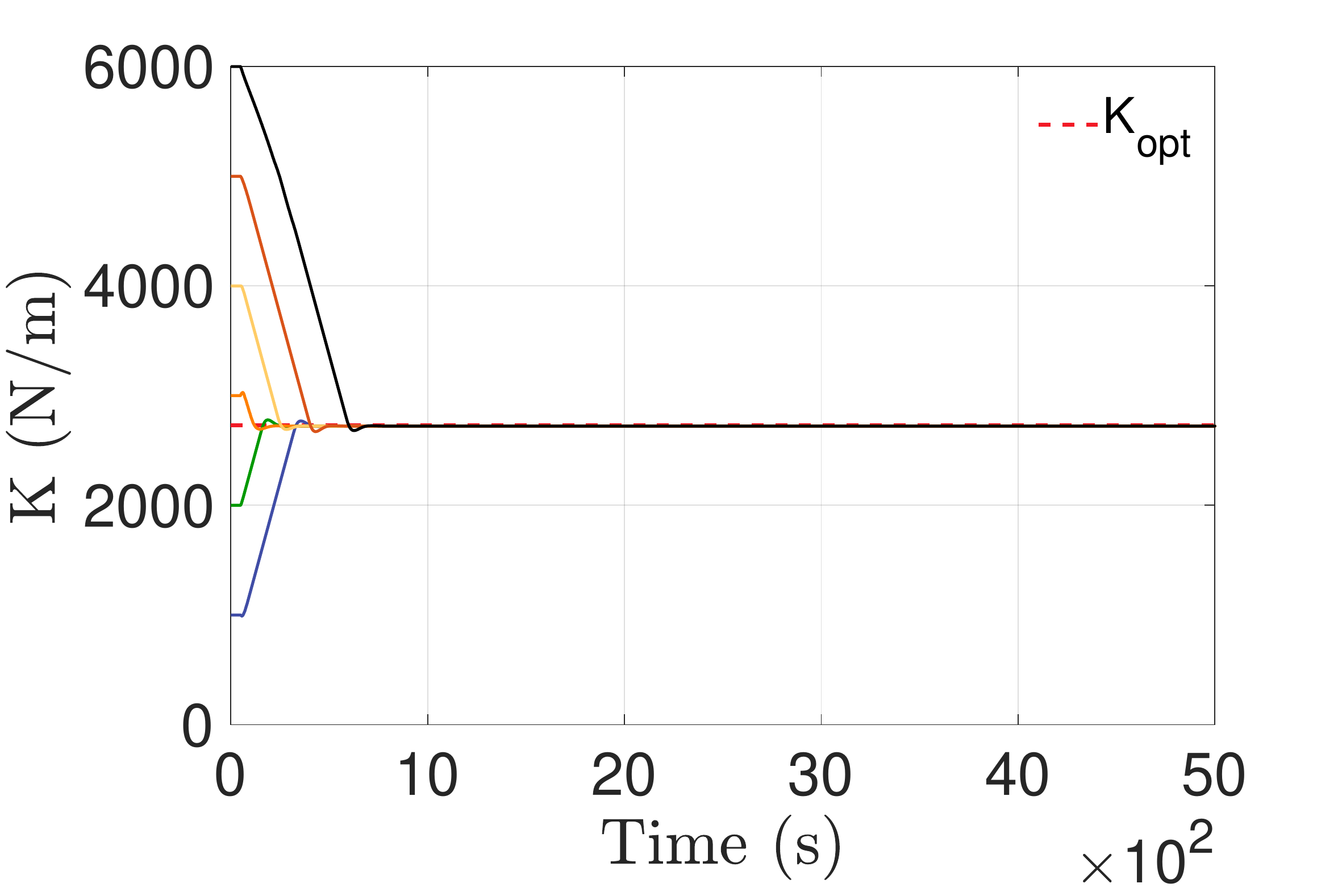}
	     \label{subfig:pesc_k}
	     }
    \caption{Optimization of reactive coefficient $K$ at a fixed value of resistive coefficient $C = C_\text{opt}$ for the mass-spring-damper system using different ES algorithms. The optimal $K_\text{opt}=2729$ N/m value is indicated by the dashed line in the plots.}
    \label{fig:msd_multi_k}
\end{figure}

\begin{figure}[]
    \centering
        \subfigure[Sliding mode ES]{
        		\includegraphics[scale=.33]{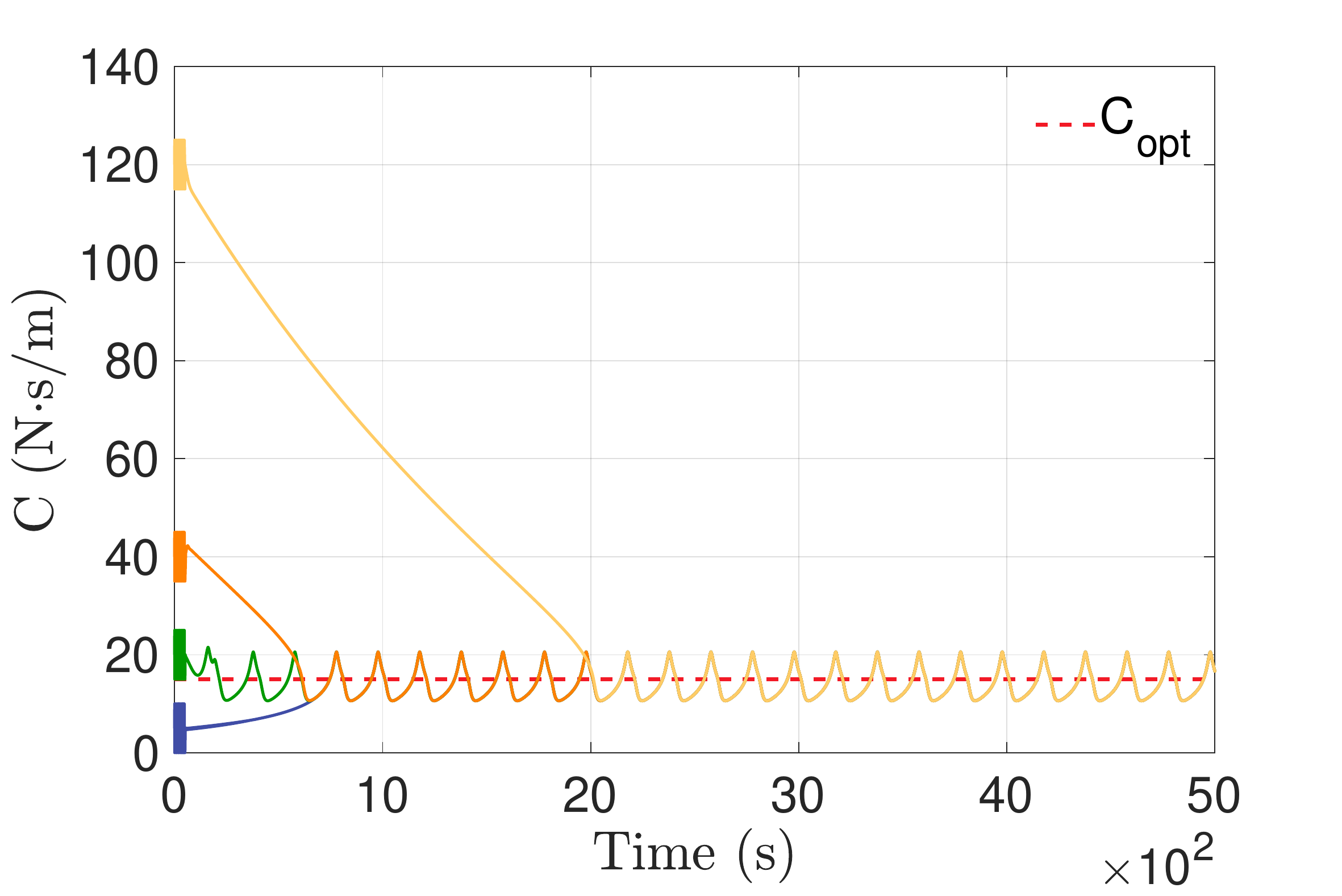}
	} 
        \subfigure[Self-driving ES]{
        		\includegraphics[scale=.33]{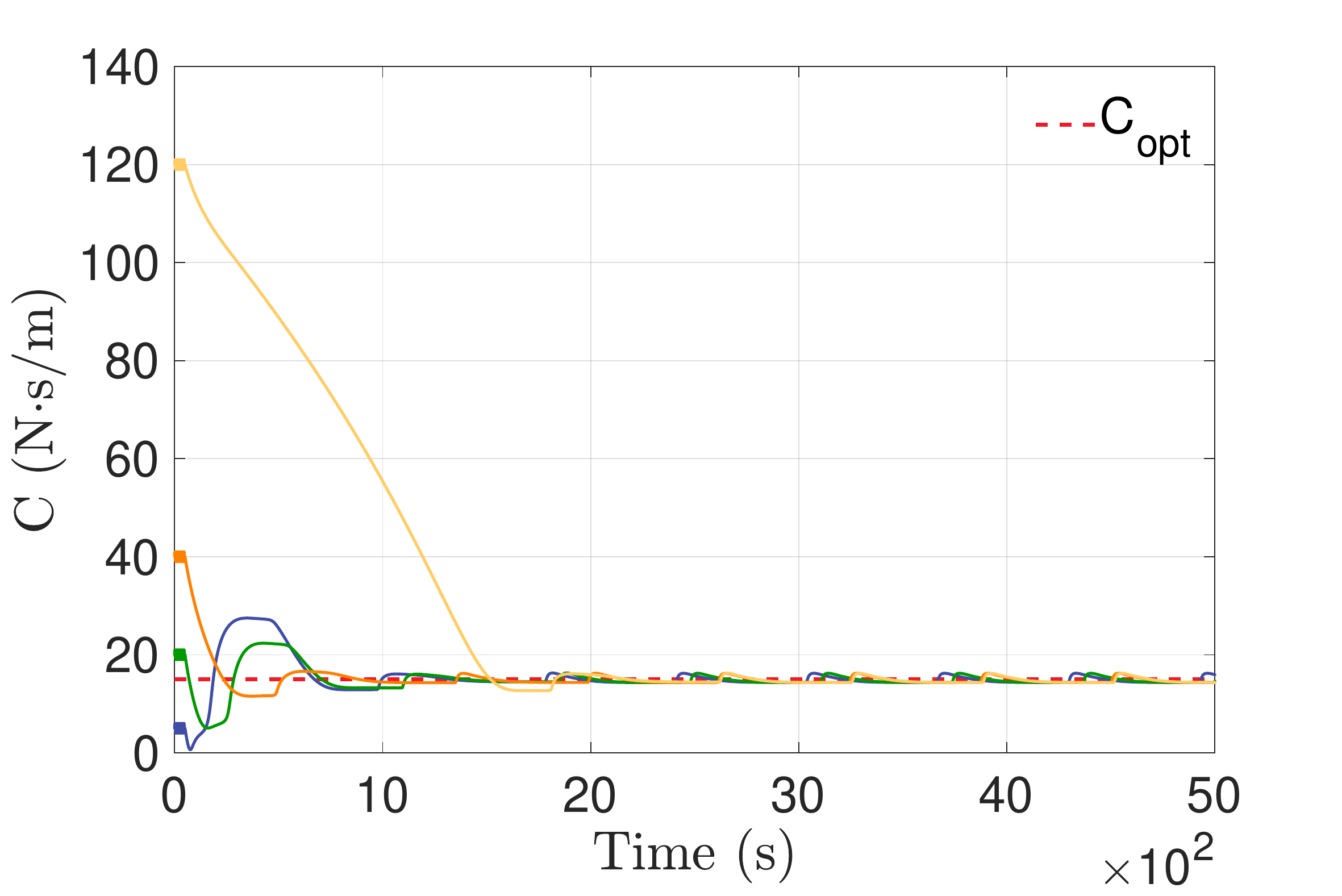}
        }
        \subfigure[Relay ES]{
        		\includegraphics[scale=.33]{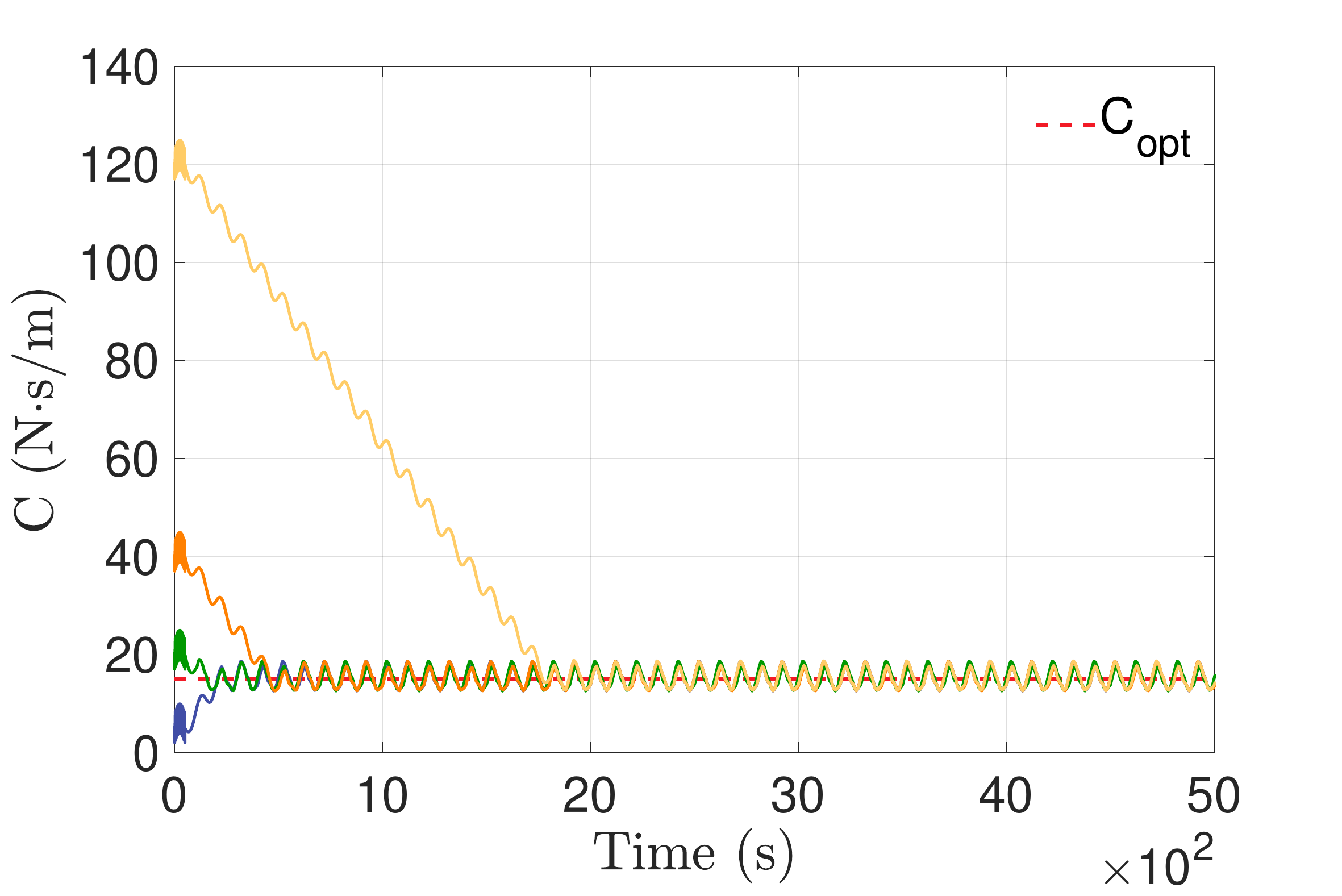}
	} 
        \subfigure[LSQ-ES]{
        		\includegraphics[scale=0.33]{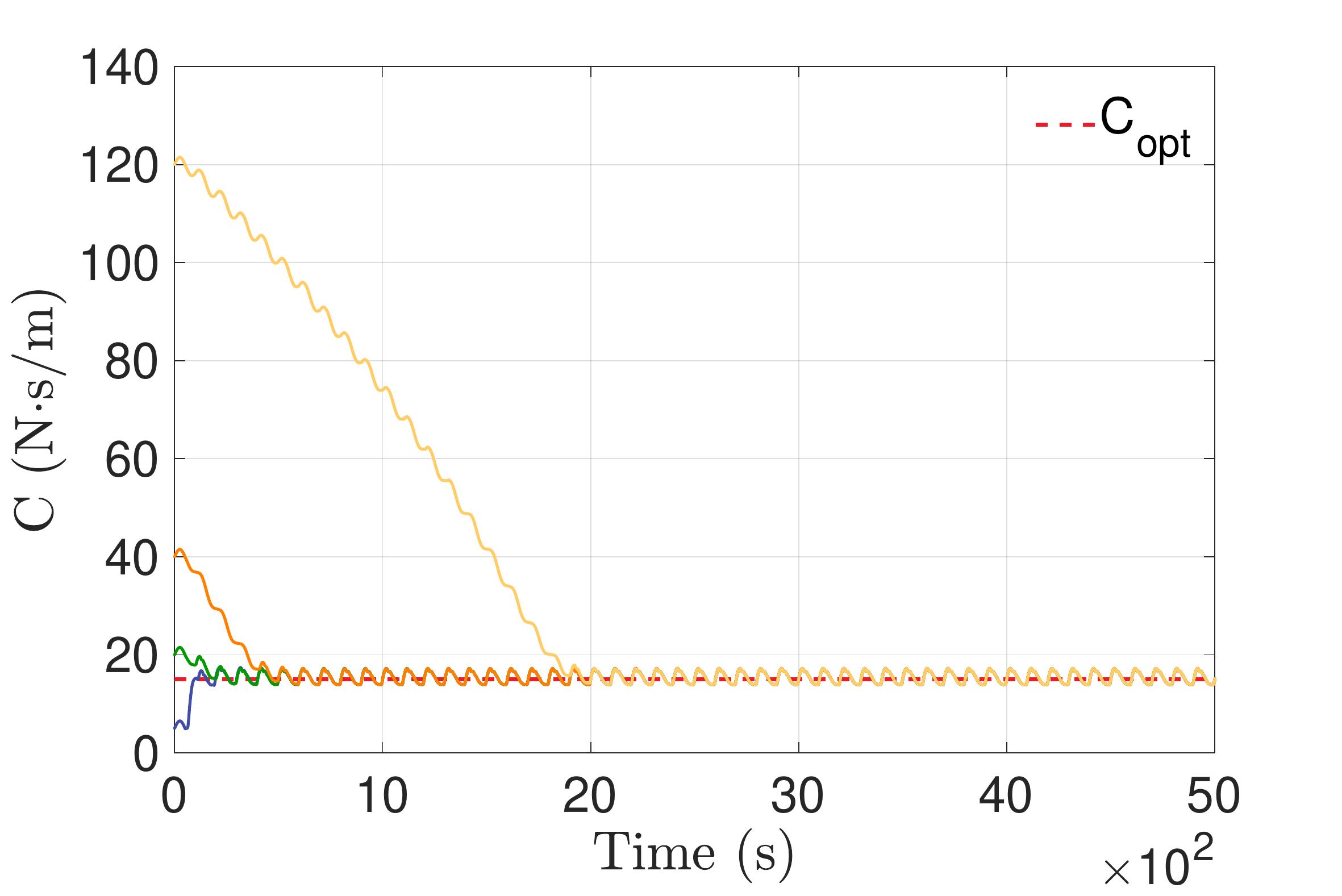}
	}
        \subfigure[Perturbation-based ES]{
        		\includegraphics[scale=0.33]{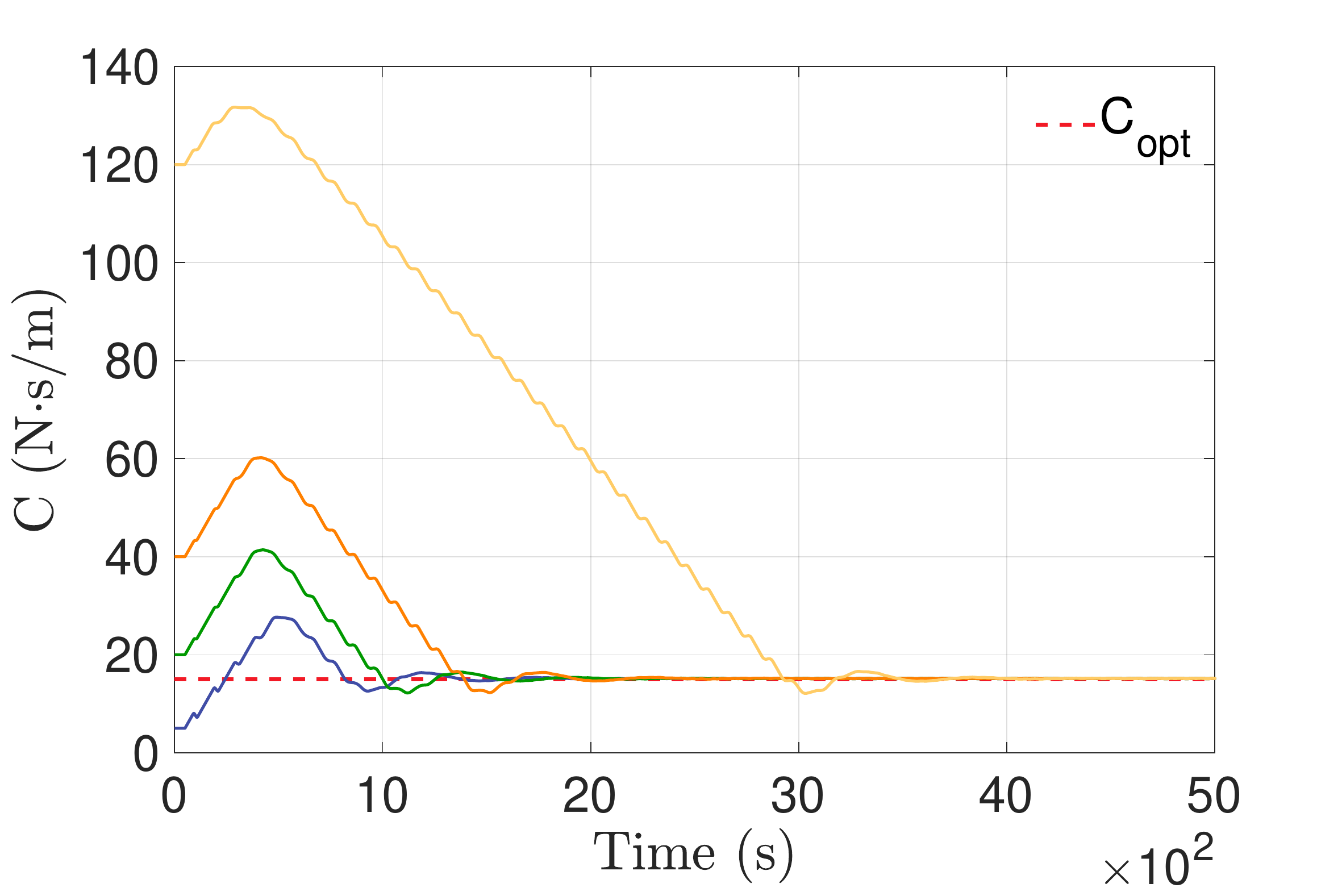}
		\label{subfig:pesc_b}
	}
    \caption{Optimization of resistive coefficient $C$ at a fixed value of reactive coefficient $K = K_\text{opt}$ for the mass-spring-damper system using different ES algorithms. The optimal $C_\text{opt} = 15$ N$\cdot$s/m value is indicated by the dashed line in the plots.}
    \label{fig:msd_multi_b}
\end{figure}

The mass-spring-damper system of Sec.~\ref{sec_msd} is considered here using the parameters tabulated in Table~\ref{tab:msd_parameters}. 
The table also lists the optimal values of $K_\text{opt}$ and $C_\text{opt}$ coefficients, obtained analytically.
We begin with single-parameter optimization of either reactive coefficient $K$ or resistive coefficient $C$, by keeping the other fixed at its optimal value. The optimization results for $K$ with different initial values are shown in Fig. \ref{fig:msd_multi_k}, whereas Fig. \ref{fig:msd_multi_b} shows the optimization results for $C$. As can be seen in both figures, all ES algorithms convergence to the theory-predicted optimal value. 

\begin{figure}[]
    \centering
        \subfigure[Sliding mode ES]{
        		\includegraphics[scale=.33]{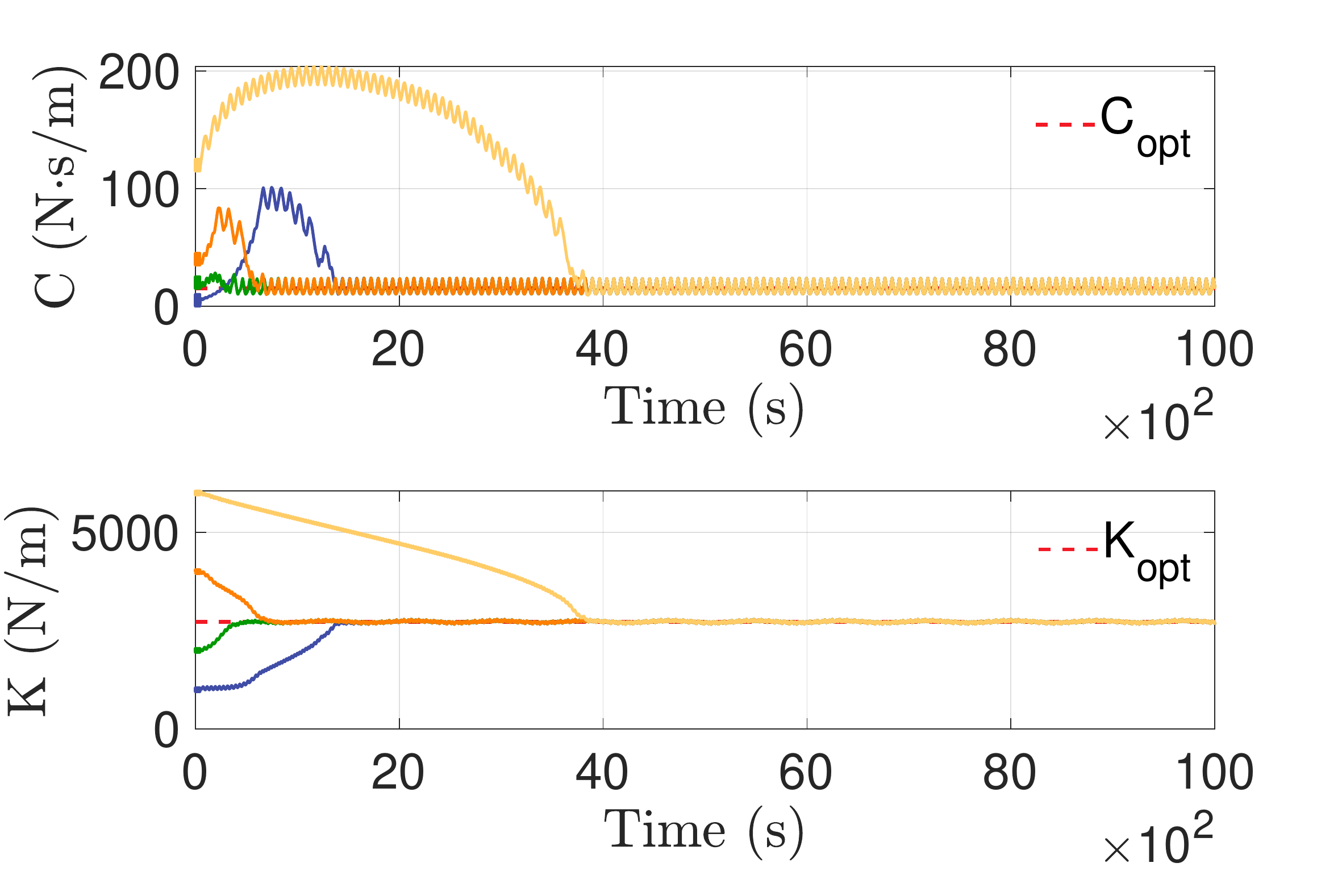}
	} 
        \subfigure[Relay ES]{
        		\includegraphics[scale=.33]{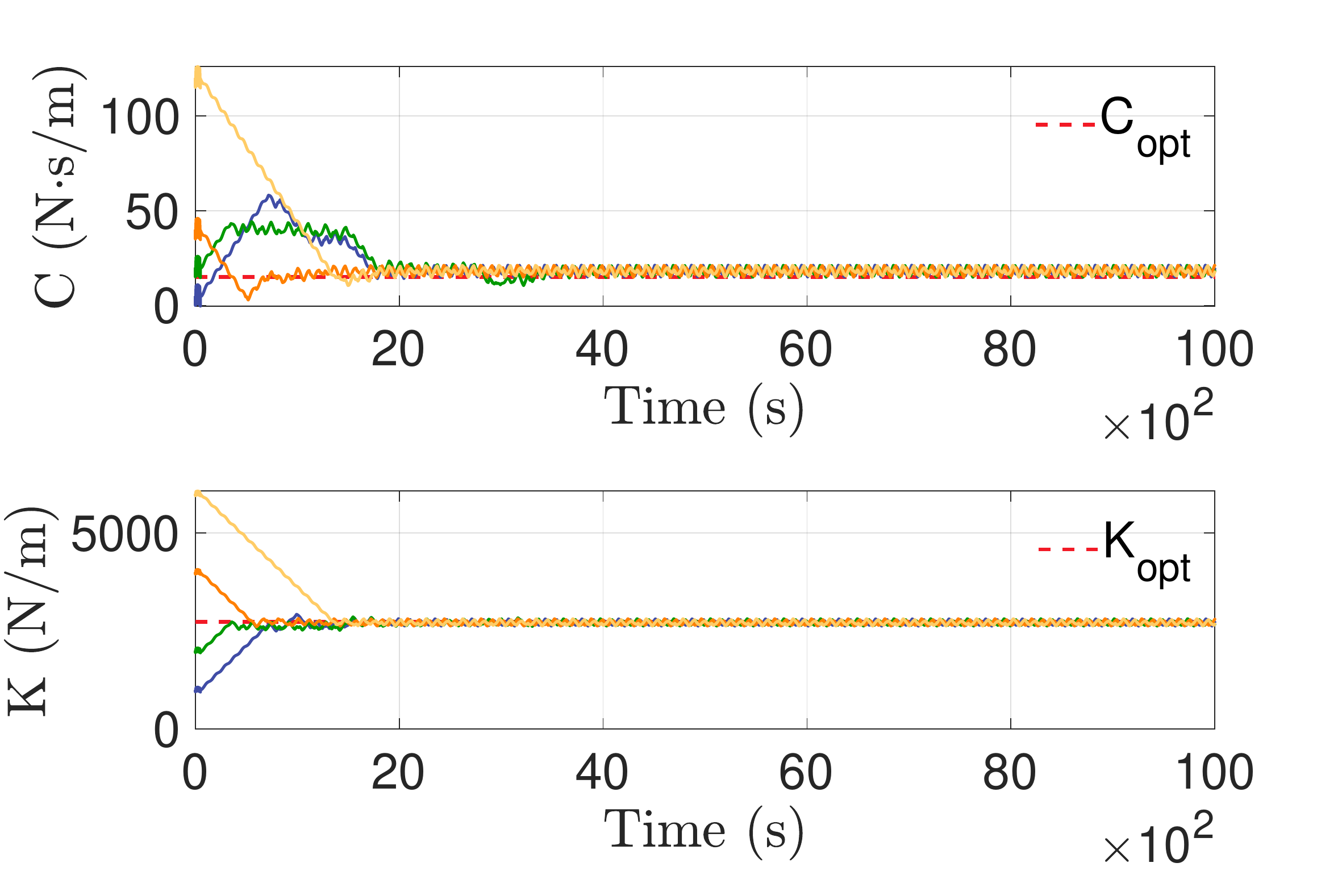}
	}
        \subfigure[LSQ-ES]{
        		\includegraphics[scale=.33]{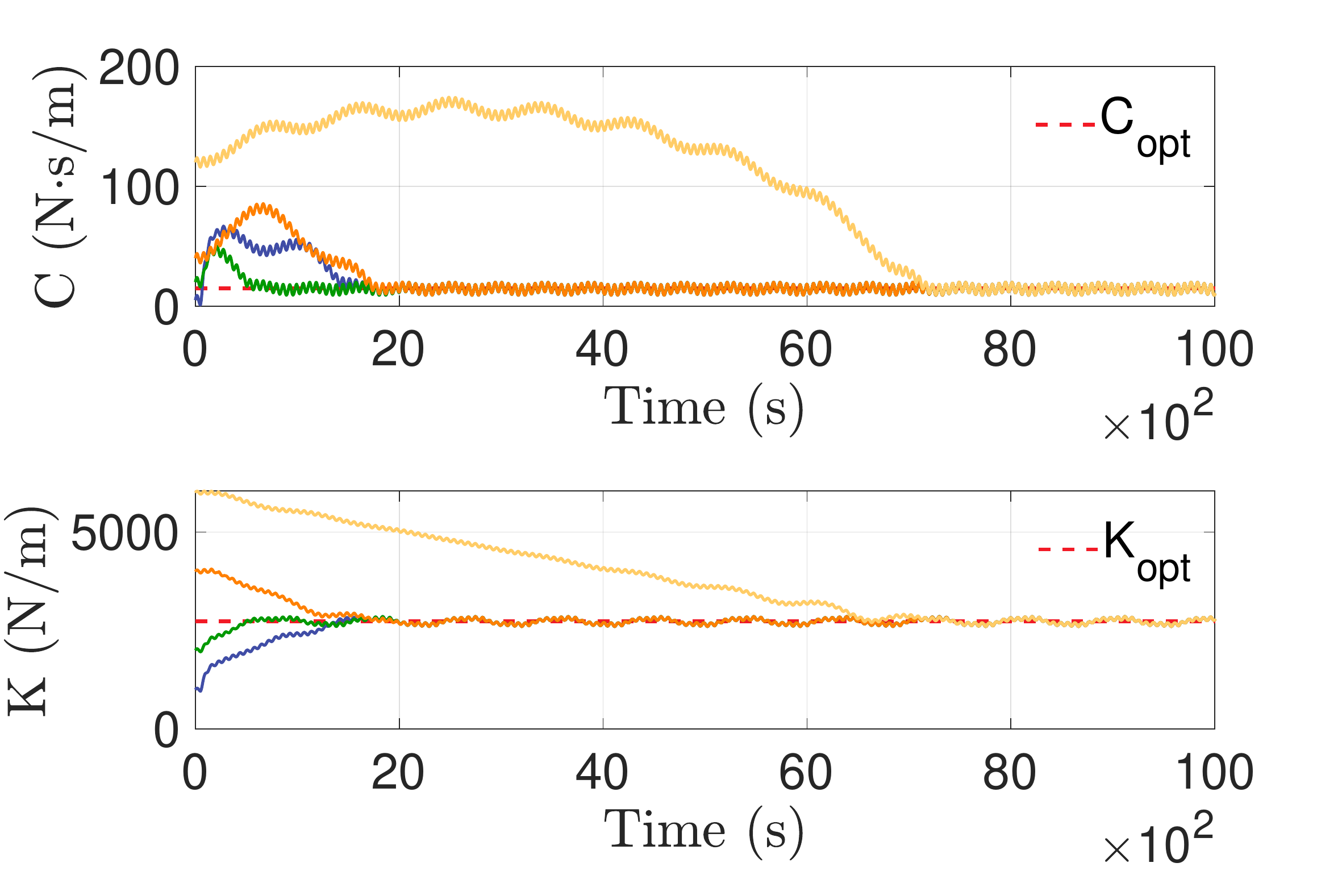}
	 } 
         \subfigure[Perturbation-based ES]{
         	\includegraphics[scale=0.33]{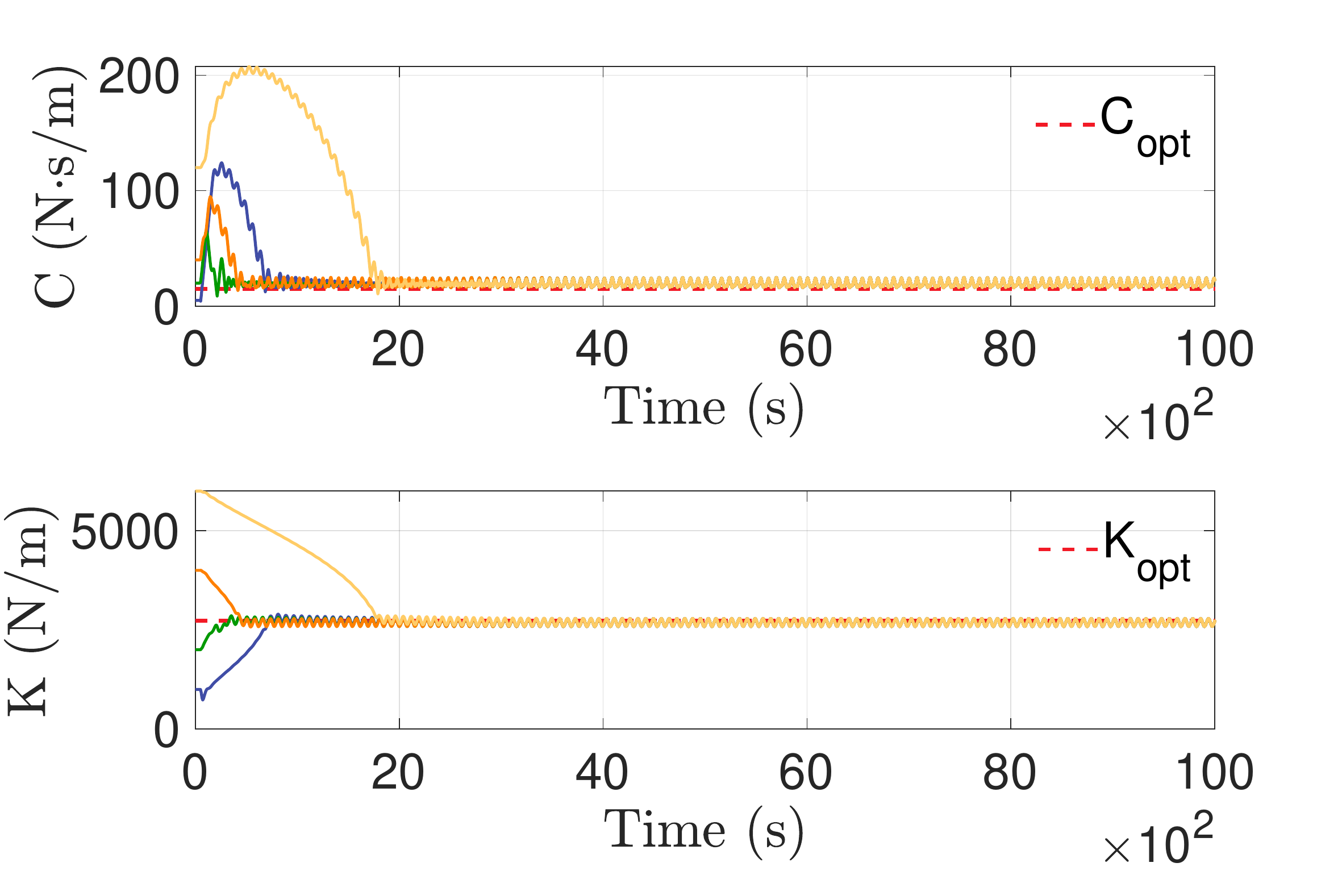}
		\label{subfig:pesc_k_b}
	}
    \caption{Optimization of reactive and resistive coefficients, $K$ and $C$, respectively, for the mass-spring-damper system using different ES algorithms. The  optimal $K_\text{opt}=2729$ N/m and $C_\text{opt} = 15$ N$\cdot$s/m values are indicated by dashed lines in the plots.}
    \label{fig:msd_multi_k_b}
\end{figure}

\begin{figure}[]
    \centering
    \subfigure[Perturbation-based ES] {
	    \includegraphics[scale=0.33]{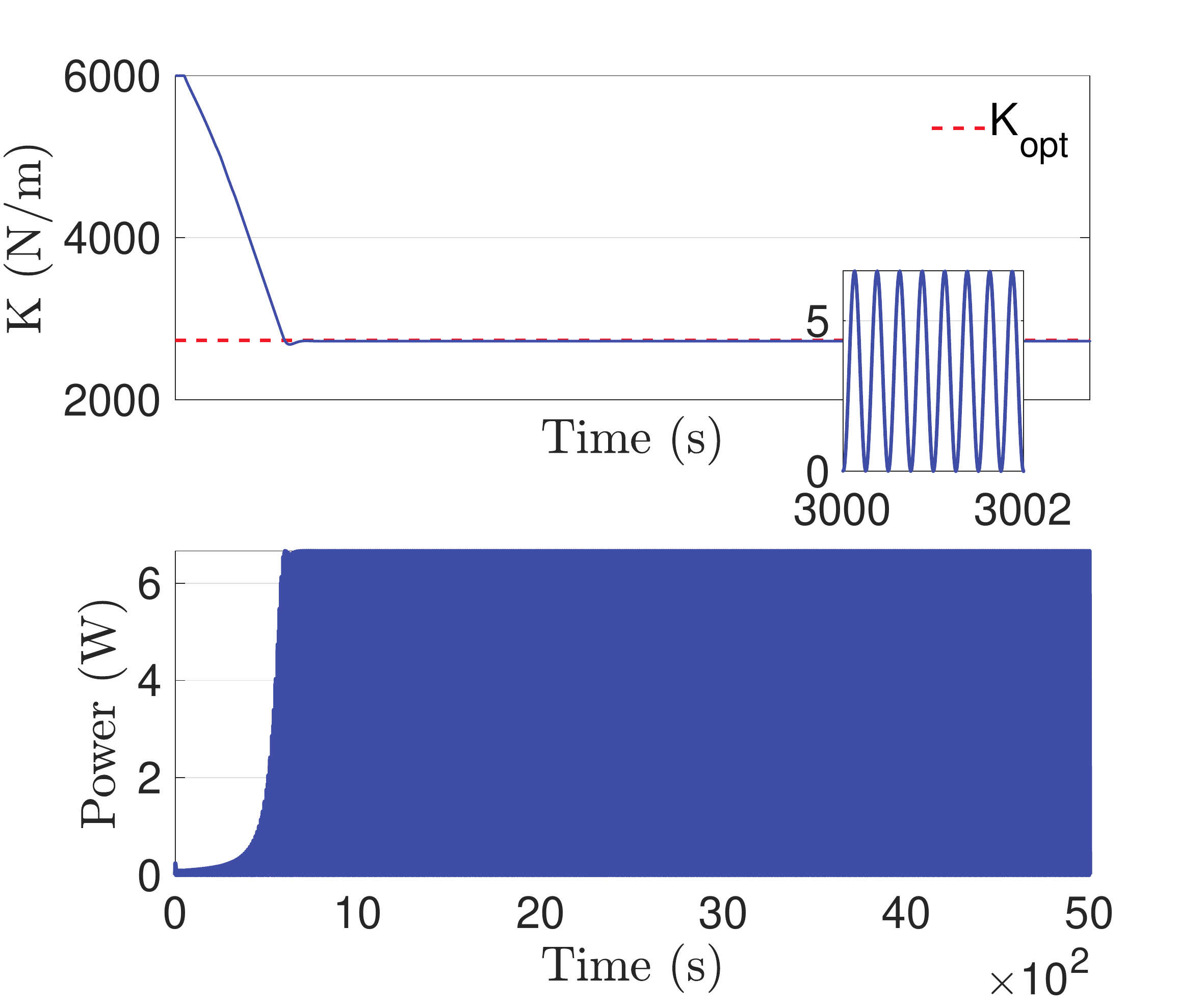}
	    \label{subfig:pesc_power}
    }
    \subfigure[Sliding mode ES] {
	    \includegraphics[scale=0.33]{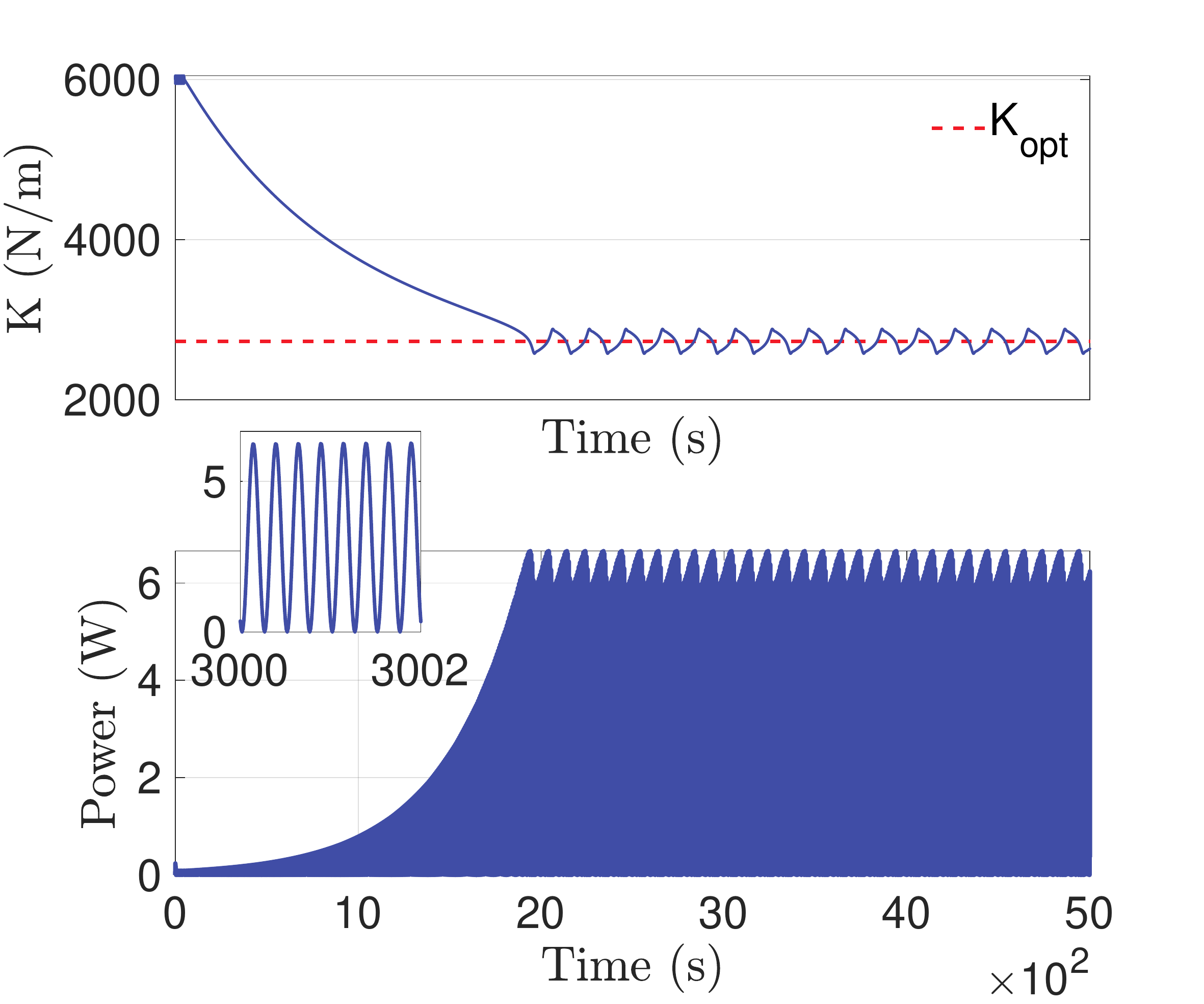}
	    \label{subfig:smesc_power}
    }
    \caption{Extracted power $P(t)$ during reactive coefficient optimization using~\subref{subfig:pesc_power} perturbation-based ES, and~\subref{subfig:smesc_power} sliding mode ES.}
    \label{fig:msd_power_sm_pesc}
\end{figure}

Next,  we perform simultaneous optimization of both PTO coefficients. The results are shown in Fig.~\ref{fig:msd_multi_k_b}
for four ES algorithms; the self-driving ES method did not converge reliably for the two-parameter optimization problem (divergent data not presented). Based upon the results obtained in this section, we make some remarks about different algorithms:

\begin{itemize}

  \item The sliding mode ES proved quite robust, and did not require re-tuning of parameters for varying conditions such as frequency and/or amplitude of the external force. Moreover, the number of parameters to tune are relatively small. A drawback of the scheme is that the steady-state solution oscillates near the optima, as seen in Fig.~\ref{subfig:smesc_msd_k}. This however, results in a negligible variation 
in the PTO power, as shown in Fig.~\ref{fig:msd_power_sm_pesc}, which compares the extracted power as a function of time using perturbation-based and sliding mode ES methods. The former controller produces negligible oscillation in the steady-state solution of $K$ coefficient.

  \item The self-driving ES achieves convergence to the optimal value without steady-state oscillations. It is relatively robust, but the main drawback includes tuning of a large number of parameters. Also, it did not converge reliably for the multi-parameter optimization problem using the algorithm described in Sec.~\ref{sec_sd_algo}.
  
 \item The relay ES is also relatively insensitive to the frequency and/or amplitude variation of the external force. It is simple to tune as well. However, it also results in steady-state oscillations in the solution. 
 
  \item The LSQ-ES proved more robust than the relay ES, having roughly the same number of parameters to tune. It also results in less oscillations in the steady-state solution compared to the relay ES, although they are not completely eliminated. 
   
  \item Perturbation-based ES is one of the more popular ES schemes used in the literature. The main advantage of this method is its stability, although the choice of  cut-off frequencies of the filters largely dictates its performance. The method can achieve a controlled amount of oscillation in the converged solution, as seen in Figs.~\ref{subfig:pesc_k}~\ref{subfig:pesc_b}, and~\ref{subfig:pesc_k_b}. The rate of convergence of this method also depends upon the frequency of the external perturbation signal used to estimate the gradient. 
\end{itemize}

%% file: CylinderRegularWavesResults.tex
\begin{figure}[]
 \centering
 \includegraphics[scale=0.4]{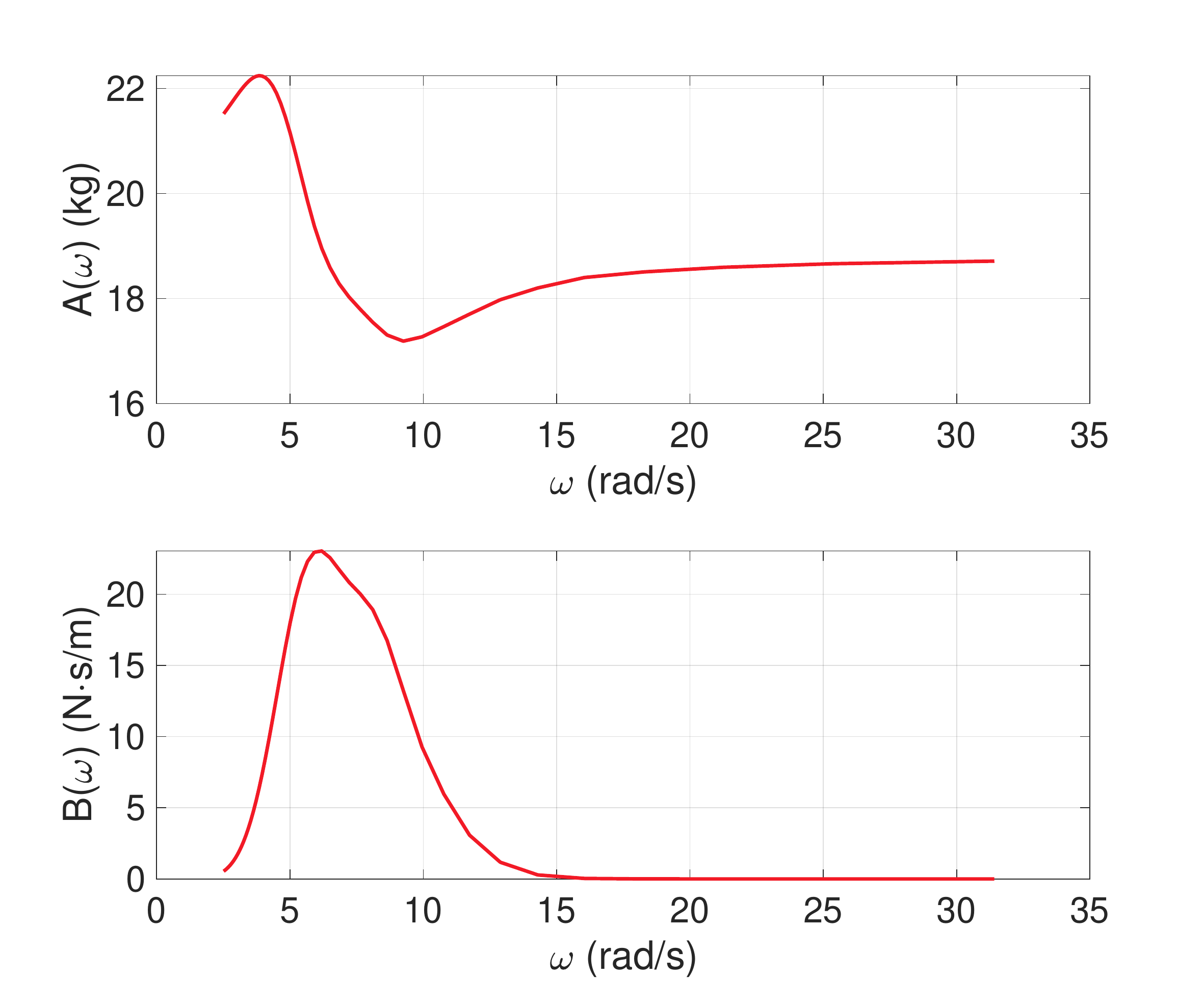}
 \caption{Frequency-dependent added-mass $A(\omega)$, and radiation damping $B(\omega)$ for the two-dimensional cylindrical buoy. The coefficients are obtained using BEM-based ANSYS AQWA software.}
 \label{fig:A_B_Cyl}
\end{figure}

\begin{figure}[]
    \centering
        \subfigure[Sea state Reg.1]{
        		\includegraphics[scale=.32]{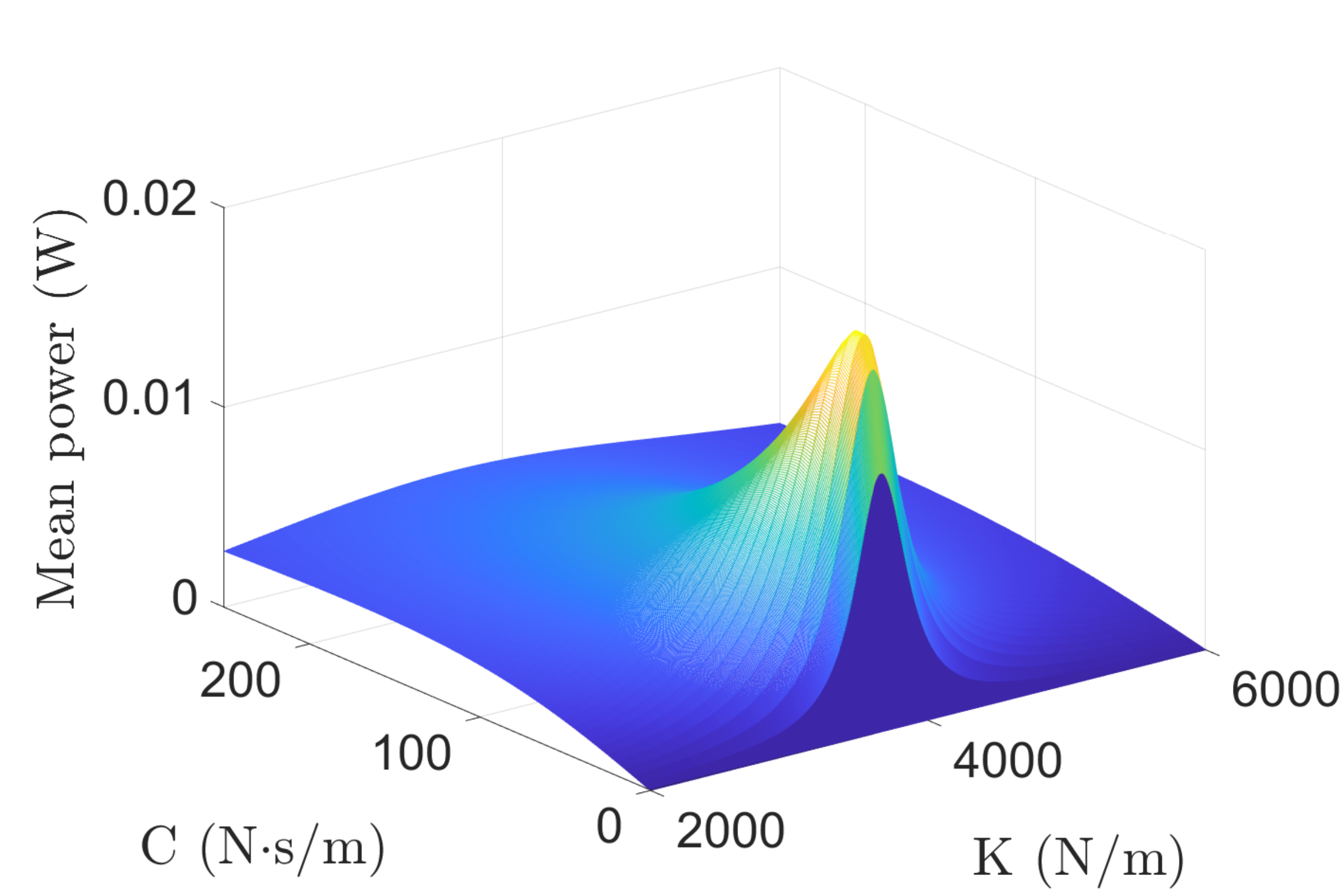}
		\label{subfig:reg1}
	} 
        \subfigure[Sea state Reg.2]{
        		\includegraphics[scale=.32]{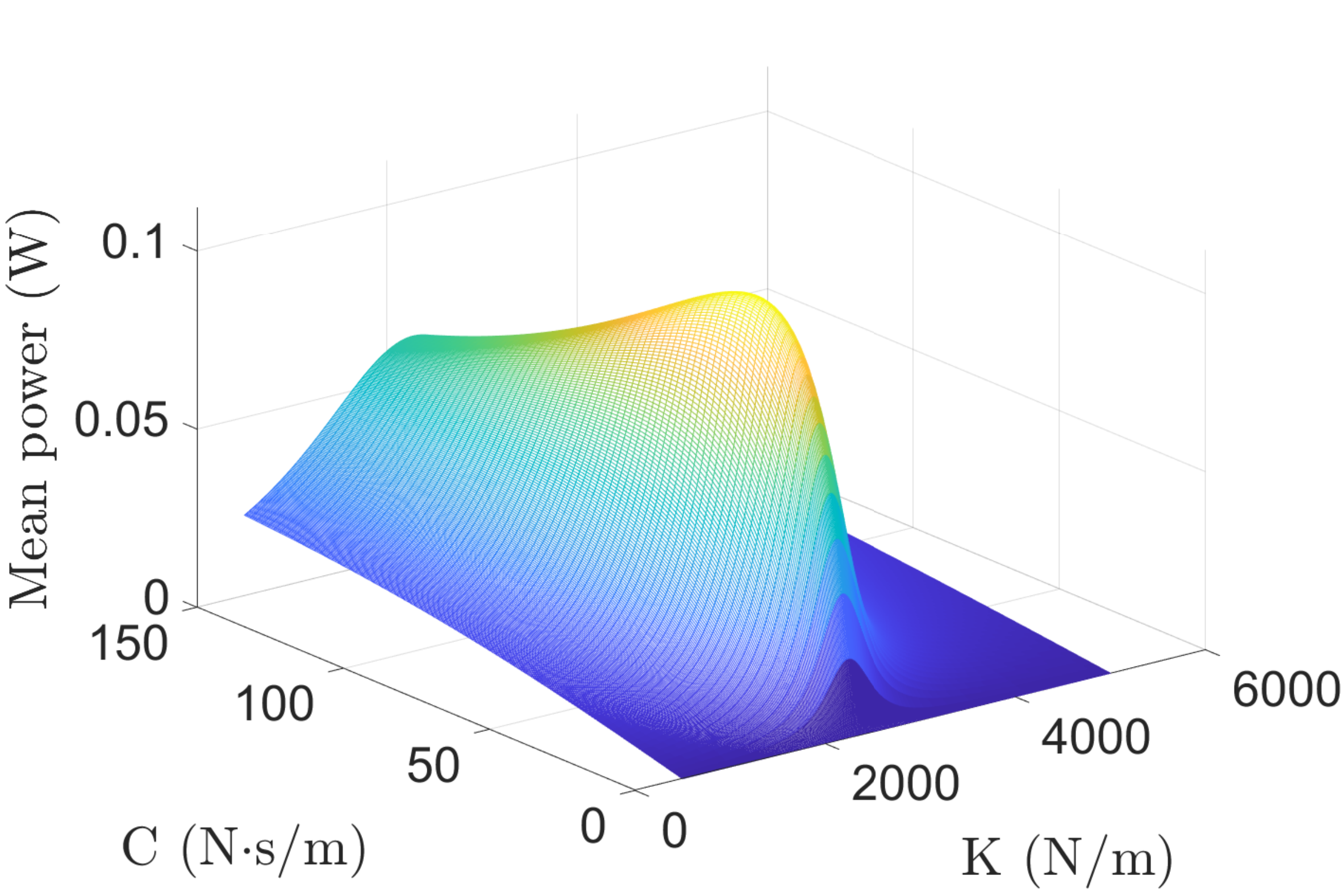}
		\label{subfig:reg2}
	}
        \subfigure[Sea state Reg.3]{
        		\includegraphics[scale=0.32]{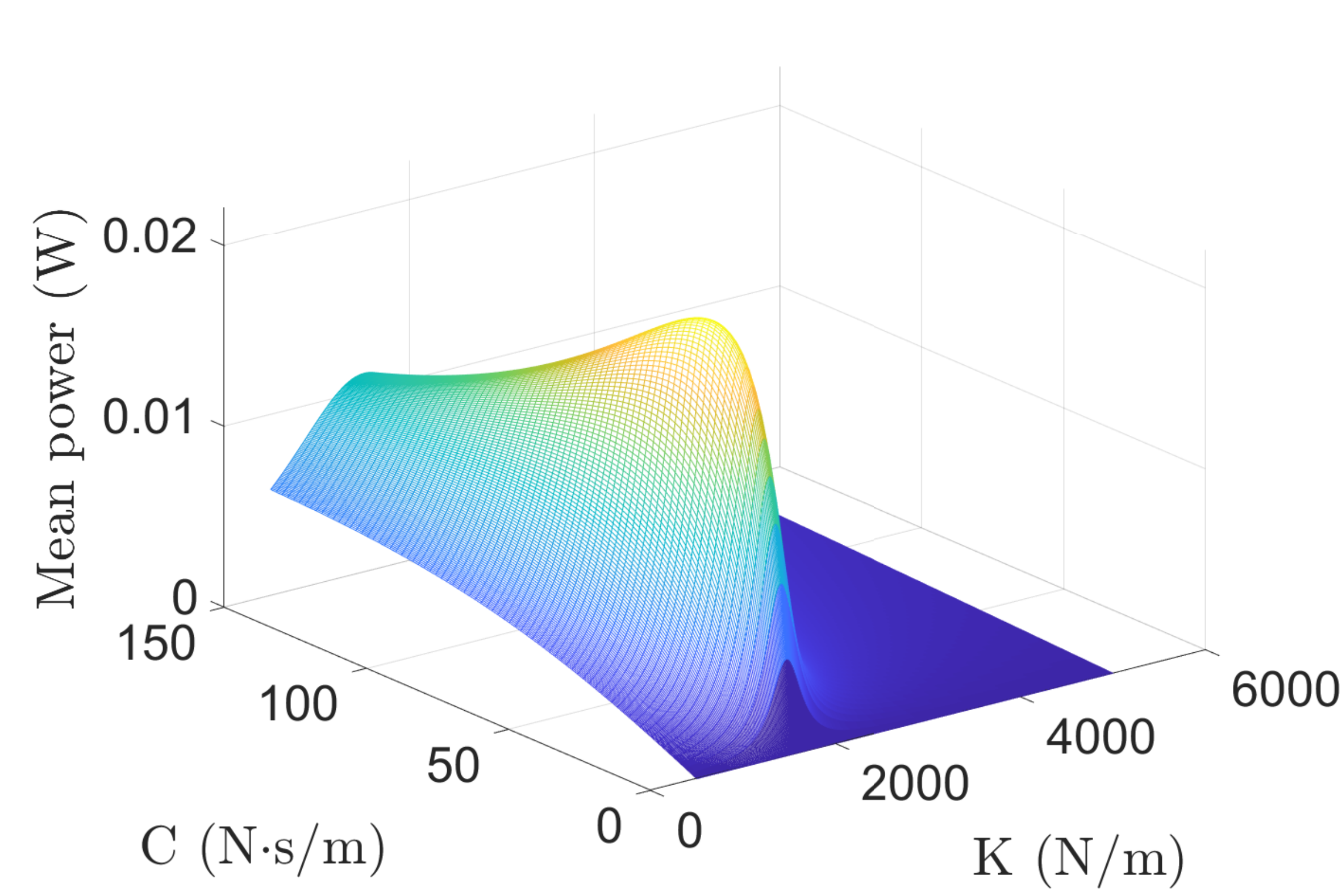}
		\label{subfig:reg3}
	}
    \caption{Power vs. PTO coefficients reference-to-output map for a two-dimensional cylinder subject to regular waves. The optimal PTO coefficients are: \subref{subfig:reg1} $K_\text{opt} = 3720$ N/m and $C_\text{opt}=18$ N$\cdot$s/m; \subref{subfig:reg2} $K_\text{opt} = 2290$ N/m and $C_\text{opt}=34$ N$\cdot$s/m; and \subref{subfig:reg3} $K_\text{opt} = 1530$ N/m and $C_\text{opt} = 30$ N$\cdot$s/m.}
    \label{fig:cyl_maps_reg}
\end{figure}

Next, we consider a two-dimensional cylindrical buoy of Sec.~\ref{sec_hull_waves} subject to regular waves. Three regular waves of different height and time period are considered (Table~\ref{tab:sea_states}). Fig.~\ref{fig:A_B_Cyl} plots the frequency-dependent added mass $A(\omega)$ and frequency-dependent radiation damping $B(\omega)$ for the two-dimensional cylinder; the plotted values can be used to estimate the optimal reactive and resistive coefficients of the PTO mechanism, using Eqs.~\eqref{eqn_kopt_pa} and \eqref{eqn_copt_pa}, respectively. Since the theoretical formula ignores the viscous drag force, we perform a brute-force search of the parametric space to find the optimal values of the resistive coefficient. Fig.~\ref{fig:cyl_maps_reg} plots power vs. coefficients reference-to-output map, and Table~\ref{tab:cyl_opt_coefs_reg} lists their optimal values for three different sea states. The tabulated values confirm that the theoretical estimates of the optimal reactive coefficients are quite accurate, whereas the optimal resistive coefficients are under-predicted.  

\begin{table}[]
 \centering
 \caption{Optimal PTO coefficients for a two-dimensional cylindrical buoy subject to regular waves using impedance-matching control theory and through a  brute-force search. Units: $\cT$ is in s, $\cH$ is in m, $K$ is in N/m, and $C$ is in N$\cdot$s/m.}
  \rowcolors{2}{}{gray!10}
	\begin{tabular}{c c c c c c c}
	\toprule
	Sea state ID   & $\cT$  & $\cH$   & $K_\text{opt,map}$ & $C_\text{opt,map}$ & $K_\text{opt,analytical}$ & $C_\text{opt,analytical}$\\ 
	\midrule
	Reg.1 &  0.625 &  0.01      &  3720             &  18                &  3717                      &  9                          \\ 
	Reg.2 &  0.8     &  0.02      &  2290             &  34                &  2302                      &  20                         \\ 
	Reg.3 &  1        &  0.0075  &  1530             &  30                &  1534                      &  21                         \\ 
	\bottomrule
\end{tabular}
\label{tab:cyl_opt_coefs_reg}
\end{table}

\begin{figure}[]
    \centering
        \subfigure[Sliding mode ES]{
        		\includegraphics[scale=.32]{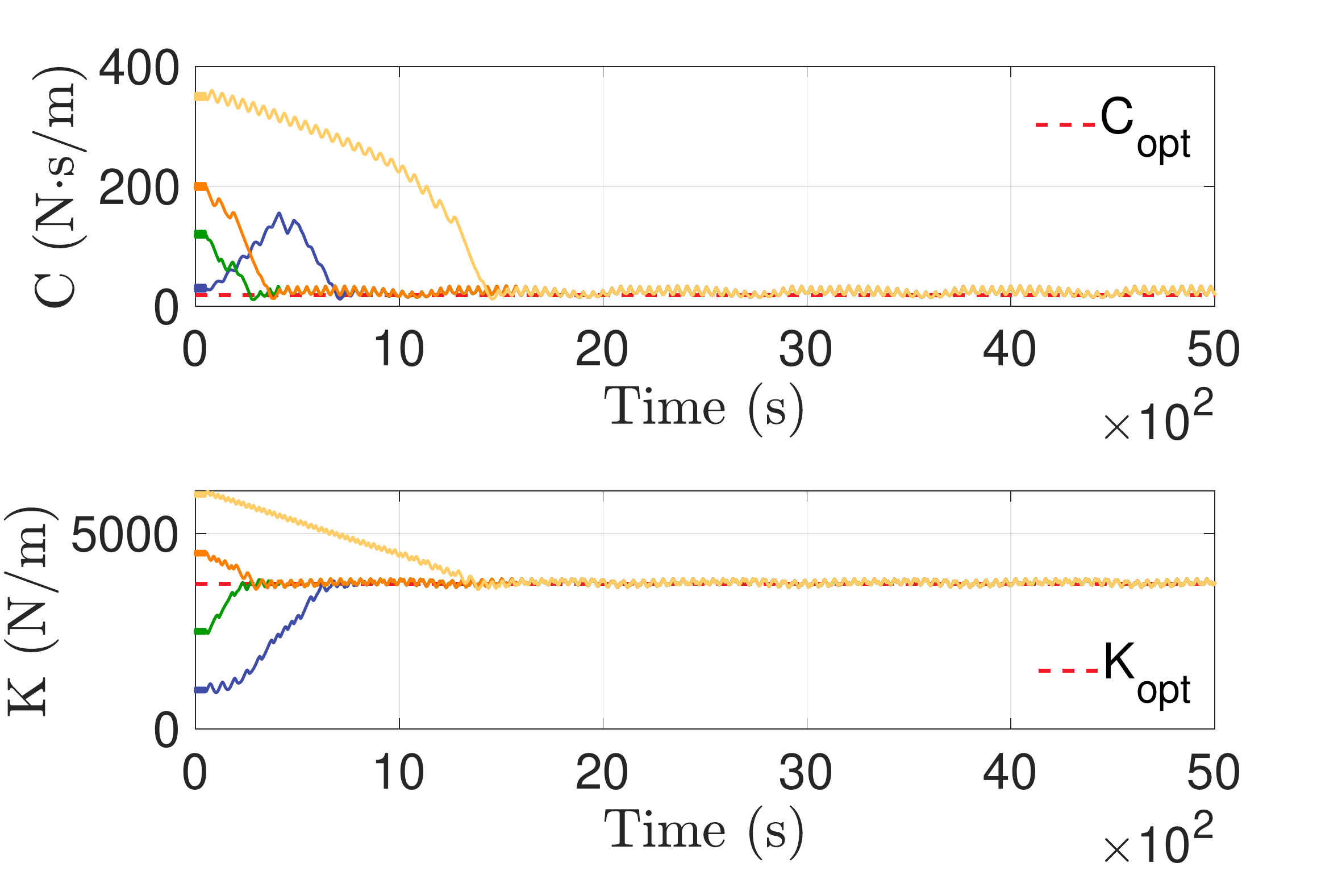}
	} 
       \subfigure[Relay ES]{
       		\includegraphics[scale=.32]{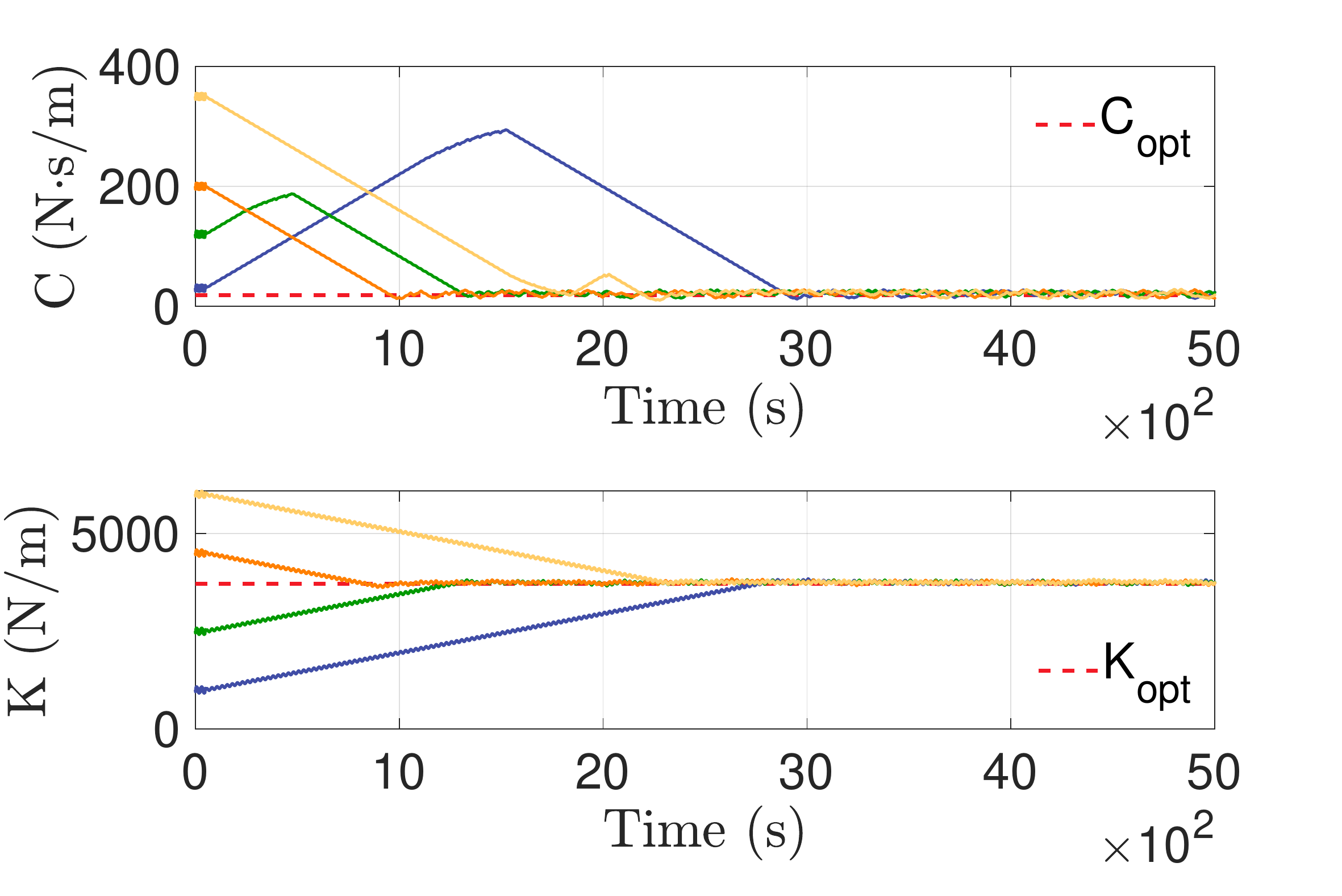}
	}
        \subfigure[LSQ-ES]{
        		\includegraphics[scale=.32]{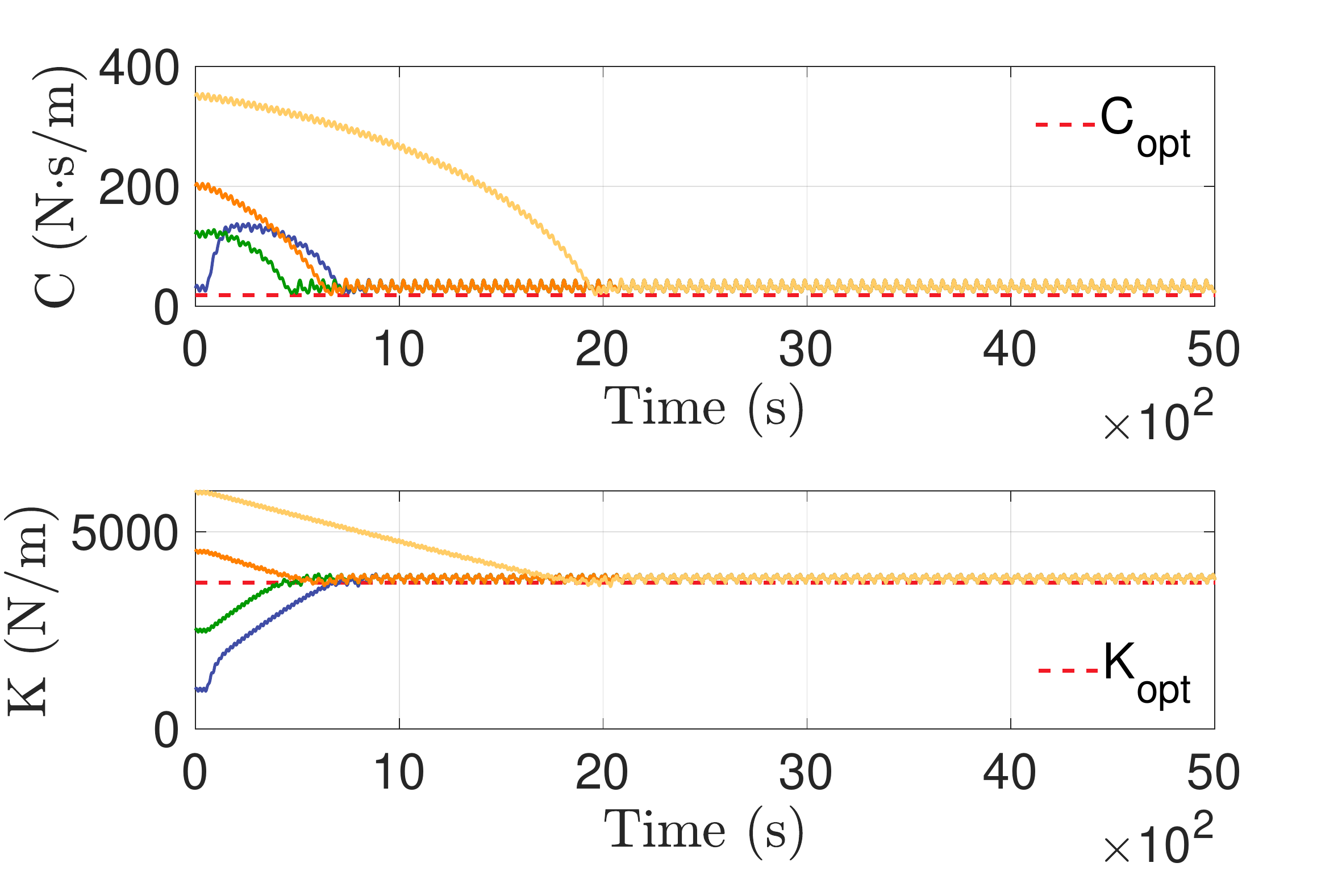}
	} 
        \subfigure[Perturbation-based ES]{
        		\includegraphics[scale=0.32]{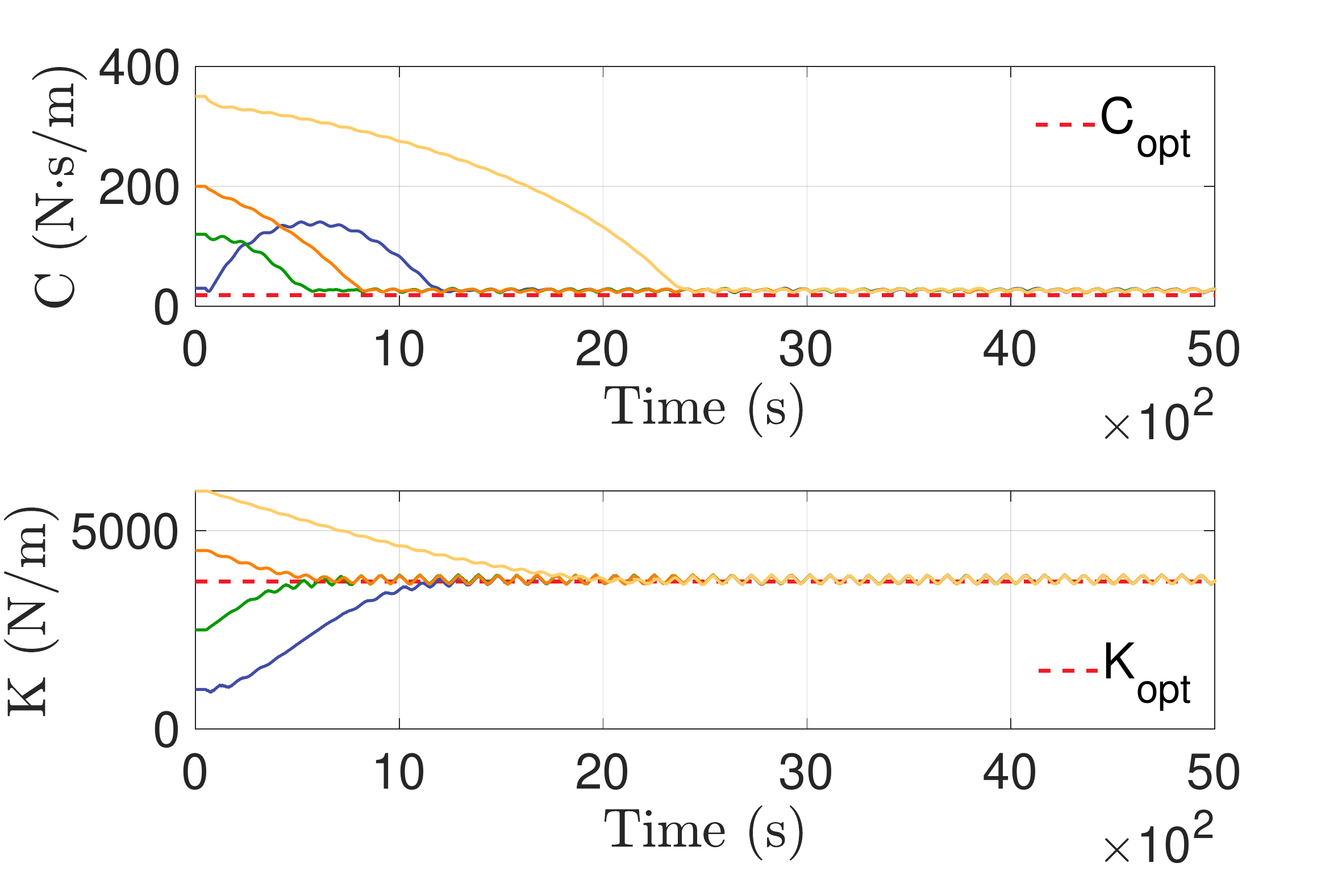}
	} 
    \caption{Optimization of reactive and resistive coefficients, $K$ and $C$, respectively, for the cylindrical buoy in regular sea ``Reg.1" using different ES algorithms. The  optimal $K_\text{opt}=3720$ N/m and $C_\text{opt} = 18$ N$\cdot$s/m values are indicated by dashed lines in the plots.}
    \label{fig:cyl_multi_k_b_reg1}
\end{figure}

\begin{figure}
    \centering
       \subfigure[]{
       		\includegraphics[scale=.32]{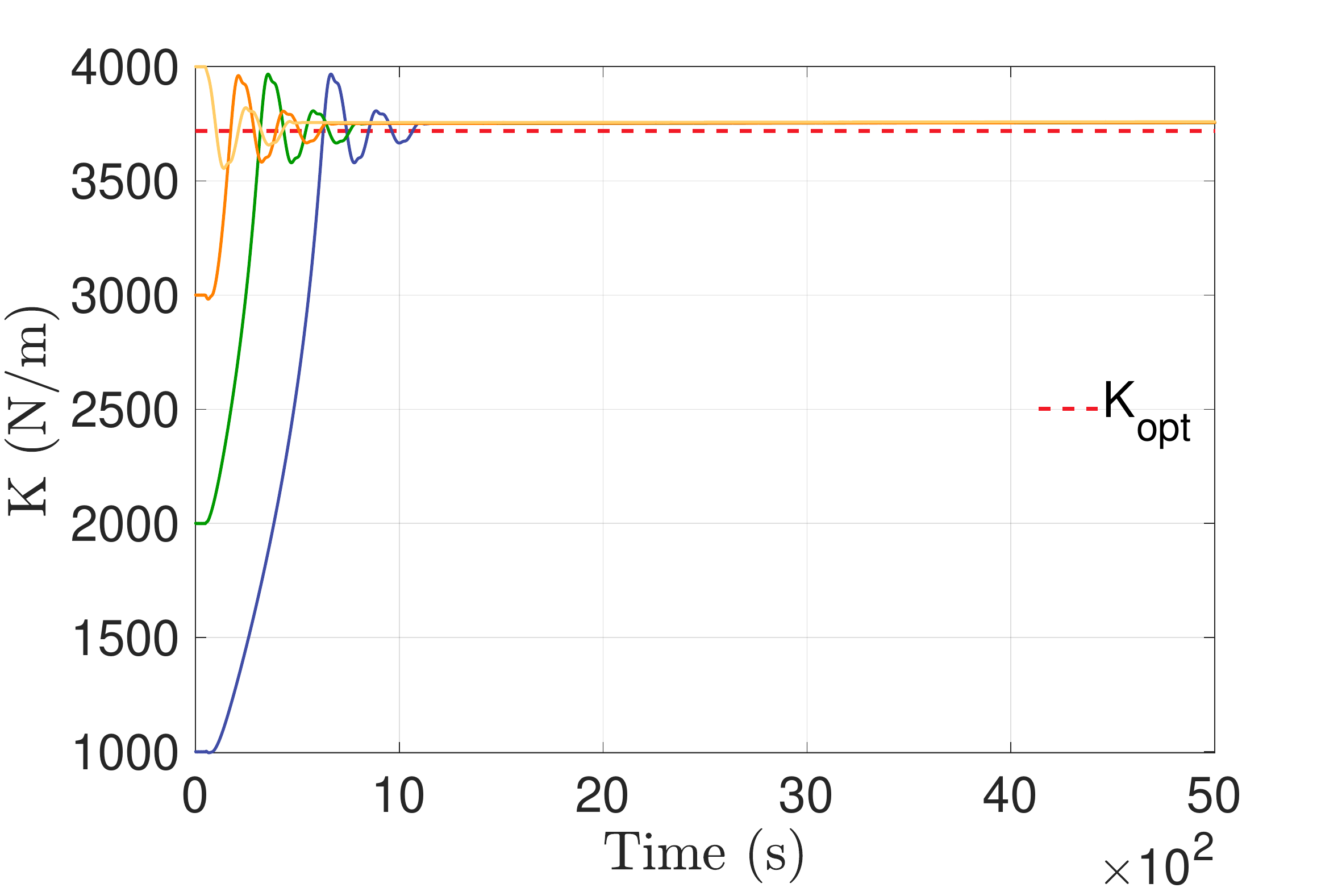}
		\label{subfig:sd_k_reg1}
	}
	\subfigure[]{
       		\includegraphics[scale=.32]{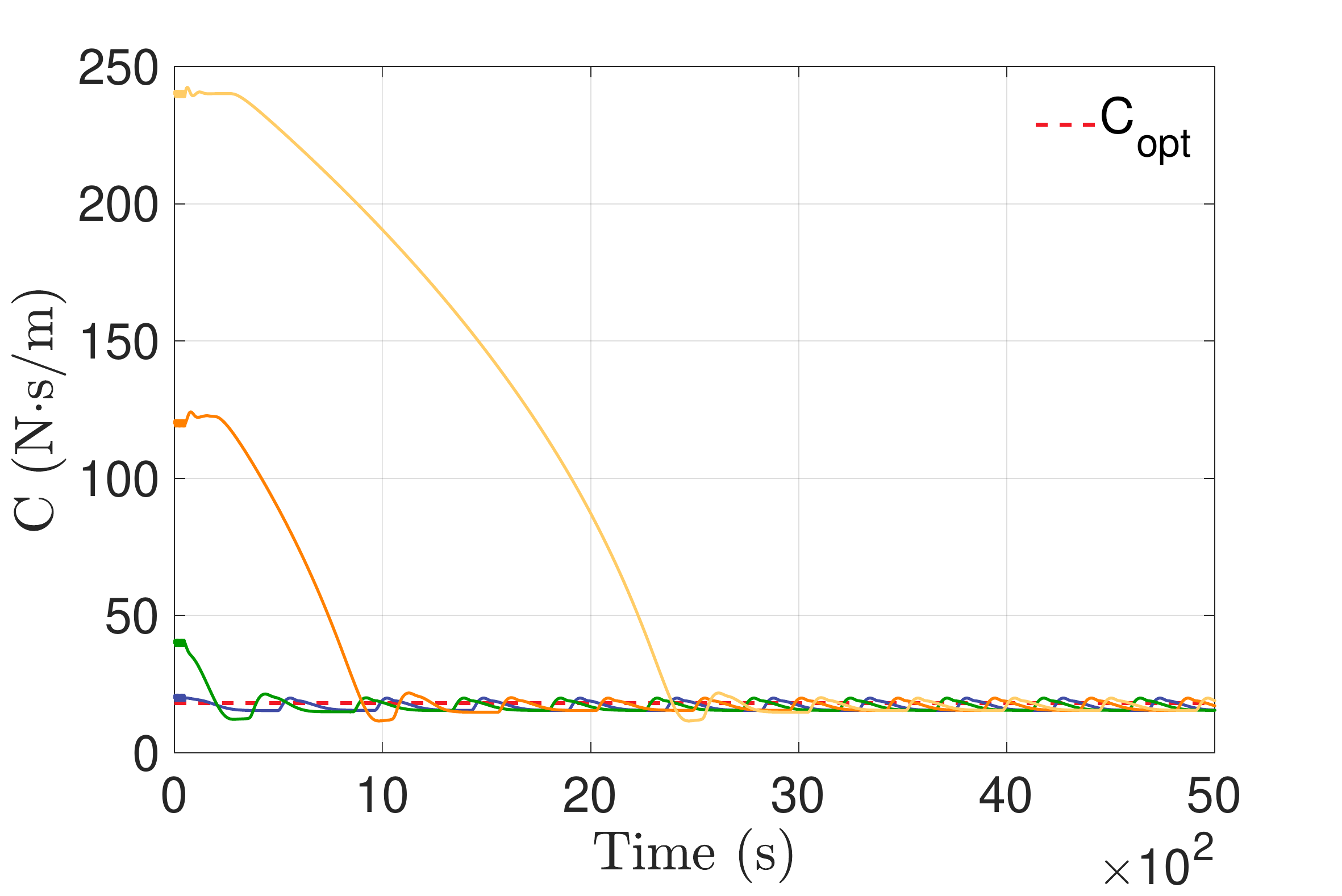}
		\label{subfig:sd_b_reg1}
	}
    \caption{Optimization of~\subref{subfig:sd_k_reg1} reactive PTO coefficient $K$; and~\subref{subfig:sd_b_reg1} resistive PTO coefficient $C$ using self-driving ES algorithm for cylindrical buoy in the regular sea ``Reg.1". The optimal $K_\text{opt}=3720$ N/m and $C_\text{opt} = 18$ N$\cdot$s/m values are indicated by dashed lines in the plots.}
    \label{fig:cyl_sd_k_b_reg1}
\end{figure}

Fig.~\ref{fig:cyl_multi_k_b_reg1} shows the convergence history of $K$ and $C$ coefficients using four ES algorithms for the regular sea state ``Reg.1". As can be seen in the figure, all four algorithms converge to the optimum values of the PTO coefficients. Moreover, their convergence behavior is similar to the mechanical oscillator problem of Sec.~\ref{sec_msd}. Similar to the mass-spring-damper case, the self-driving ES method did not converge for the two-parameter optimization problem. However, oscillation-free steady-state solutions are obtained, when self-driving ES is used to optimize either $K$ or $C$. The single-parameter optimization results using self-driving ES method are shown in Fig.~\ref{fig:cyl_sd_k_b_reg1}. 

\begin{figure}[]
    \centering
        \subfigure[Sliding mode ES]{
        		\includegraphics[scale=.32]{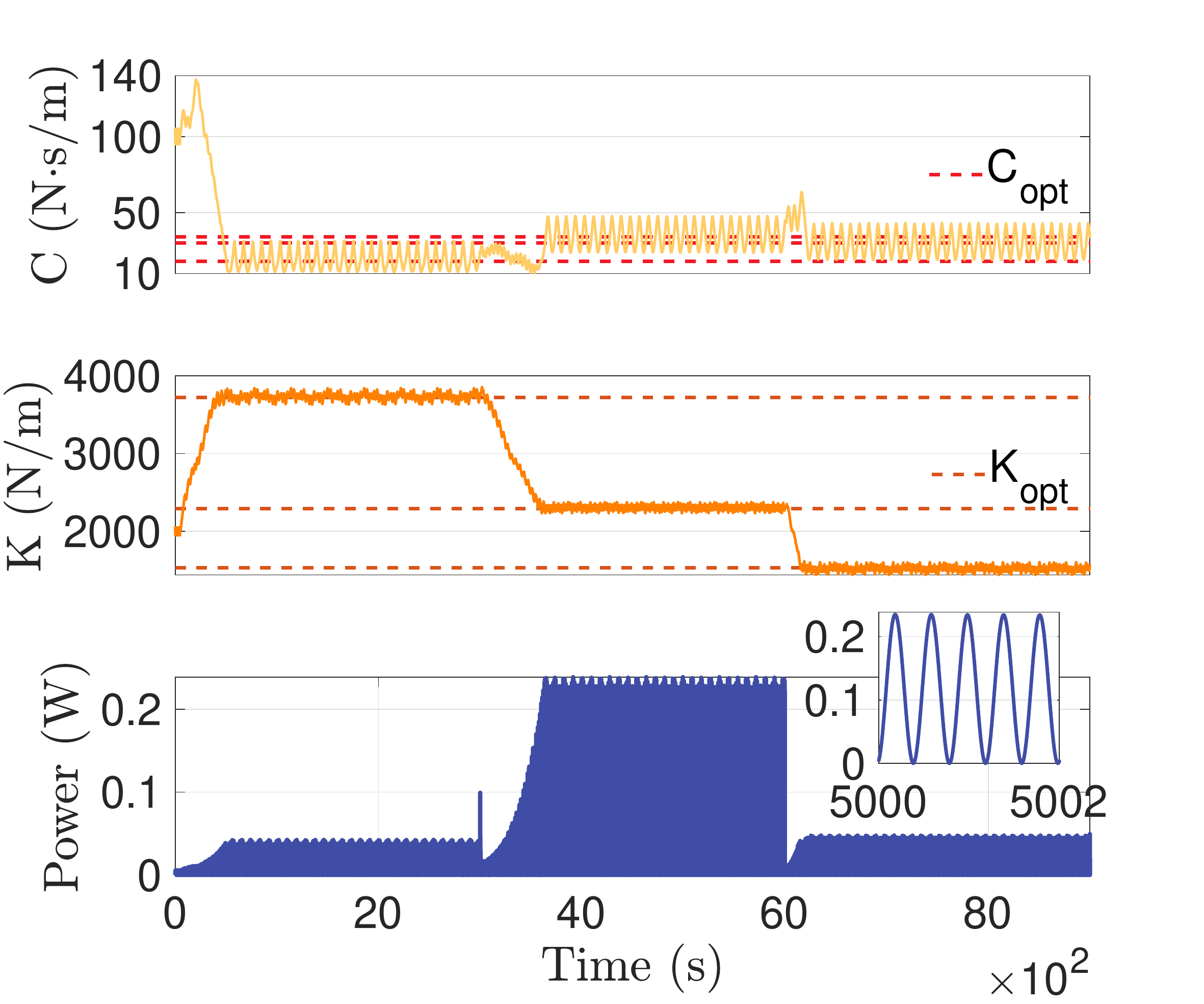}
	} 
       \subfigure[Relay ES]{
       		\includegraphics[scale=.32]{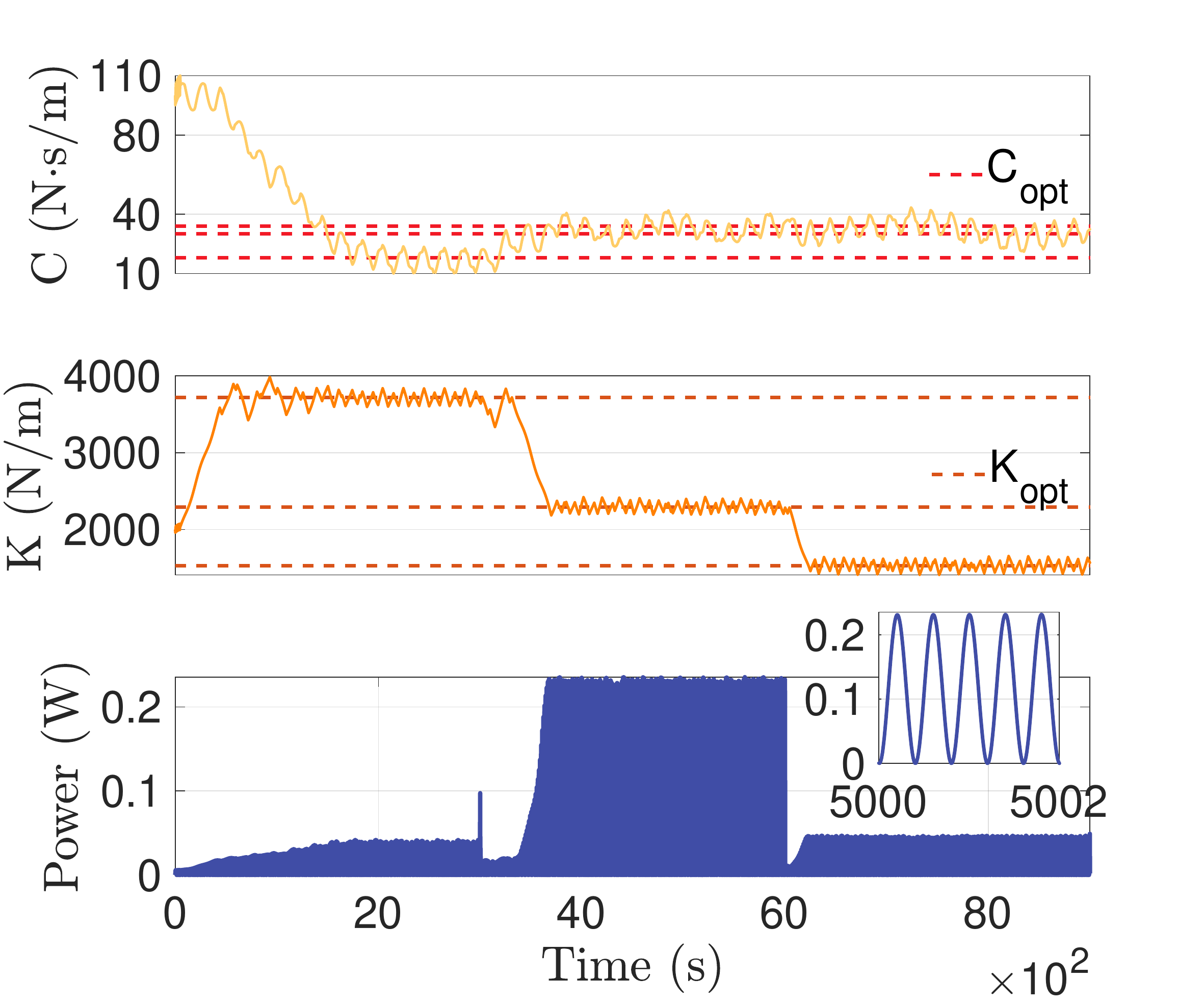}
	}
        \subfigure[LSQ-ES]{
        		\includegraphics[scale=.32]{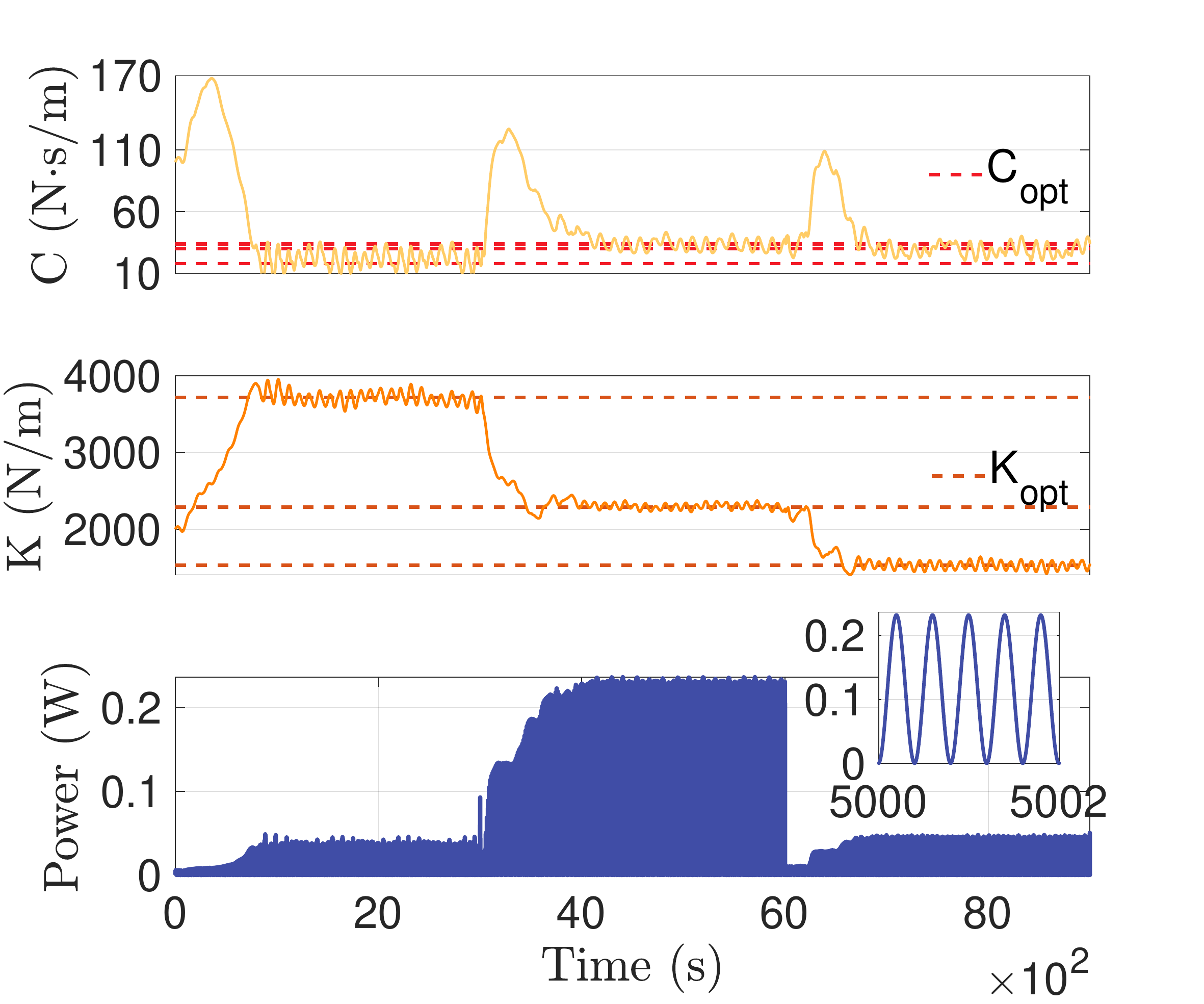}
	} 
        \subfigure[Perturbation-based ES]{
        		\includegraphics[scale=0.32]{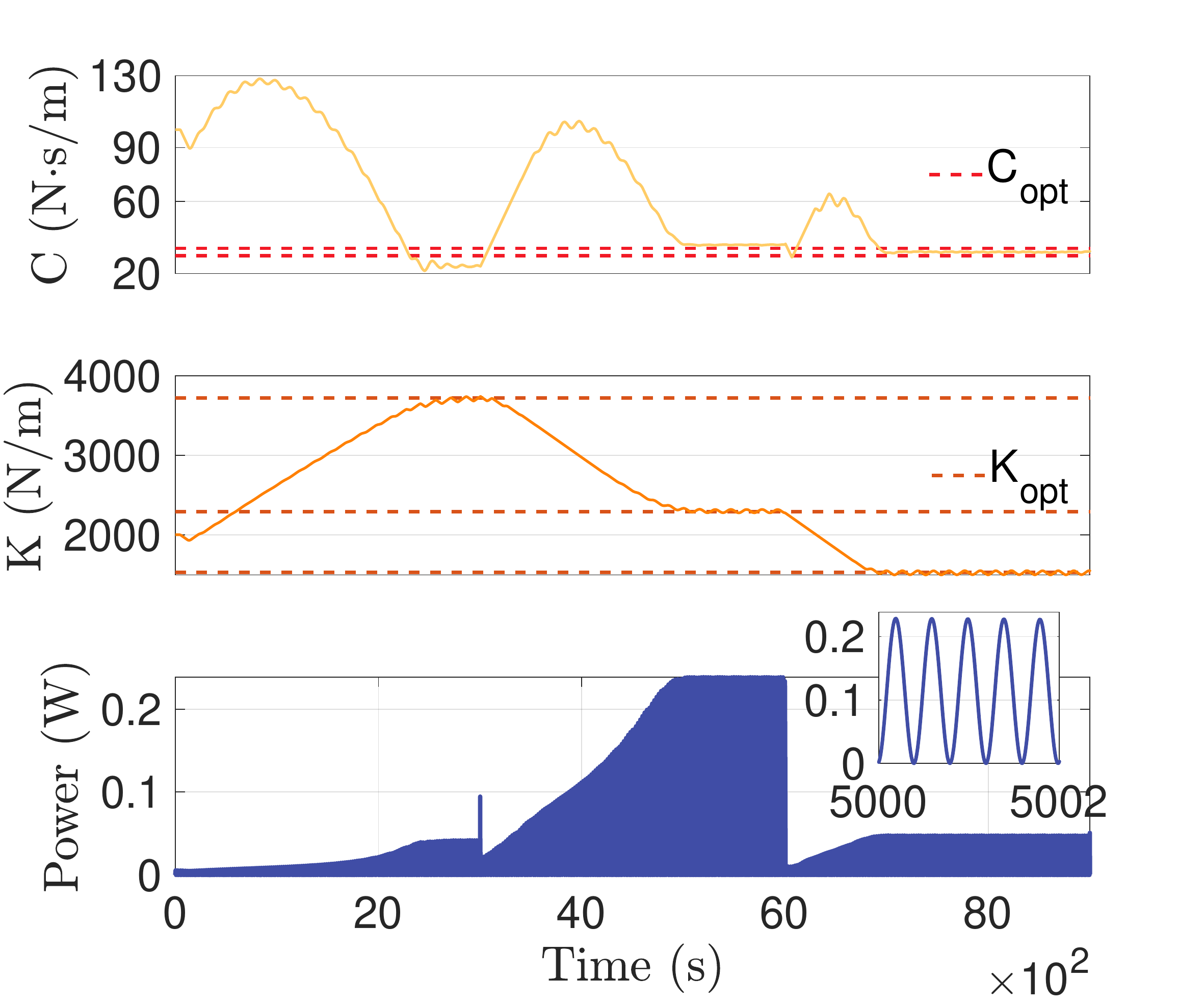}
	} 
    \caption{Optimization of reactive and resistive coefficients, $K$ and $C$, respectively, for the cylindrical buoy subject to changing sea states using different ES algorithms. The optimal $K_\text{opt}$ and $C_\text{opt}$ values in three different sea states from Table~\ref{tab:cyl_opt_coefs_reg} are indicated by dashed lines in the plots.}
    \label{fig:adaptive_cyl_k_b}
\end{figure}

Next, we test the adaptive capability of ES algorithms. Starting with sea state ``Reg.1", the wave conditions for the cylindrical buoy are changed to ``Reg.2", and then subsequently to ``Reg.3". Fig.~\ref{fig:adaptive_cyl_k_b} shows that all four ES algorithms can reliably adapt to the changing wave conditions, and adjust the PTO coefficients automatically to achieve optimal performance in each sea state. The results also confirm that extremum-seeking control algorithms do not require any wave forecasting/prediction information to attain the optimum. Indeed, the performance function used in the extremum-seeking algorithm requires only on-board instrumentation to estimate the absorbed PTO power. Thus, ES can be used as a \emph{causal} controller for WECs.

%% file: CylinderIrregularWavesResults.tex
 \begin{table}[]
 \centering
 \caption{Optimal PTO coefficients for a two-dimensional cylindrical buoy subject to irregular waves using impedance-matching control theory and through a brute-force search. Units: $\cT_{\rm p}$ is in s, $\cH_{\rm s}$ is in m, $K$ is in N/m, and $C$ is in N$\cdot$s/m.}
  \rowcolors{2}{}{gray!10}
	\begin{tabular}{c c c c c}
	\toprule
	Sea state ID   & $\cT_{\rm p}$  & $\cH_{\rm s}$   & $K_\text{opt,map}$ & $C_\text{opt,map}$ \\ 
	\midrule
	Irreg.1 &  0.625 &  0.01      &  3440             &  32          \\ 
	Irreg.2 &  0.8     &  0.02      &  2170             &  44          \\ 
	Irreg.3 &  1        &  0.0075  &  1480             &  40          \\ 
	\bottomrule
\end{tabular}
\label{tab:cyl_opt_coefs_irreg}
\end{table}
 
\begin{figure}[]
    \centering
        \subfigure[Sea state Irreg.1]{
        		\includegraphics[scale=.32]{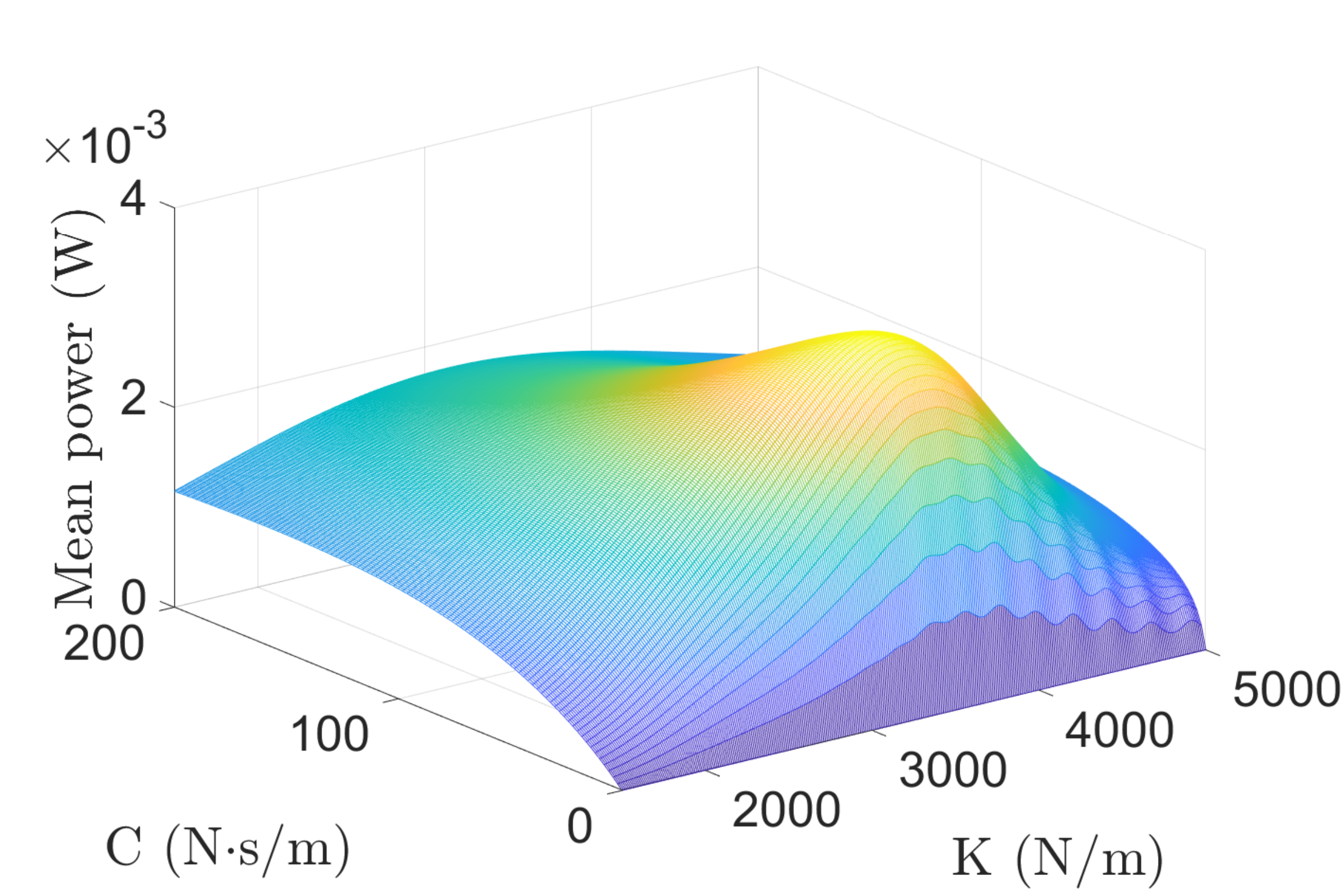}
		\label{subfig:irreg1}
	} 
        \subfigure[Sea state Irreg.2]{
        		\includegraphics[scale=.32]{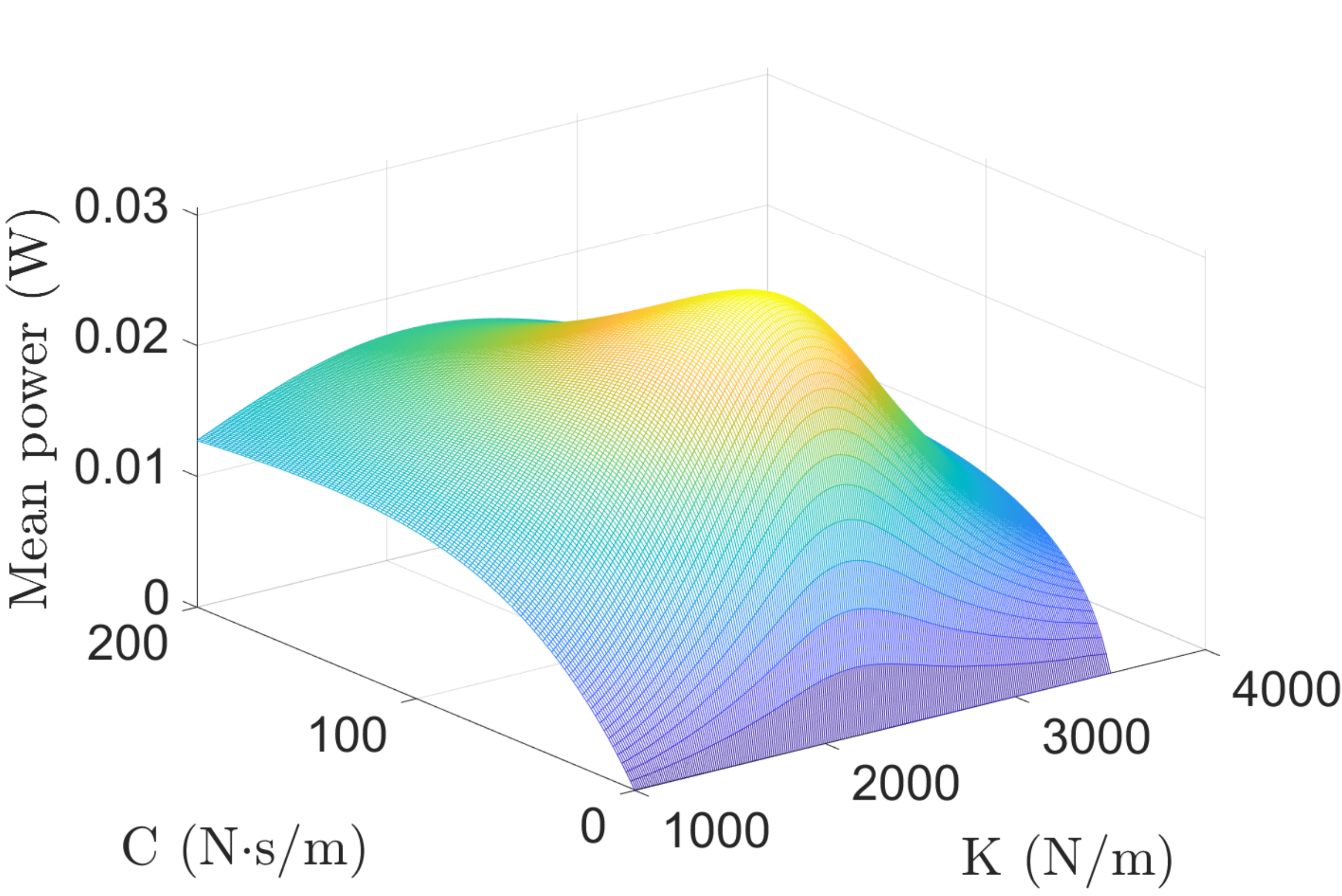}
		\label{subfig:irreg2}
	}
        \subfigure[Sea state Irreg.3]{
        		\includegraphics[scale=0.32]{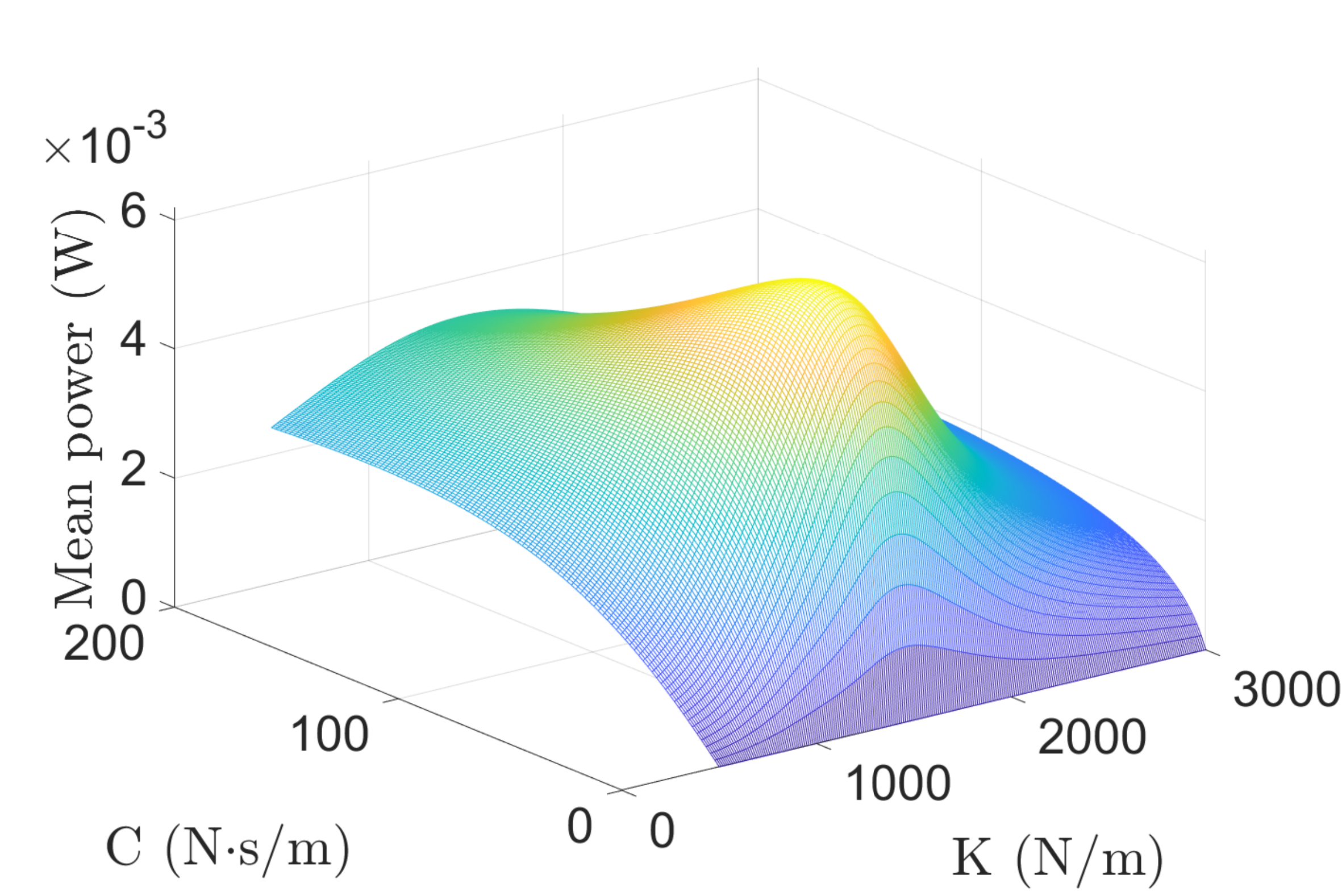}
		\label{subfig:irreg3}
	}
    \caption{Power vs. PTO coefficients reference-to-output map for a two-dimensional cylinder operating in an irregular sea. The optimal PTO coefficients are: \subref{subfig:irreg1} $K_\text{opt} = 3440$ N/m and $C_\text{opt}=32$ N$\cdot$s/m; \subref{subfig:reg2} $K_\text{opt} = 2170$ N/m and $C_\text{opt}=44$ N$\cdot$s/m; and \subref{subfig:reg3} $K_\text{opt} = 1480$ N/m and $C_\text{opt} = 40$ N$\cdot$s/m.}
    \label{fig:cyl_maps_irreg}
\end{figure}

 A WEC device subject to irregular waves has multiple frequencies in its dynamics. Consequently, a closed-form solution for energy-maximizing PTO coefficients is difficult to obtain analytically. To verify the controller results, we perform a brute-force search of the parametric space to find the optimal values of the PTO coefficients for three irregular sea states.  Fig.~\ref{fig:cyl_maps_irreg} plots power vs. coefficients reference-to-output map, and Table~\ref{tab:cyl_opt_coefs_irreg} lists their optimal values. 

\begin{figure}
    \centering
        \subfigure[Sliding mode ES]{
        		\includegraphics[scale=.33]{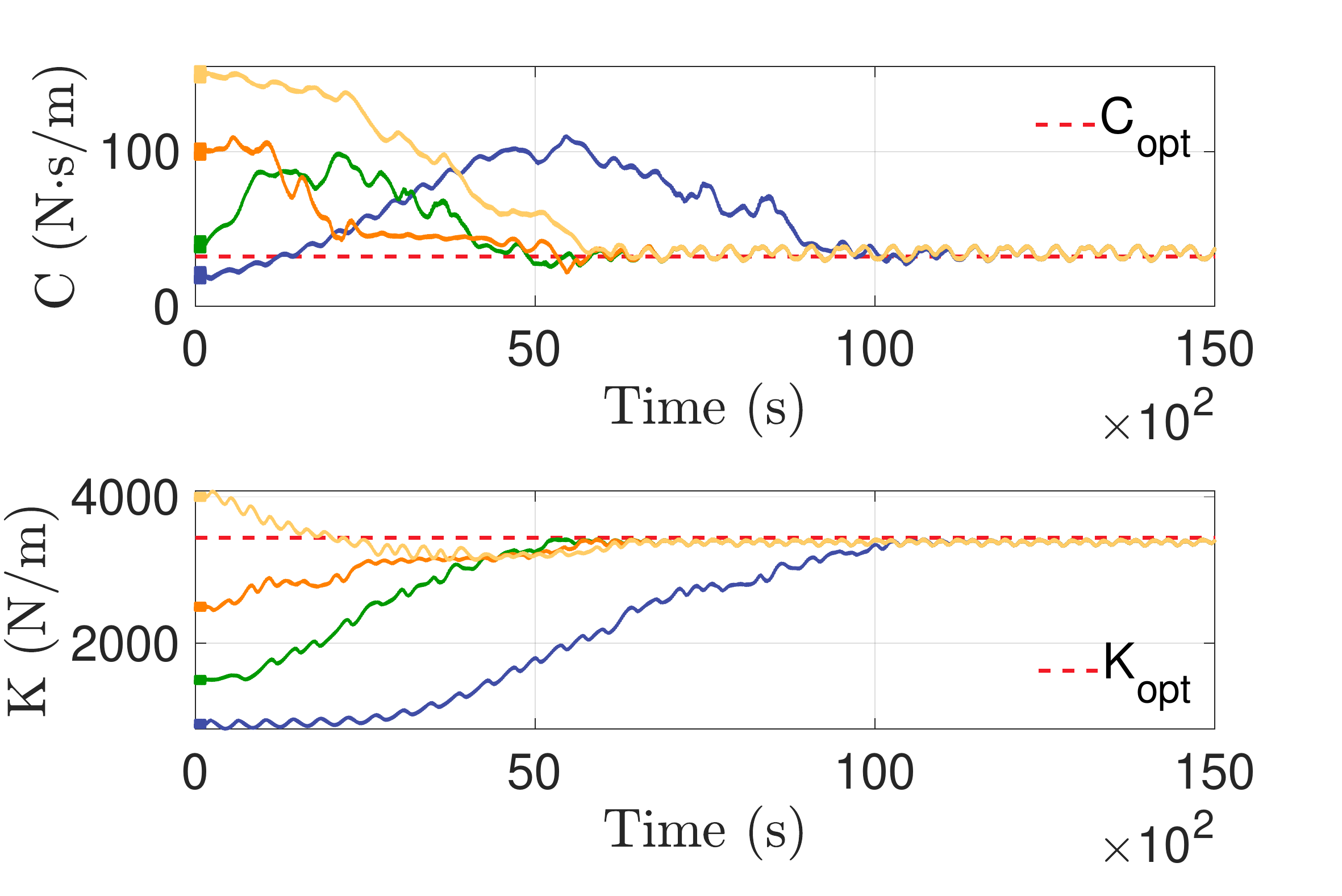}
        } 
        \subfigure[Relay ES]{
        		\includegraphics[scale=.33]{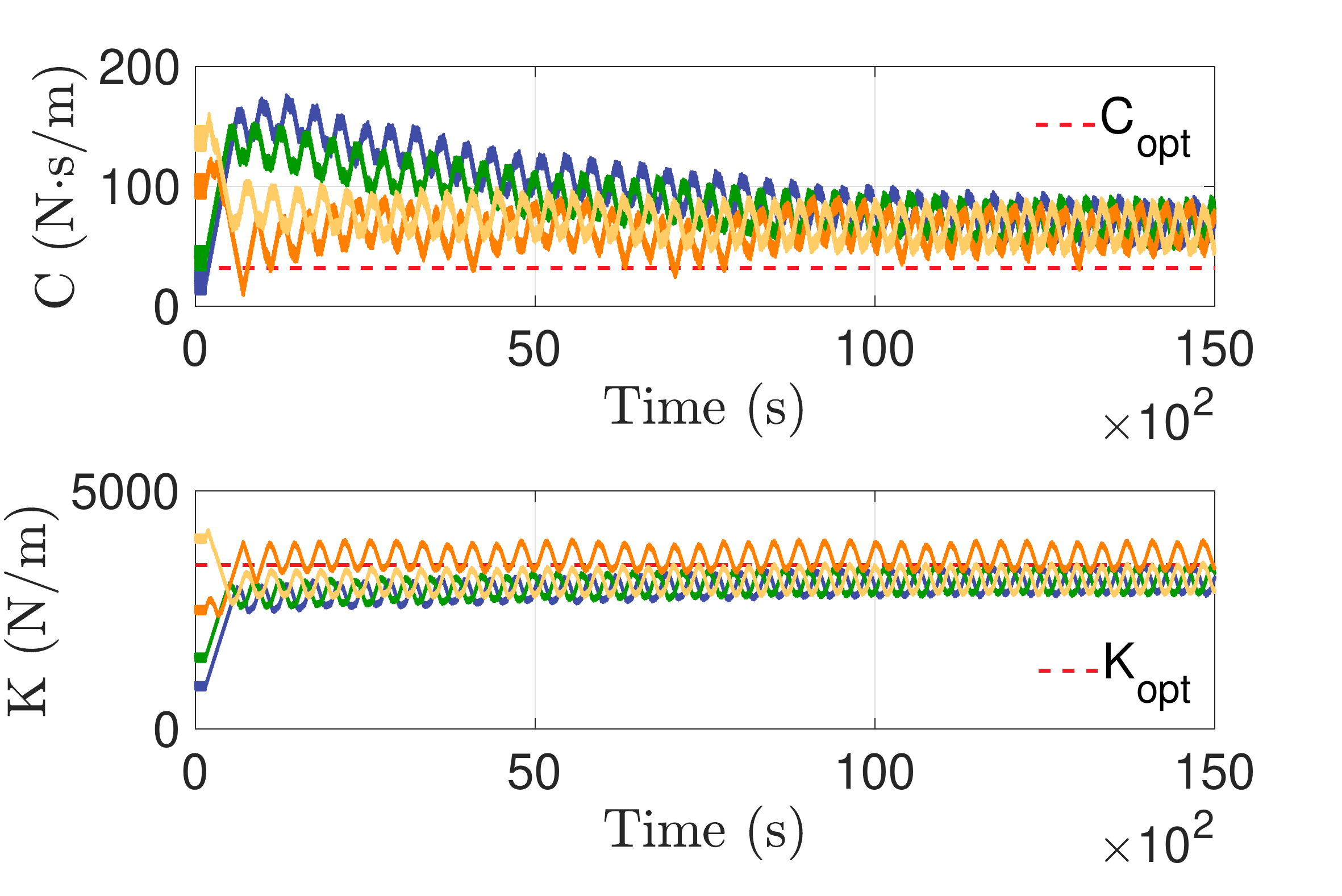}
	} 
        \subfigure[LSQ-ES]{
        		\includegraphics[scale=0.33]{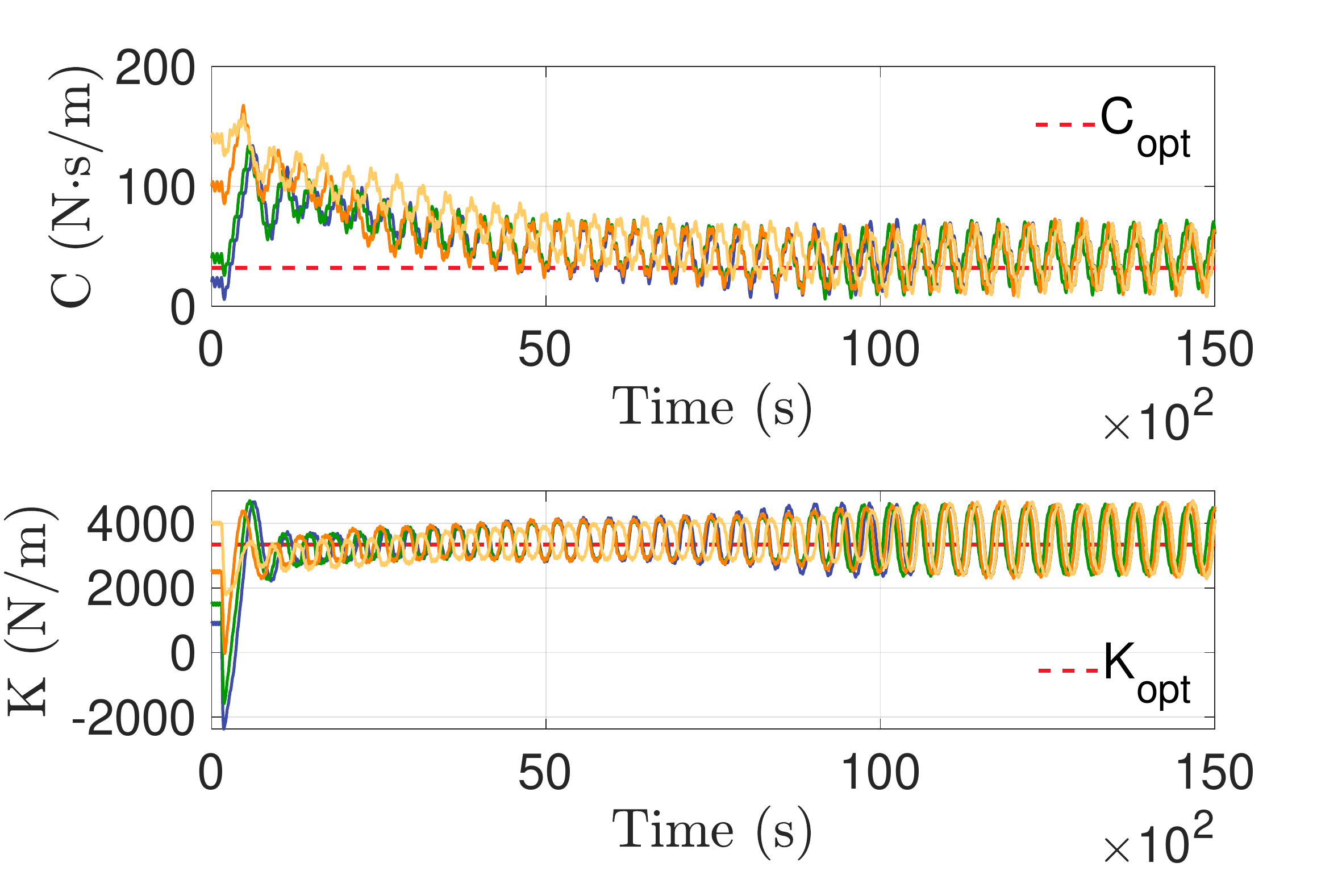}
	}
        \subfigure[Perturbation-based ES]{
        		\includegraphics[scale=0.33]{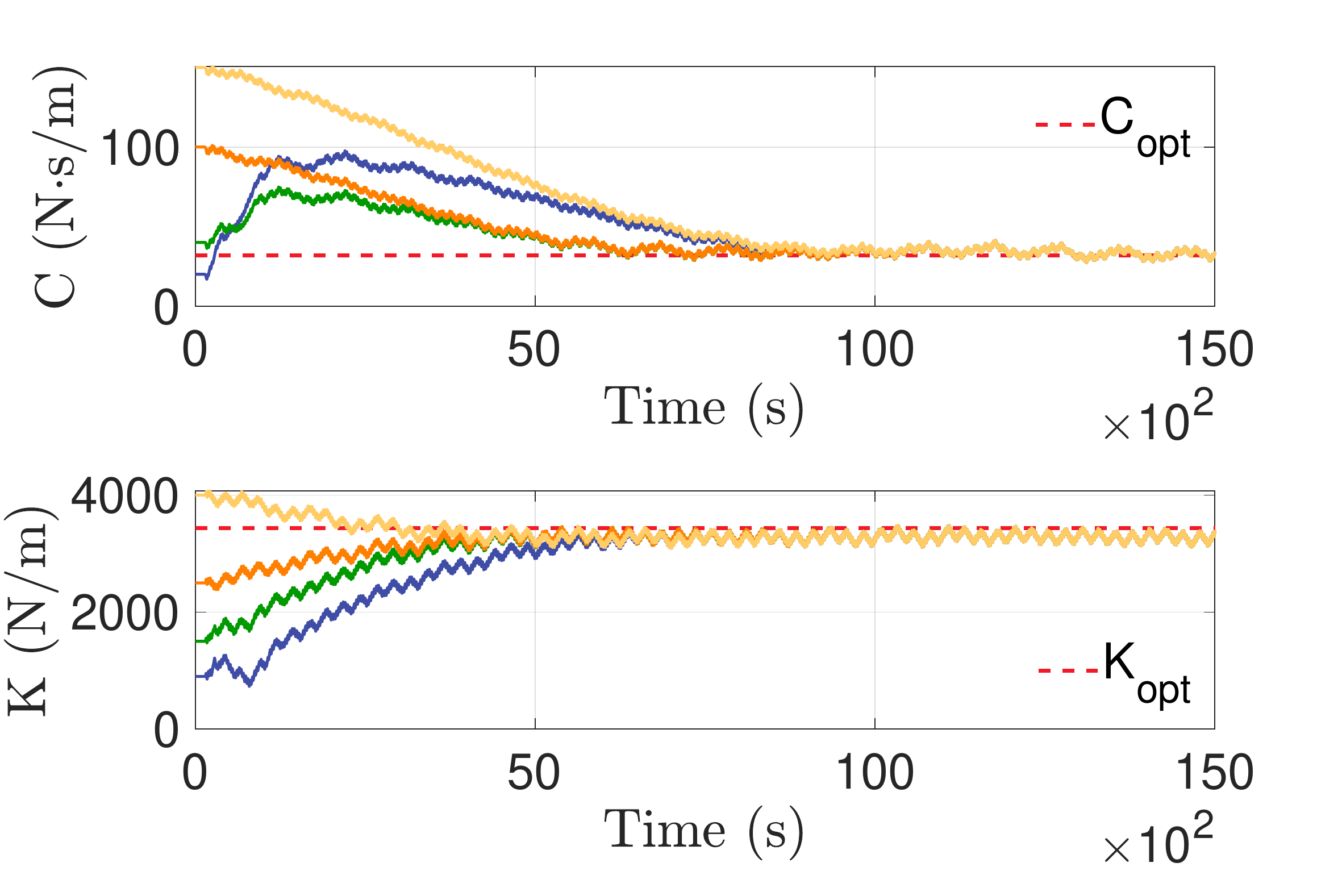}
	}
    \caption{Optimization of reactive and resistive coefficients, $K$ and $C$, respectively, for the cylindrical buoy operating in the irregular sea state ``Irreg.1" using different ES algorithms. The  optimal $K_\text{opt}=3440$ N/m and $C_\text{opt} = 32$ N$\cdot$s/m values are indicated by dashed lines in the plots.}
    \label{fig:cyl_multi_k_b_irreg1}
\end{figure}

\begin{figure}
    \centering
        \subfigure[Sliding mode ES]{
        		\includegraphics[scale=0.33]{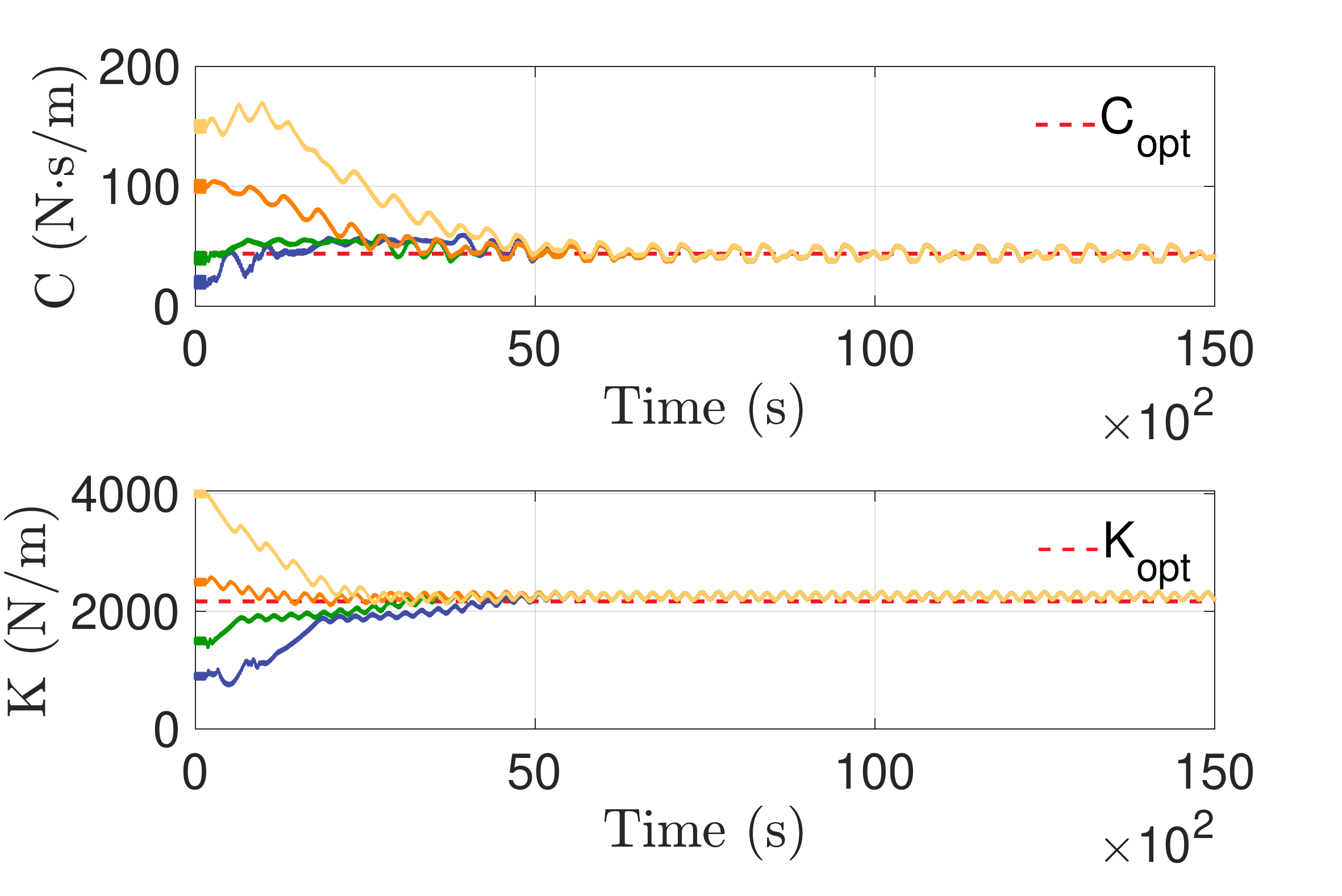}
        } 
        \subfigure[Relay ES]{
        		\includegraphics[scale=0.33]{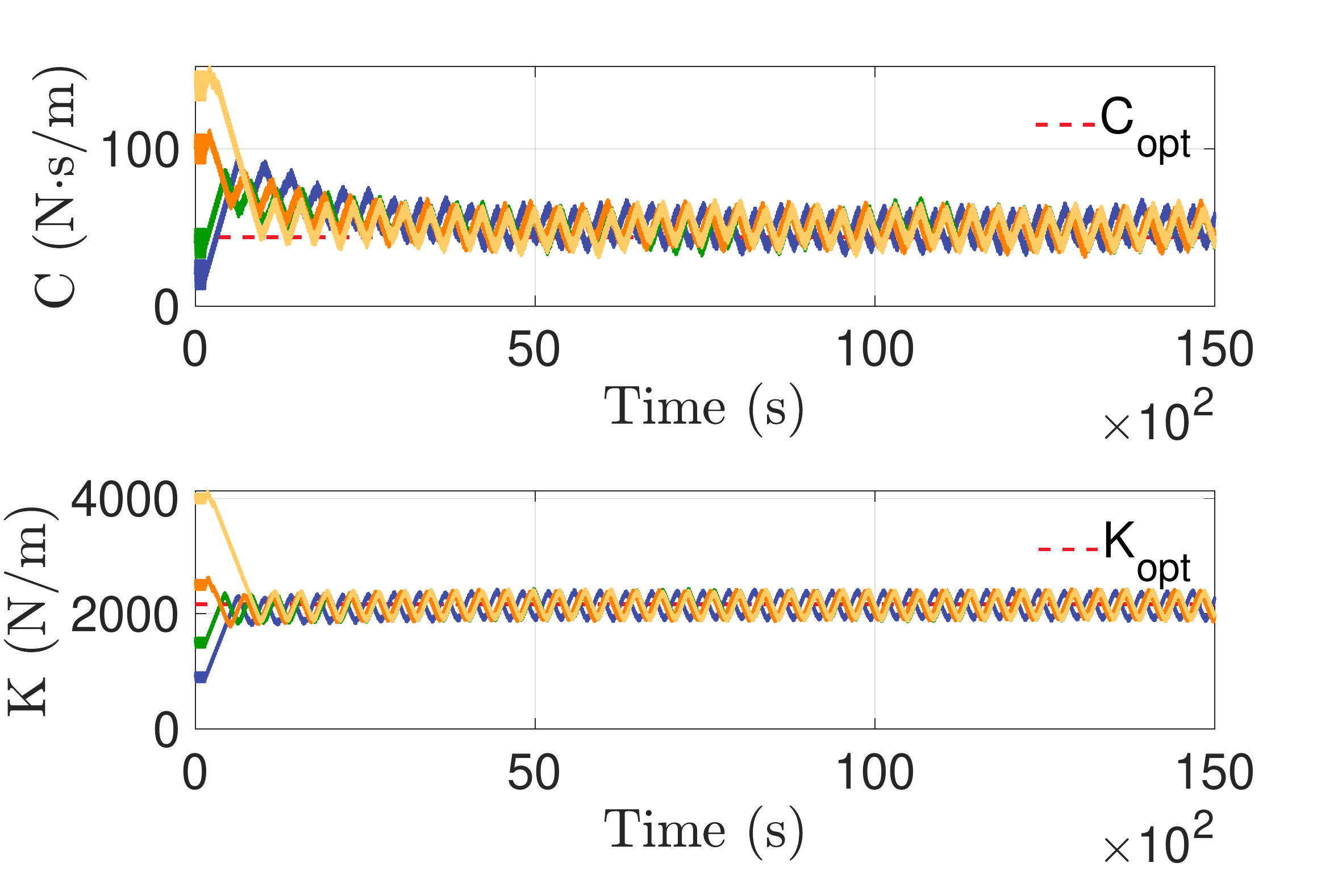}
	} 
        \subfigure[LSQ-ES]{
        		\includegraphics[scale=0.33]{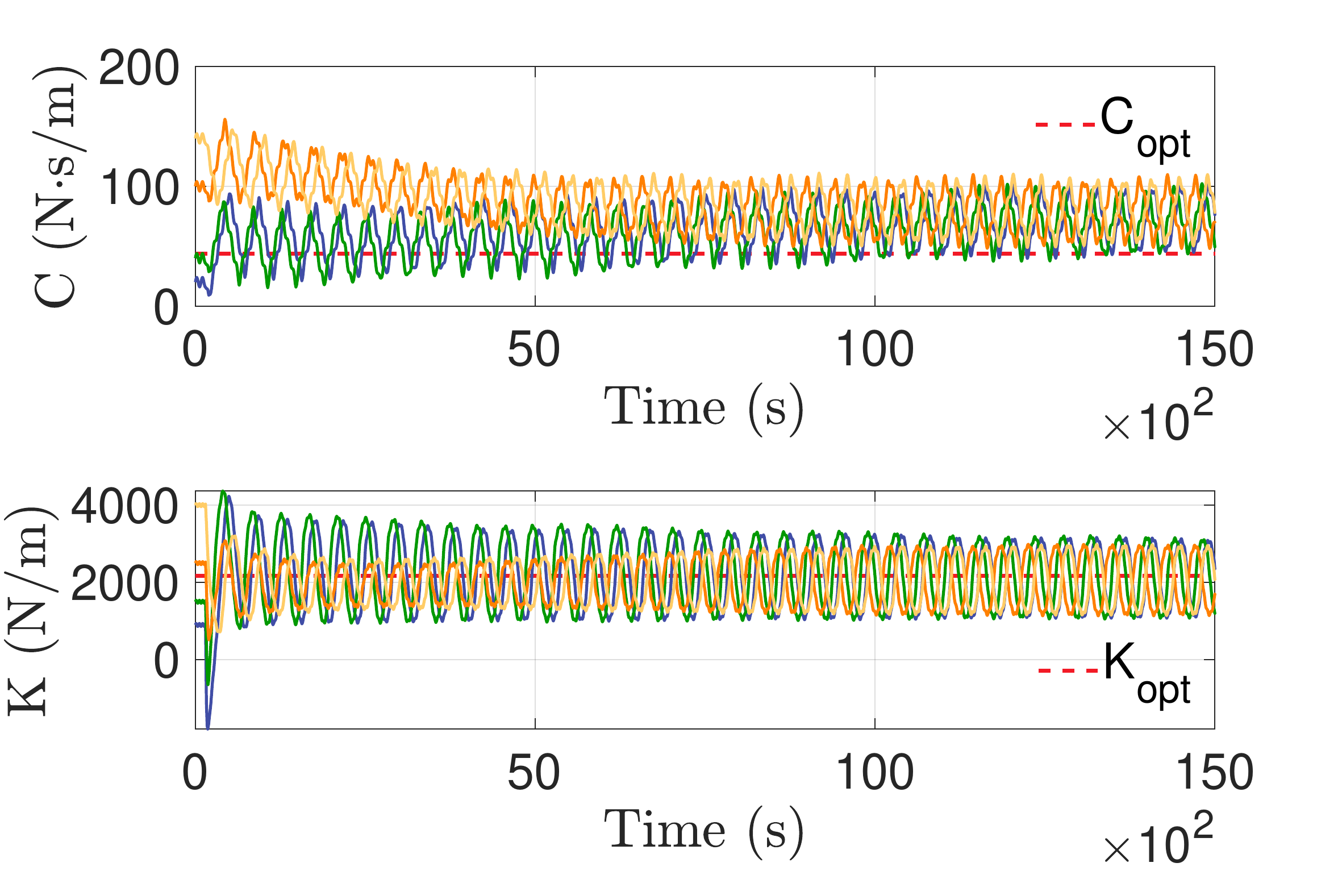}
	}
        \subfigure[Perturbation-based ES]{
        		\includegraphics[scale=0.33]{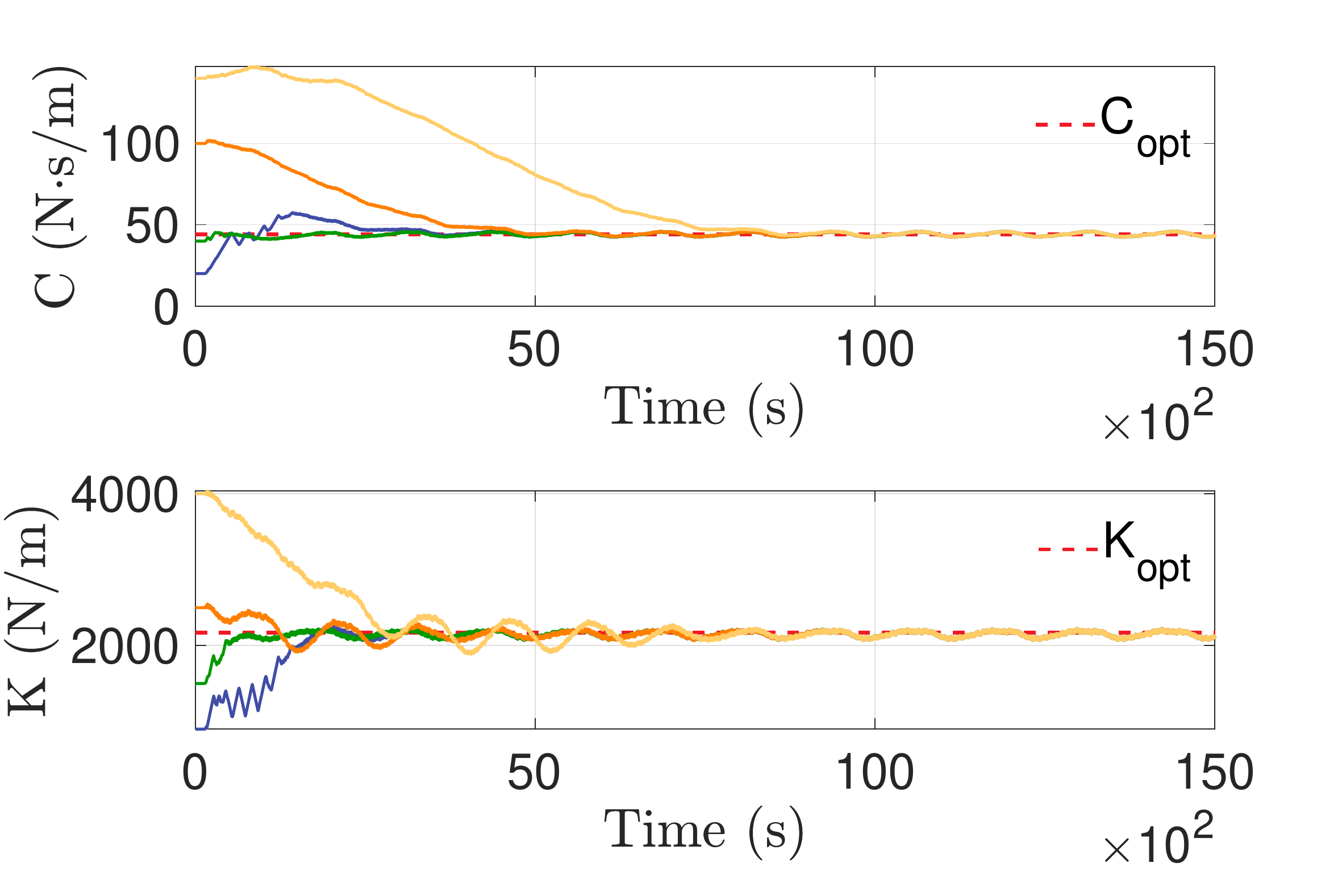}
	}
    \caption{Optimization of reactive and resistive coefficients, $K$ and $C$, respectively, for the cylindrical buoy operating in the irregular sea state ``Irreg.2" using different ES algorithms. The  optimal $K_\text{opt}=2170$ N/m and $C_\text{opt} = 44$ N$\cdot$s/m values are indicated by dashed lines in the plots.}
    \label{fig:cyl_multi_k_b_irreg2}
\end{figure}

\begin{figure}
    \centering
        \subfigure[Sliding mode ES]{
        		\includegraphics[scale=0.33]{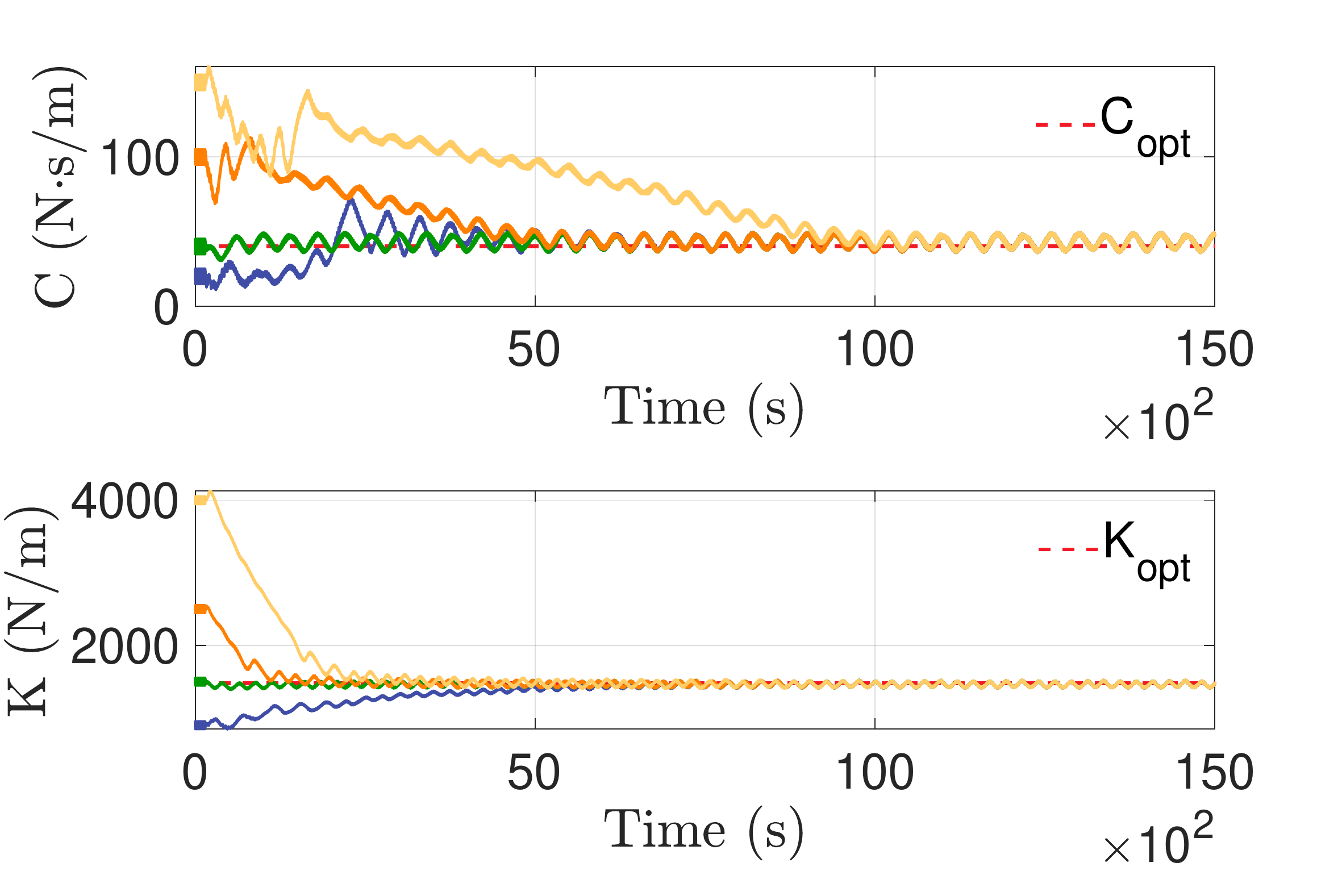}
        } 
        \subfigure[Relay ES] {
        		\includegraphics[scale=0.33]{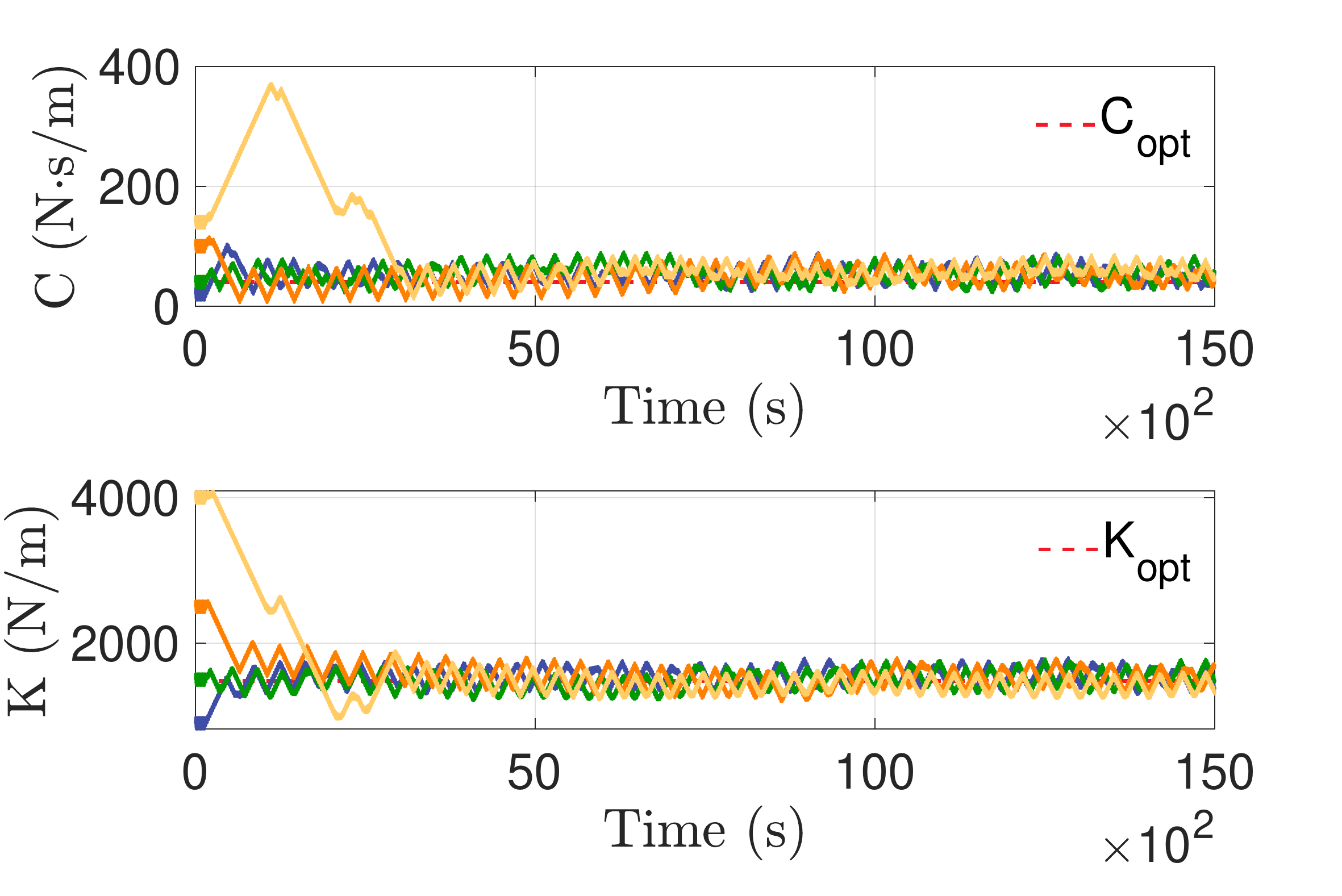}
	} 
        \subfigure[LSQ-ES]{
        		\includegraphics[scale=0.33]{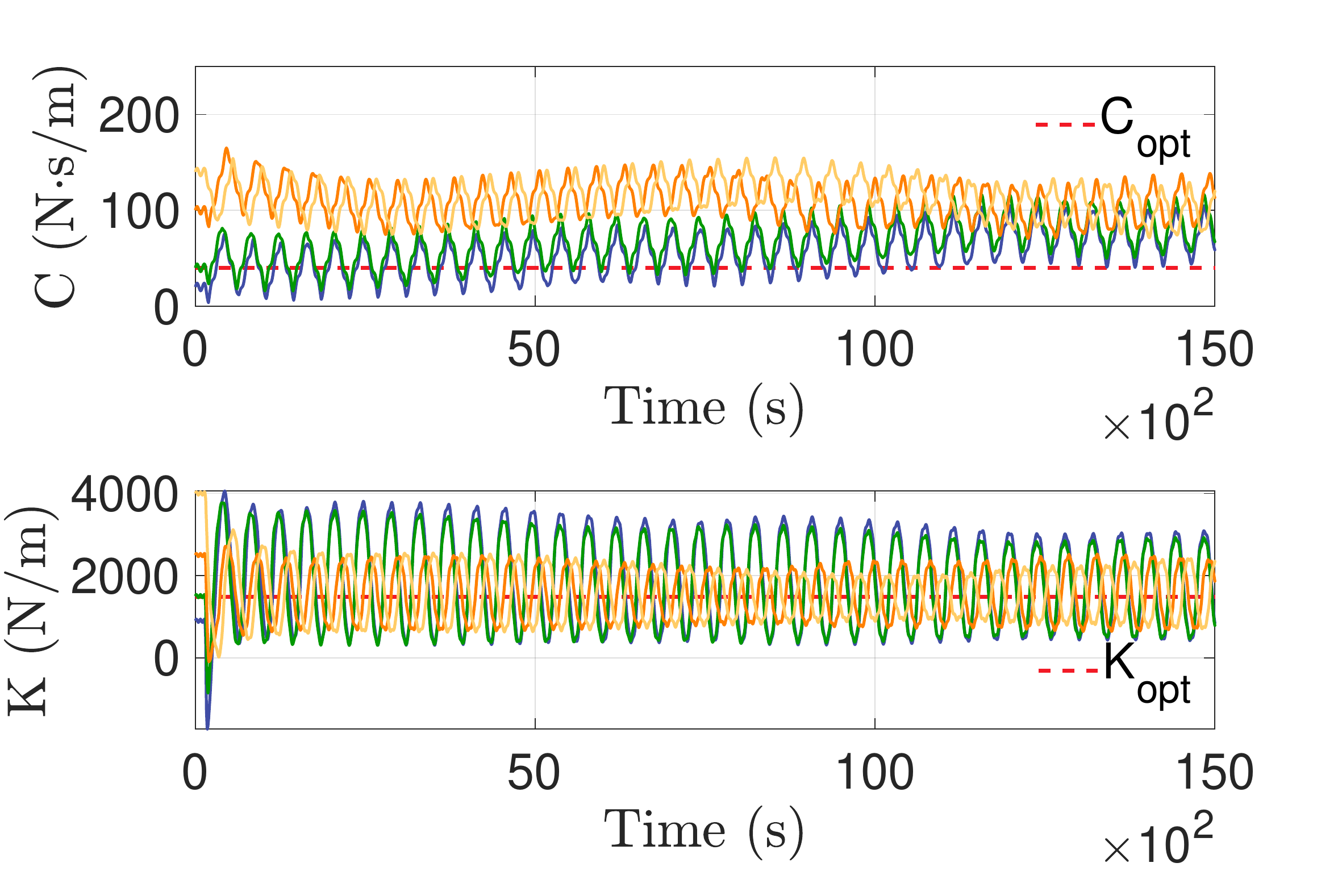}
	}
        \subfigure[Perturbation-based ES]{
        		\includegraphics[scale=0.33]{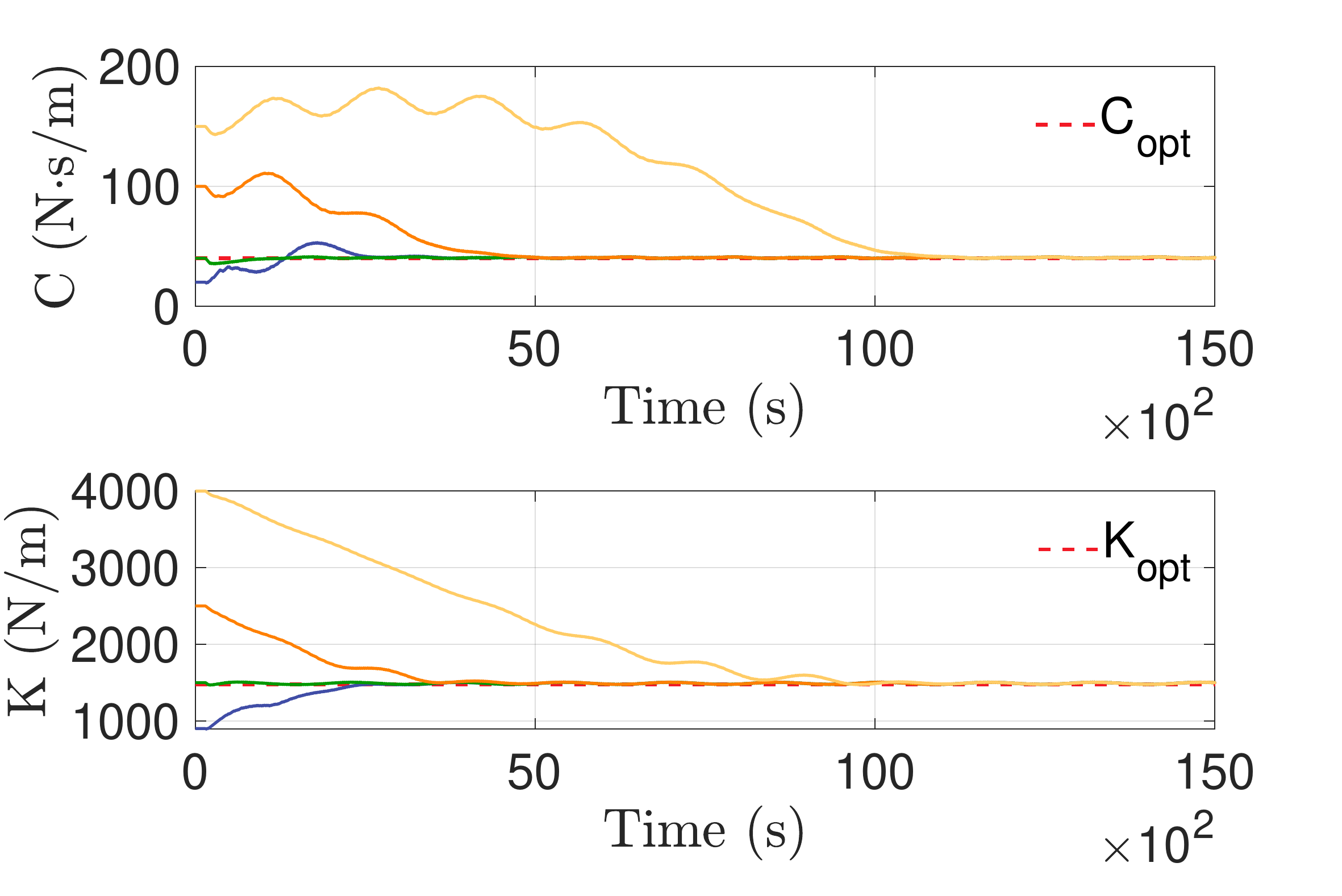}
	}
    \caption{Optimization of reactive and resistive coefficients, $K$ and $C$, respectively, for the cylindrical buoy operating in the irregular sea state ``Irreg.3" using different ES algorithms. The  optimal $K_\text{opt}=1480$ N/m and $C_\text{opt} = 40$ N$\cdot$s/m values are indicated by dashed lines in the plots.}
    \label{fig:cyl_multi_k_b_irreg3}
\end{figure}

Figs.~\ref{fig:cyl_multi_k_b_irreg1},~\ref{fig:cyl_multi_k_b_irreg2}, and~\ref{fig:cyl_multi_k_b_irreg3} show the convergence history of $K$ and $C$ coefficients using four ES algorithms for sea states ``Irreg.1", ``Irreg.2", and ``Irreg.3", respectively.  From the figures, it can be seen that the sliding mode and perturbation-based ES methods outperform the relay and LSQ-ES algorithms. The latter two methods display large oscillations in the steady-state solution. Between the relay and LSQ methods, the former has a better convergence rate. This can be attributed to the fact that the relay ES algorithm does not use the magnitude of the performance gradient, which is noisier compared to its regular sea state counterpart when computed through data buffers. However for regular waves, LSQ-ES convergence is better compared to the relay ES scheme because a more accurate gradient is available in this case; see Fig.~\ref{fig:cyl_multi_k_b_reg1}. 


%% file: SphericalPointAbsorberResults.tex
Intuitively speaking, as we are employing model-free ES schemes to optimize mechanical oscillators in this work, a particular device geometry 
should not matter for the algorithmic success. However, to verify that the model-free ES algorithms also converge for a different device geometry (and consequently, for a different hydrodynamical system), we perform ES optimization of a spherical buoy described in Sec.~\ref{sec_hull_waves}. Sea states ``Reg.1", and ``Irreg.1" are considered for the sliding mode and perturbation-based ES algorithms. 

\begin{figure}[]
    \centering
        \subfigure[Sea state Reg.1]{
        		\includegraphics[scale=0.33]{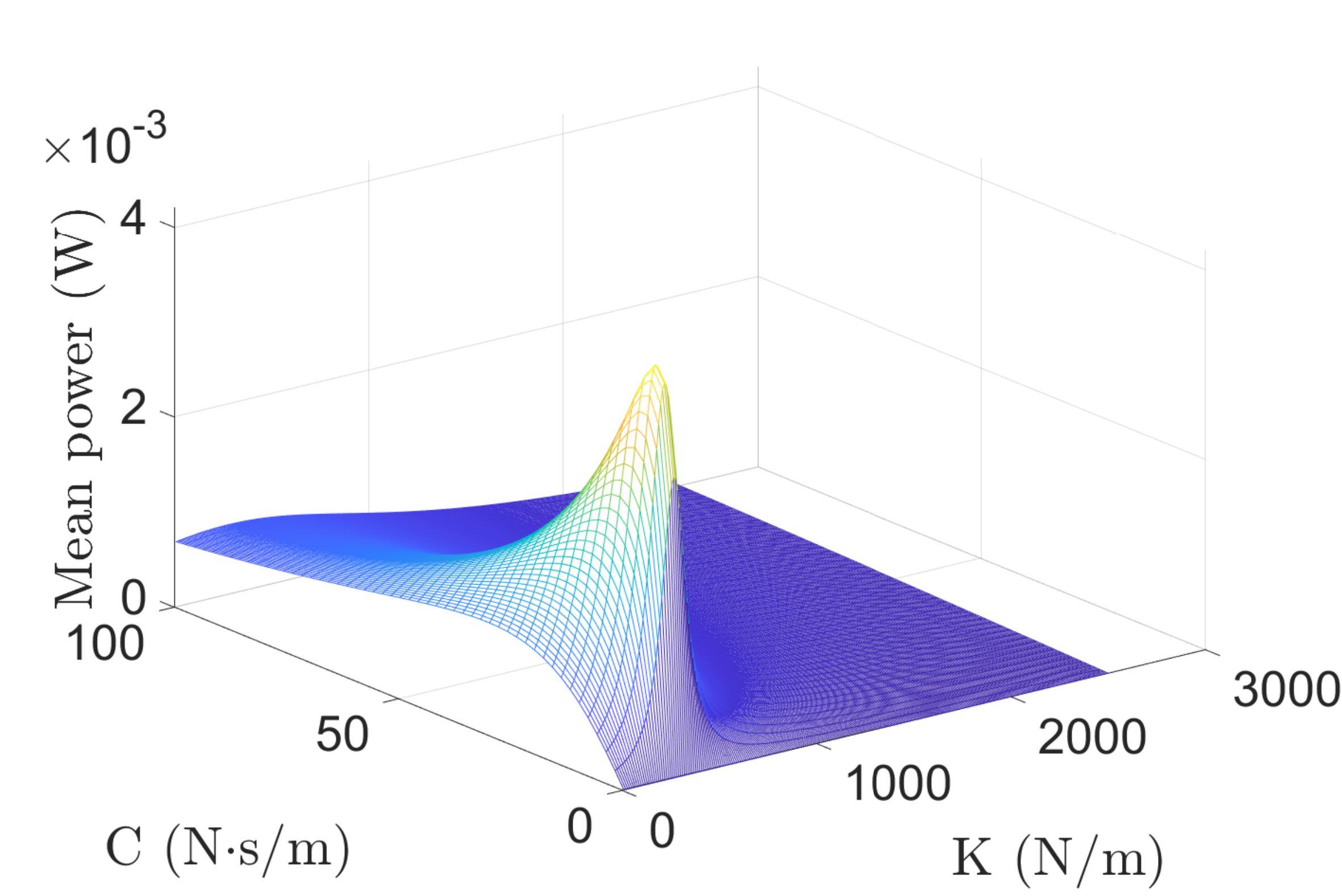}
		\label{subfig:map_sphere_reg1}
	} 
	\subfigure[Sea state Irreg.1]{
        		\includegraphics[scale=0.33]{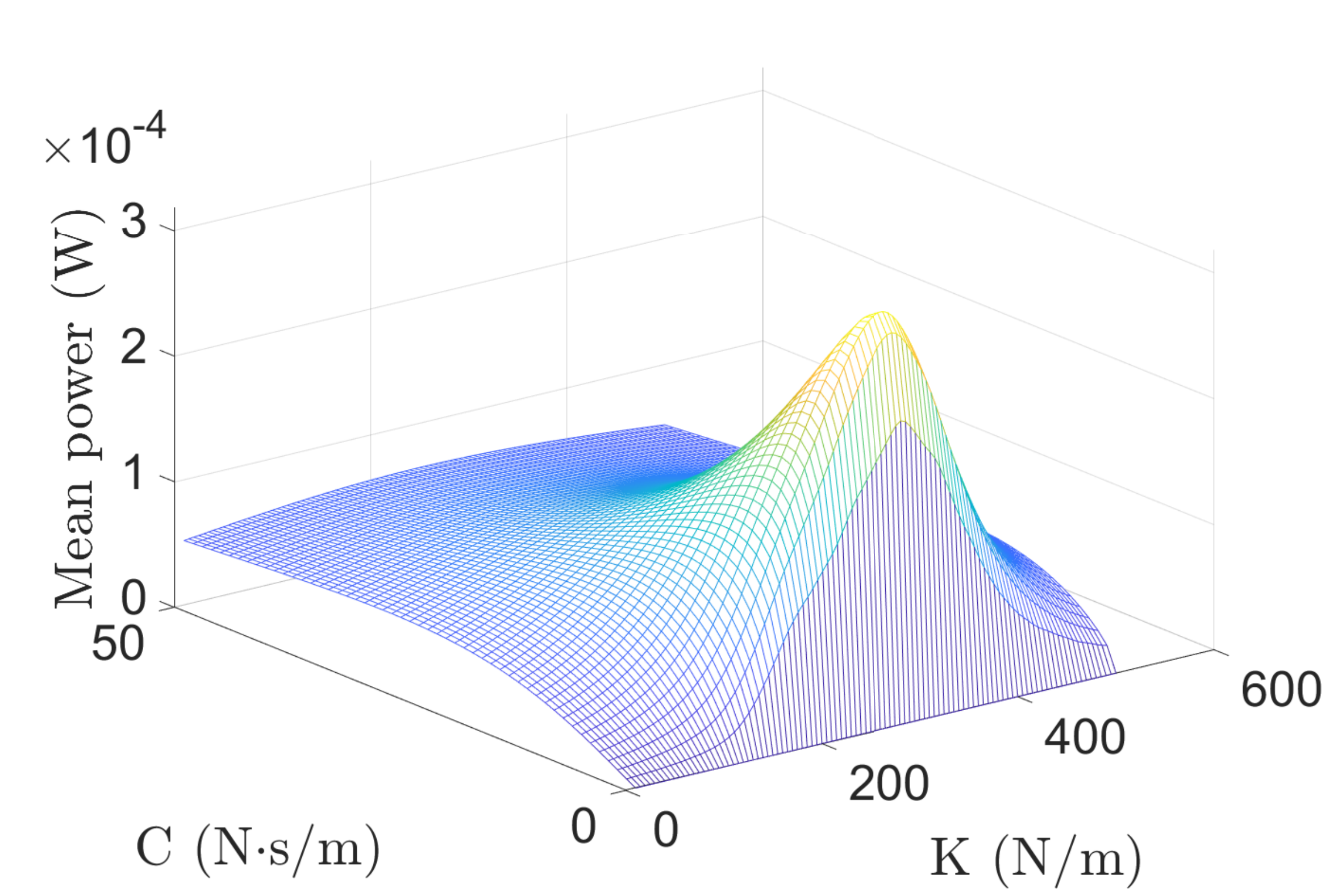}
		\label{subfig:map_sphere_irreg1}
	} 
	\subfigure[Sliding mode ES for Reg.1]{
        		\includegraphics[scale=0.33]{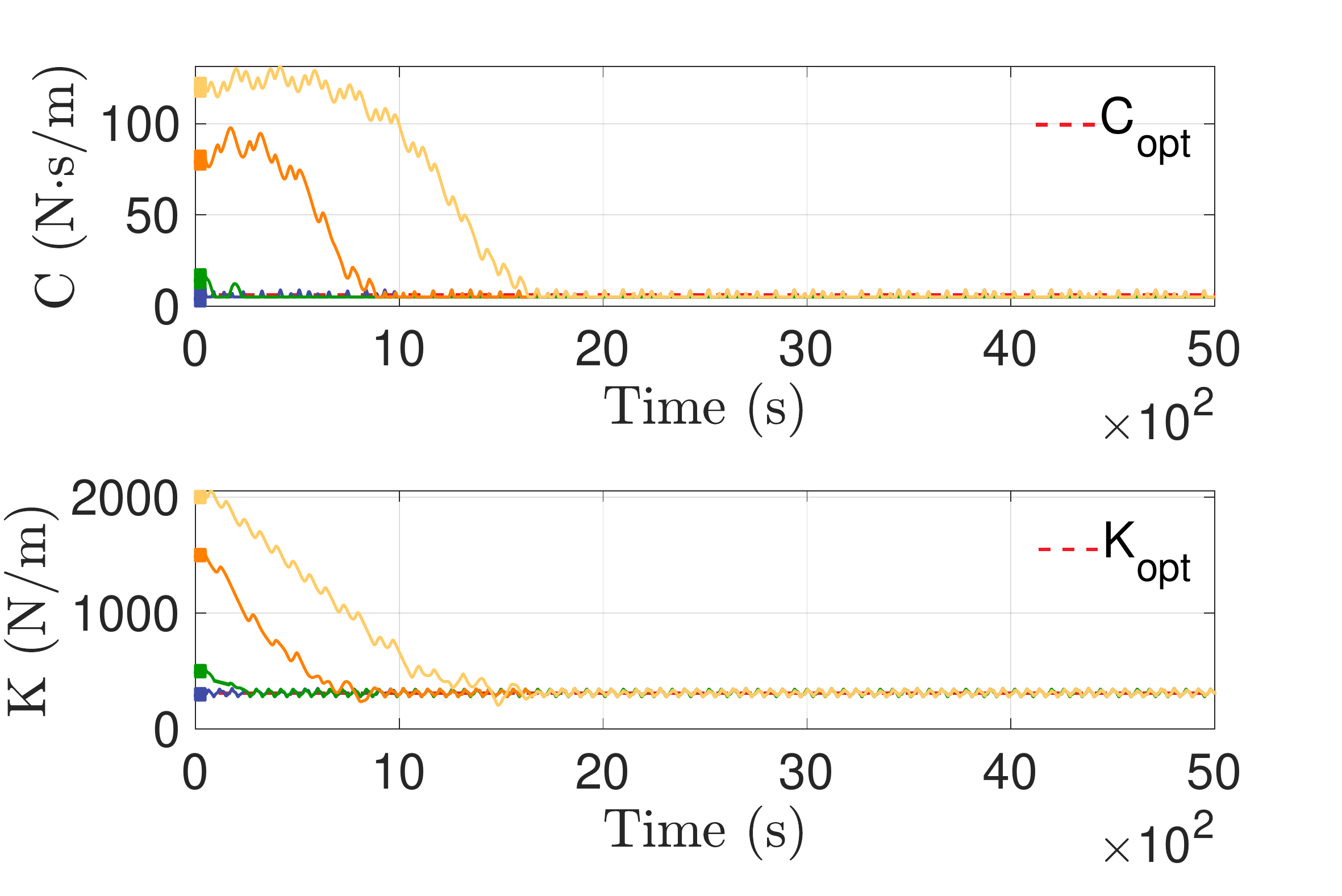}
	}
	\subfigure[Sliding mode ES for Irreg.1]{
        		\includegraphics[scale=.33]{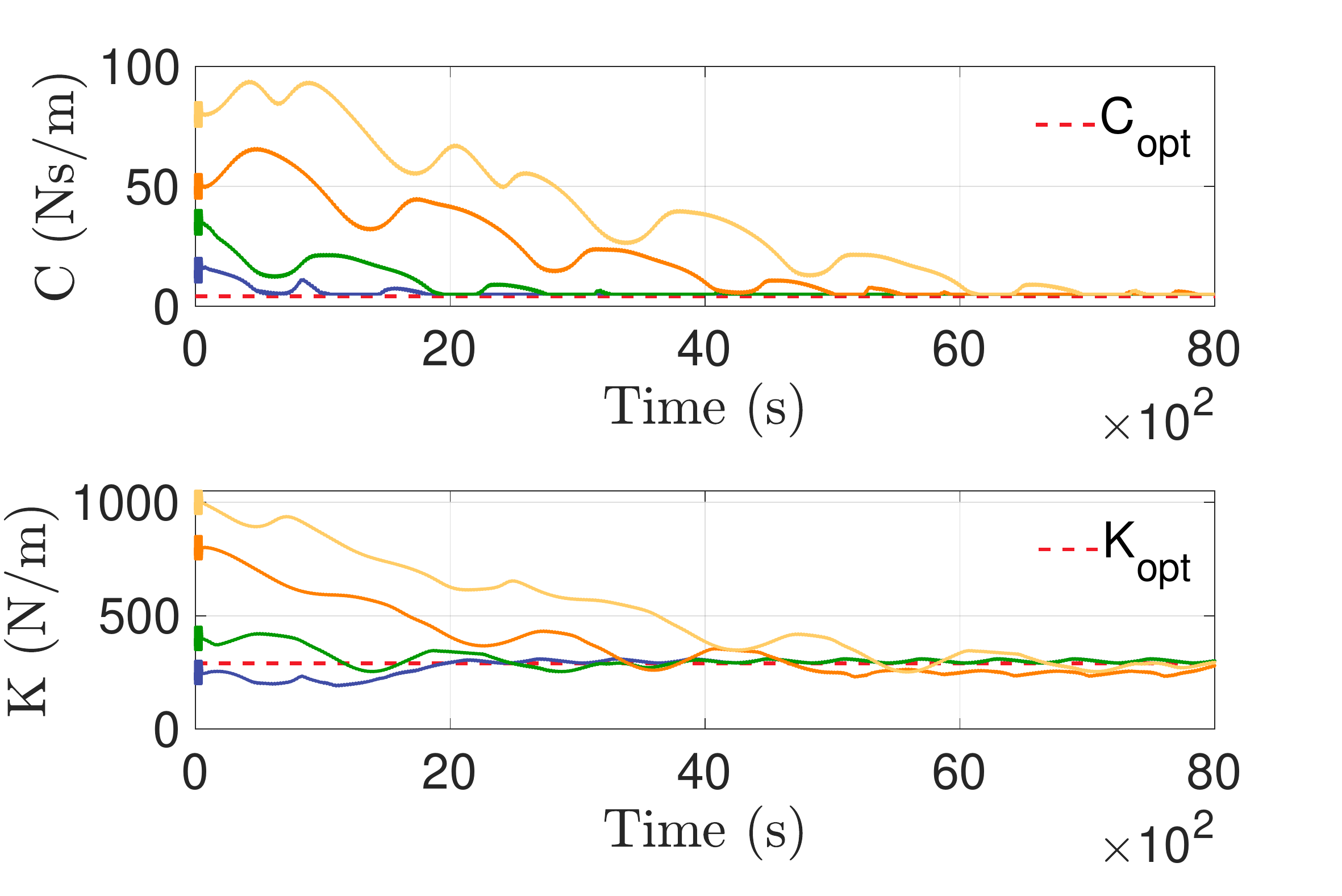}
        } 
	\subfigure[Perturbation-based ES for Reg.1]{
        		\includegraphics[scale=0.33]{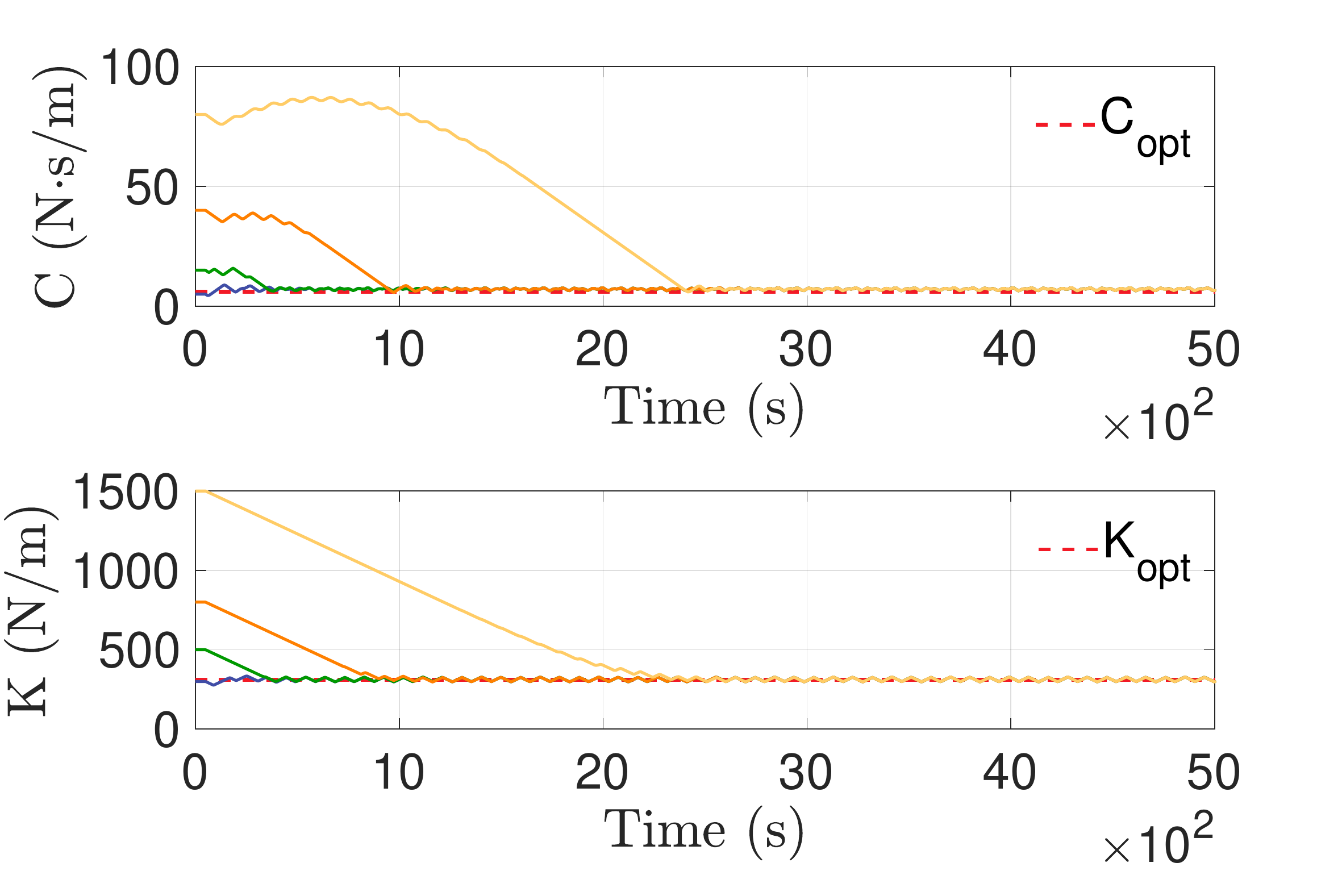}
	} 
        \subfigure[Perturbation-based ES for Irreg.1]{
        		\includegraphics[scale=0.33]{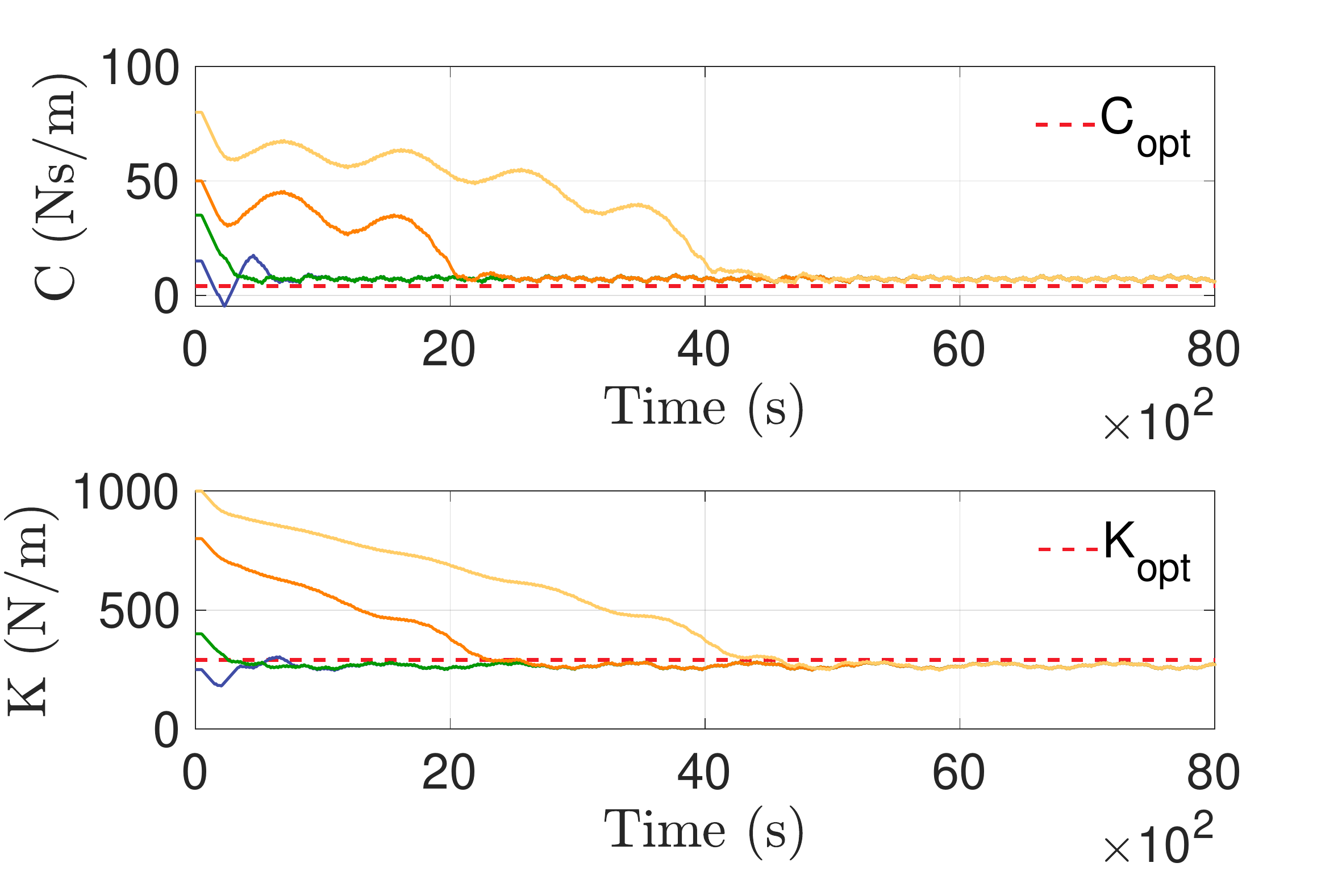}
	}
    \caption{Optimization of reactive and resistive coefficients, $K$ and $C$, respectively, for the spherical buoy operating in regular and irregular sea states ``Reg.1", and ``Irreg.1" using different sliding mode and perturbation-based ES algorithms. For sea state ``Reg.1", the optimal PTO coefficients are $K_\text{opt,map} = 310$ N/m and $C_\text{opt,map} = 6$ N$\cdot$s/m; see~\subref{subfig:map_sphere_reg1}. The  optimal PTO coefficients for sea state ``Irreg 1" are $K_\text{opt,map} = 290$ N/m and $C_\text{opt,map} = 4$ N$\cdot$s/m; see~\subref{subfig:map_sphere_irreg1}. }
    \label{fig:sphere_reg1_irreg1}
\end{figure}

Fig.~\ref{fig:sphere_reg1_irreg1} shows the convergence history of PTO coefficients $K$ and $C$ using sliding mode ES (middle row), and perturbation-based ES (bottom row) algorithms. As can be seen in the figure, steady-state convergence towards performance-optimal solution is achieved with both algorithms (convergence data not shown for relay and LSQ schemes for brevity).  The optimal values obtained from the ES optimization can be confirmed from power vs. PTO coefficients reference-to-output maps (top row).

%% file: Conclusions.tex
In this study, we systematically investigated the feasibility of ES control for wave energy converters to improve their conversion efficiency. Five different ES schemes were tested for heaving WECs: (i) sliding mode ES; (ii) relay ES; (iii) least-squares gradient ES; (iv) self-driving ES; and (v) perturbation-based ES. The optimization problem of wave energy absorption in WECs was formulated in terms of finding the optimal PTO coefficients. Alternatively, the ES optimization problem can also be formulated in terms of finding the optimal PTO force directly, as typically done in model predictive control (MPC) of WECs~\cite{faedo2017optimal}. Direct optimization of PTO force using ES control is deferred to future endeavors. 

The performance function for the ES control was defined as the power absorbed by the PTO system over a given period of time, which could be measured through on-board instrumentation, and does not require any wave measurements. The optimization results were verified against analytical solutions and the extremum of reference-to-output maps. The numerical results demonstrate that except for the self-driving ES algorithm, the other four ES schemes reliably converge for the two-parameter optimization problem. The self-driving ES is more suitable for optimizing a single-parameter of the problem, or when the objective is to find the optimal control force directly, rather than to optimize the gains of the control law.   An advantage of the self-driving ES scheme is that it leads to oscillation-free steady-state solution. The results also show that for an irregular sea state, the sliding mode and perturbation-based ES schemes have better convergence to the optimum in comparison to other ES schemes. The least-squares ES scheme performs better than the relay ES scheme, whenever the gradient estimation through data acquisition is smooth and accurate. This can be concluded by comparing the convergence history of LSQ-ES and relay ES for regular and irregular sea states; LSQ performed better than relay in the case of regular waves and the converse is true for irregular waves. For all ES schemes, the convergence of PTO coefficients towards the performance-optimal values are tested for widely different initial values, in order to avoid bias towards the extremum. We also demonstrated the adaptive capability of ES control by considering a case in which the ES controller adapts to the new extremum automatically amidst changes in the simulated wave conditions.

All extremum-seeking schemes achieve optimum within a single simulation. This allows for a possibility of using model-free ES algorithms within nonlinear computational fluid dynamics (CFD) framework to simulate wave-structure interaction of WECs. In the CFD literature, evolution-based optimization strategies (e.g. genetic algorithm) are predominantly used for solving optimization problems. However, such evolutionary strategies typically require a large number of function evaluations, which can be prohibitively expensive for fully-resolved wave structure interaction problems~\cite{khedkar2020inertial,nangia2019dlm}. We shall consider such an approach in the future.

%% file: Appendix.tex
\section{Power components in performance function} \label{sec_appendix}

\begin{figure}[ht]
    \centering
    \text{Regular sea state Reg.1} \\
        \subfigure[Only resistive power]{
        		\includegraphics[scale=0.33]{figs/CYL_Reg1_perturbation_K_B_multi.pdf}
        } 
        \subfigure[Both reactive and resistive powers] {
        		\includegraphics[scale=0.33]{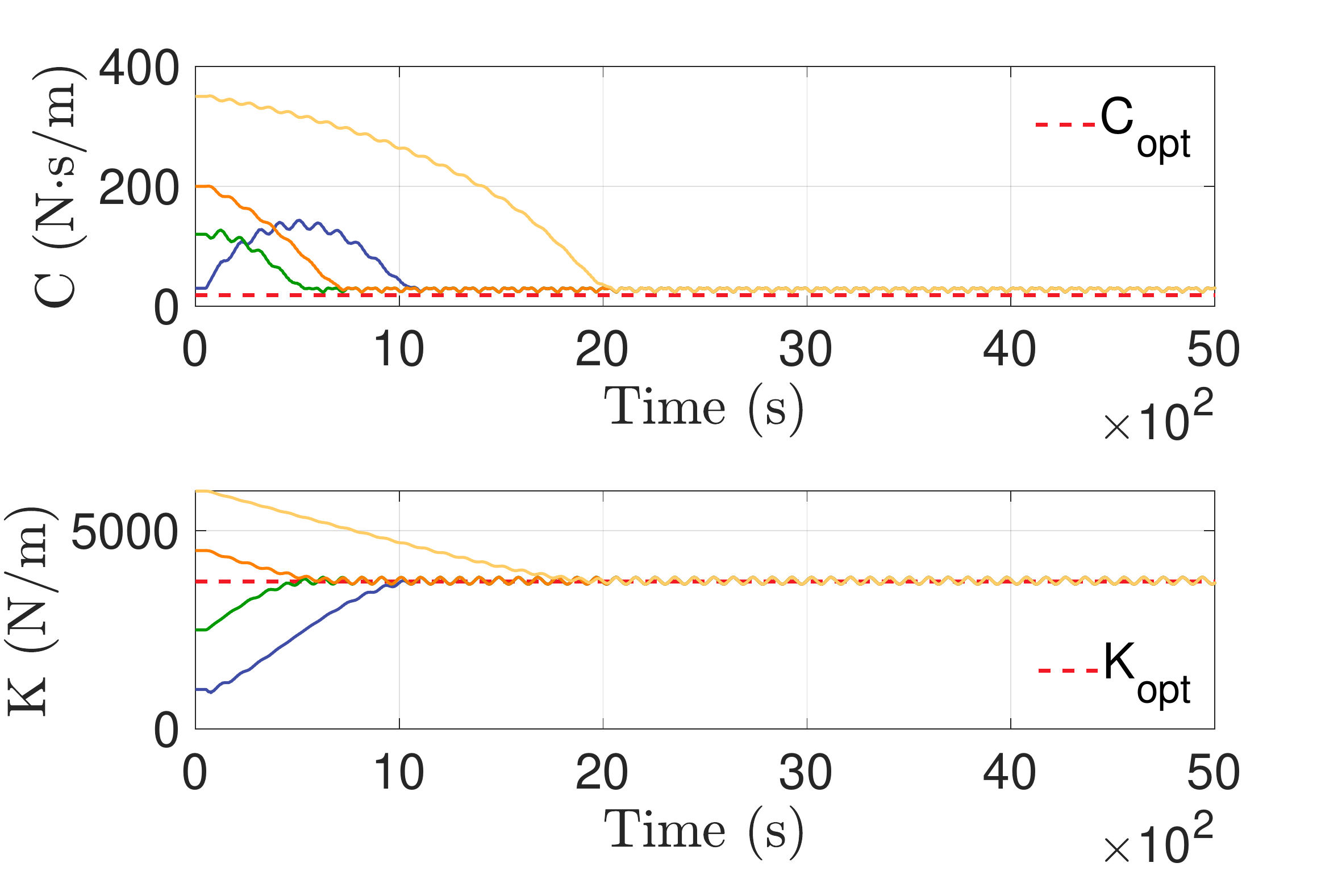}
	} \\ 
	\text{Irregular sea state Irreg.1} \\
        \subfigure[Only resistive power]{
        		\includegraphics[scale=0.33]{figs/CYL_Irreg1_perturbation_K_B_multi.pdf}
        } 
        \subfigure[Both reactive and resistive powers] {
        		\includegraphics[scale=0.33]{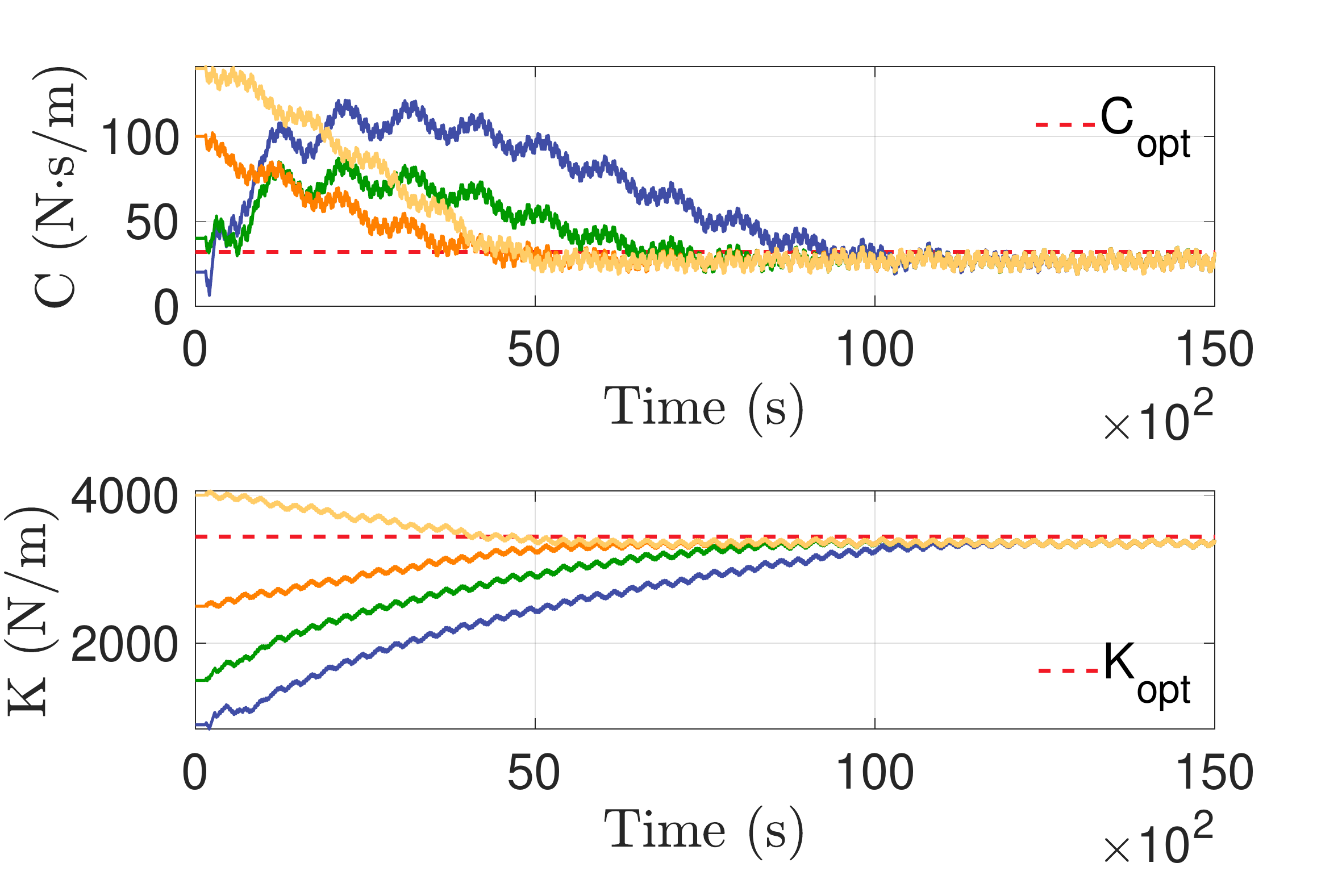}
	} 	
    \caption{Optimization of reactive and resistive coefficients, $K$ and $C$, respectively, for the cylindrical buoy in regular  ``Reg.1" (top row), and irregular ``Irreg.1" (bottom row) waves, using different power components in the performance function. The  optimal $K_\text{opt}=3720$ N/m and $C_\text{opt} = 18$ N$\cdot$s/m values for regular waves, and $K_\text{opt}=3440$ N/m and $C_\text{opt} = 32$ N$\cdot$s/m values for irregular waves are indicated by dashed lines in the plots.}
    \label{fig:cyl_res_react_reg_irreg}
\end{figure}

Here we numerically verify that including the reactive component of power in the performance function does not affect the optimal values of the PTO coefficients obtained from an ES control scheme. We demonstrate this by comparing results in Fig.~\ref{fig:cyl_res_react_reg_irreg} for the cylindrical buoy subject to regular and irregular waves of the type,  ``Reg.1" and  ``Irreg.1", respectively. The perturbation-based ES method is used here. As can be seen in the figure, the inclusion of the reactive power term in the performance function, does not affect the final outcome of the optimization. This was also demonstrated theoretically in Eq.~\eqref{eqn_p_mpc_pd}.

